\documentclass{article}

\usepackage{arxiv}

\usepackage[utf8]{inputenc} 
\usepackage[T1]{fontenc}    
\usepackage{hyperref}       
\usepackage{url}            
\usepackage{booktabs}       
\usepackage{amsfonts}       
\usepackage{nicefrac}       
\usepackage{microtype}      
\usepackage{lipsum}
\usepackage{graphicx}
\graphicspath{ {./images/} }
\usepackage{mathtools}
\usepackage{xcolor}
\usepackage{bm}
\usepackage{rotating}

\ifpdf
  \DeclareGraphicsExtensions{.eps,.pdf,.png,.jpg}
\else
  \DeclareGraphicsExtensions{.eps}
\fi

\usepackage{algorithm}
\usepackage{algpseudocode}
\usepackage{threeparttable}

\usepackage{caption}
\usepackage{subcaption}
\usepackage{amssymb}
\usepackage{amsmath}
\usepackage{multirow}
\usepackage{siunitx}

\usepackage[shortlabels]{enumitem}
\setlist[enumerate]{leftmargin=.5in}
\setlist[itemize]{leftmargin=.5in}

\newenvironment{codefont}{\fontfamily{lmtt}\selectfont}{\par}
\DeclareTextFontCommand{\codetext}{\codefont}

\DeclareMathOperator{\atantwo}{atan2}

\newtheorem{definition}{Definition}

\title{The tail wags the distribution: Only sample the tails for efficient reliability analysis}

\author{
 Promit Chakroborty \\
  Department of Civil and Systems Engineering\\
  Johns Hopkins University\\
  Baltimore, MD 21218 \\
  \texttt{pchakro1@jhu.edu} \\
   \And
 Michael D. Shields \\
  Department of Civil and Systems Engineering\\
  Johns Hopkins University\\
  Baltimore, MD 21218 \\
  \texttt{michael.shields@jhu.edu} \\
}

\begin{document}
\maketitle

\begin{abstract}
To ensure that real-world infrastructure is safe and durable, systems are designed to not fail for any but the most rarely occurring parameter values. By only happening deep in the tails of the parameter distribution, failure probabilities are kept small. At the same time, it is essential to understand the risk associated with the failure of a system, no matter how unlikely. However, estimating such small failure probabilities is challenging; numerous system performance evaluations are necessary to produce even a single system state corresponding to failure, and each such evaluation is usually significantly computationally expensive. To alleviate this difficulty, we propose the Tail Stratified Sampling (TSS) estimator - an intuitive stratified sampling estimator for the failure probability that successively refines the tails of the system parameter distribution, enabling direct sampling of the tails, where failure is expected to occur. The most general construction of TSS is presented, highlighting its versatility and robustness for a variety of applications. The intuitions behind the formulation are explained, followed by a discussion of the theoretical and practical benefits of the method. Various details of the implementation are presented. The performance of the algorithm is then showcased through a host of analytical examples with varying failure domain geometries and failure probabilities as well as multiple numerical case studies of moderate and high dimensionality. To conclude, a qualitative comparison of TSS against the existing foundational variance-reduction methods for reliability analysis is presented, along with suggestions for future developments.
\end{abstract}


\section{Introduction \& Background}
\label{section:Introduction}

Reliability analysis is the formal process of estimating failure probabilities for engineering systems given their input uncertainties. It is a critical branch of engineering that guides the safety and serviceability of systems ranging from aircraft to nuclear power plants. Given the criticality of these systems, it is essential that their probability of failure be sufficiently low to avoid the catastrophic consequences of failure. But evaluating probability of failure is a daunting task with potentially huge computational burden that requires simulating system failures that often occur under extreme conditions having low probability of occurrence associated with the tails of the input probability distribution. This paper introduces an efficient Monte Carlo technique to estimate probability of failure by concentrating samples in the region of these extreme values, through a careful stratification of the tail region of the input distribution.

\subsection{Problem Statement}
\label{section:problem_statement}

Given the random vector $\mathbf{X}$ defined on the sample space $\Omega$ such that $\mathbf{X}:\Omega \to \mathbb{R}^d$ and having probability density function $f_{\mathbf{X}}(\mathbf{x})$, we define the performance function of an engineering system by $g(\mathbf{x})$ such that $g(\mathbf{x}) \leq 0$ corresponds to system failure, where $ \mathbf{x} $ is a realization of $ \mathbf{X} $. Evaluating the performance function $g(\mathbf{x})$ typically requires performing expensive numerical simulations of the highly complex and nonlinear system. We define the \textit{failure domain}, $\Omega_\mathcal{F}$, as the subset of $\Omega$ such that $g(\mathbf{x}) \leq 0$, i.e. $\Omega_\mathcal{F} = \{\mathbf{x} \in \Omega: g(\mathbf{x}) \leq 0\}$. This allows us to compute the probability of failure as
\begin{equation}
    P_{\mathcal{F}} = \int_{\Omega_{\mathcal{F}}}f_{\mathbf{X}}(\mathbf{x})d\mathbf{x}, 
    \label{eqn:PF}
\end{equation}
which can be simplified as follows to account explicitly for the failure condition $g(\mathbf{x}) \leq 0$:
\begin{equation}
    \begin{aligned}
    P_{\mathcal{F}} & = \int_{g(\mathbf{x}) \leq 0}f_{\mathbf{X}}(\mathbf{x})d\mathbf{x}\\
    & = \int_{\Omega} \text{I}_{\left\{g(\mathbf{X}) \leq 0\right\}} (\mathbf{x}) f_{\mathbf{X}}(\mathbf{x})d\mathbf{x}\\
    & = \mathbb{E}_f \left[\text{I}_{\Omega_{\mathcal{F}}} (\mathbf{X}) \right],
    \end{aligned}
    \label{eqn:PF_exp}
\end{equation}
where $ \text{I}_{\Omega_\mathcal{F}} (\mathbf{x}) = \text{I}_{\left\{ g(\mathbf{X}) \leq 0\right\}} (\mathbf{x})$ is the indicator function taking value 1 if $g(\mathbf{x}) \leq 0$ and zero otherwise. 

\subsection{Challenges and Previous Solutions}
\label{section:challenges_and_past_literature}

Numerous challenges arise in probability of failure estimation that derive directly from the problem statement above. The root of these challenges is that $g(\mathbf{x})$ is not available in a simple analytical form and is computationally demanding to evaluate. This makes the exact solution of the integral in Eq.~\eqref{eqn:PF} (expectation in Eq.~\eqref{eqn:PF_exp}) impossible. Instead, using Eq.~\eqref{eqn:PF_exp}, probability of failure can be estimated in a robust manner using Monte Carlo simulation by
\begin{equation}
    \hat{P}^{MC}_{\mathcal{F}}  \approx \dfrac{1}{N} \sum_{k=1}^N \text{I}_{\left\{g(\mathbf{x}) \leq 0\right\}} (\mathbf{x}_k) \quad \text{for } N \to \infty \; \text{, where } \mathbf{x}_k \sim f_{\mathbf{X}} (\mathbf{x}) \; \forall k
    \label{eqn:PF_MC}
\end{equation}

Of course, conventional Monte Carlo simulation is usually impractical because it requires a large number of simulations $N$. This is exacerbated by the fact that $P_\mathcal{F}$ is, by construction of the engineering system, very small. As previously mentioned, we require the system to have a small probability of failure to avoid the consequences of system failure. In the Monte Carlo setting, estimating small probabilities requires $N$ to be extremely large. 

Of further interest to this work is that failure is often associated with a priori unknown extreme values of the random vector $\mathbf{X}$ that have a very small probability of occurrence and are generally located in the tails of the probability distribution. 
Consequently, a great deal of effort in reliability analysis is dedicated to simply finding the failure region $\Omega_{\mathcal{F}}$, which is often deep in the tail of $f_{\mathbf{X}}(\mathbf{x})$. The challenge is compounded by the fact that, once $\Omega_{\mathcal{F}}$ has been found, probability of failure estimation requires understanding its geometry and quantifying its probability. For this reason, we endeavor to formally define the tail of a probability distribution in later sections.

Finally, probability of failure estimation is made more difficult by the fact that $\mathbf{X}$ is often high-dimensional, i.e., $d$ is large. Although this is not an intrinsic problem for Monte Carlo estimates of the form in Eq.~\eqref{eqn:PF_MC}, which are independent of dimension, it becomes a practical challenge when implementing variance reduction techniques or numerical methods that do not scale well to high dimensions. 

The aforementioned challenges have been addressed extensively in the literature, and yet they remain. Due to their extensive study, we cannot possibly provide an exhaustive review of the literature. Instead, we will focus on two primary classes of approaches: (1) Those that aim to approximate $g(\mathbf{x})$ through a simpler mathematical model; and (2) those that apply the broad class of variance reduction methods for Monte Carlo simulation. Among the first and most widely used probability of failure estimators are the First Order Reliability Method (FORM) and Second Order Reliability Method (SORM), that approximate the performance function $g(\mathbf{x})$ using a Taylor series expansion~\cite{hasofer1974exact}. Although FORM/SORM are not specifically discussed here, they both rely on identifying the most probable point (MPP) or design point -- that is, the point along the limit surface $g(\mathbf{x})=0$ with the highest probability. Efficient algorithms exist to find the MPP that can be advantageous for the proposed approach and will therefore be discussed in some detail later.  
More recently, a vast number of methods have aimed to approximate the performance function using surrogate models including Gaussian process regression / Kriging~\cite{bichon2008efficient,echard2011ak}, polynomial chaos expansions~\cite{sudret2012meta, marelli2018active}, artificial neural networks~\cite{hurtado2001neural,papadopoulos2012accelerated} and others. These methods may also be beneficial when combined with the proposed approach, but this falls beyond the scope of this work and will not be discussed further here.

Variance reduction schemes aim to reduce the variance of the Monte Carlo estimator in Eq.~\eqref{eqn:PF_MC}, thereby allowing it to be computed more efficiently with fewer samples $N$. Numerous variance reduction schemes have been developed for reliability analysis, such as subset simulation~\cite{au2001estimation}, importance sampling~\cite{au1999new}, directional simulation~\cite{bjerager1988probability}, and various stratified sampling methods including the widely used Latin hypercube sampling~\cite{olsson2003latin}. Most of these methods share the common attribute that they aim to concentrate samples in the vicinity of the failure domain, but they do so in different ways. The proposed tail stratified sampling (TSS) method shares this attribute as well. 

Subset simulation directs samples toward the failure domain by decomposing the full domain $ \Omega $ into a sequence of nested intermediate failure domains, $\Omega_{\mathcal{F}_i}, i = 0,\dots, M-1$, which transforms the probability of failure estimator into the following product of conditional probabilities:
\begin{equation}
    P_{\mathcal{F}} = \mathcal{P}\left(\bigcap_{i=0}^{M-1} \Omega_{\mathcal{F}_i}\right) = \mathcal{P}(\Omega_{\mathcal{F}_0}) \prod_{i=1}^{M-1} \mathcal{P}(\Omega_{\mathcal{F}_i}|\Omega_{\mathcal{F}_{i-1}})
    \label{eqn:subset_simulation}
\end{equation}
Each conditional probability $\mathcal{P}(\Omega_{\mathcal{F}_i}|\Omega_{\mathcal{F}_{i-1}})$
is estimated using Monte Carlo simulation by drawing conditional samples using Markov Chain Monte Carlo (MCMC) methods~\cite{papaioannou2015mcmc, wang2019hamiltonian,shields2021subset}. Although subset simulation is perhaps the most widely used Monte Carlo reliability method, it has some notable drawbacks. In particular, researchers have recently highlighted that, while it is asymptotically unbiased and statistically consistent, individual realizations of subset simulation may miss failure domains altogether -- especially when the performance function is complex and involves several disjoint failure regions~\cite{breitung2019geometry,vovrechovsky2022reliability}. Furthermore, the multiplicative decomposition employed in subset simulation results in estimates with large coefficient of variation. In fact, the variance of the estimator grows as the failure probability diminishes (i.e., as the number of subsets grows). Finally, it can struggle to converge for problems in high dimensions due to the difficulty of effective MCMC sampling in high dimensions.  As we'll see, the proposed TSS estimator relies on a similar but additive decomposition of conditional probabilities that (in most practical cases) does not require MCMC sampling, has no explicit dimension dependence, and, importantly, whose variance does not depend on the magnitude of the failure probability. 

Another widely used method, importance sampling, directs samples toward the failure domain by defining an alternate importance sampling density, $q_{\mathbf{X}}(\mathbf{x})$, whose probability density is concentrated near $\Omega_{\mathcal{F}}$. By drawing samples from $q_{\mathbf{X}}(\mathbf{x})$, the probability of failure can be computed by
\begin{equation}
\label{eqn:IS_general_defn}
\begin{aligned}
     P_{\mathcal{F}} & = \int_\Omega \text{I}_{\left\{g(\mathbf{x}) \leq 0\right\}} (\mathbf{x}) \dfrac{f_\mathbf{X}(\mathbf{x})}{q_\mathbf{X}(\mathbf{x})}q_\mathbf{X}(\mathbf{x})d\mathbf{x} \\
     & = \mathbb{E}_q\left[ \text{I}_{\left\{g(\mathbf{x}) \leq 0\right\}} (\mathbf{X}) \dfrac{f_\mathbf{X}(\mathbf{X})}{q_\mathbf{X}(\mathbf{X})} \right]
\end{aligned}
\end{equation}
leading to the Monte Carlo estimator given by
\begin{equation}
\label{eqn:IS_monte_carlo_estimator}
     \hat{P}^{IS}_{\mathcal{F}} \approx \dfrac{1}{N} \sum_{k=1}^N \text{I}_{\left\{g(\mathbf{x}) \leq 0\right\}} (\mathbf{x}_k) \dfrac{f_\mathbf{X}(\mathbf{x}_k)}{q_\mathbf{X}(\mathbf{x}_k)} = \dfrac{1}{N} \sum_{i=1}^N w_k \text{I}_{\left\{g(\mathbf{x}) \leq 0\right\}} (\mathbf{x}_k)
\end{equation}
where $w_k = \dfrac{f_\mathbf{X}(\mathbf{x}_k)}{q_\mathbf{X}(\mathbf{x}_k)}$ represent individual sample weights, and $ \mathbf{x}_k \sim q_{\mathbf{X}} (\mathbf{x}) $ $ \forall k $. Importance sampling has been widely used for a diverse miscellany of reliability problems. The innate challenge of importance sampling is defining the appropriate importance sampling density, $q_{\mathbf{X}}(\mathbf{x})$. Numerous approaches have been employed, with some recent research focused on sequential importance sampling methods that employ tempering distributions sampled with MCMC to converge toward an appropriate importance sampling density over $\Omega_{\mathcal{F}}$~\cite{papaioannou2016sequential}. Moreover, importance sampling is known to struggle for high-dimensional problems~\cite{katafygiotis2008geometric} although several works have explored how to mitigate these challenges~\cite{au2003important,wang2016cross}. Like importance sampling, the proposed TSS method relies on weighted sample evaluations, although it does not require the arbitrary specification of an alternate sampling density and, again, it has no explicit dimension dependence. Instead, the weights in the proposed method are derived from stratified sampling, as discussed next.  

\subsection{Stratified Sampling}
\label{section:stratified_sampling}

Stratified sampling~\cite{tocher1967art} is an important class of variance reduction methods that includes the ubiquitously used Latin hypercube sampling~\cite{mckay2000comparison,helton2003latin,shields2016generalization}. The stratified sampling estimator is based on the decomposition of the sample space $\Omega$ into a partition of $M$ disjoint strata $\Omega_i$ such that $\bigcup_{i=1}^M \Omega_i = \Omega$ and $\Omega_i \cap \Omega_j = \emptyset, \forall i \neq j$. Employing the Law of Total Probability, the stratified sampling estimator for $P_{\mathcal{F}}$ can be expressed as
\begin{equation}
    P_{\mathcal{F}} = \sum_{i=1}^M \mathcal{P}\left(\Omega_{\mathcal{F}}\cap \Omega_i  \right) = \sum_{i=1}^M \mathcal{P}(\Omega_{\mathcal{F}} | \Omega_i) \mathcal{P}(\Omega_i) 
\end{equation}
where $\mathcal{P}(\Omega_{\mathcal{F}} | \Omega_i)$ denotes the probability of a failure event given that the event lies in $\Omega_i$ and $\mathcal{P}(\Omega_i)$ is the probability of 
$\Omega_i$. Given an arbitrary stratification of the sample space, the Monte Carlo estimator can be derived as 
\begin{equation}
    \hat{P}^{SS}_{\mathcal{F}} \approx \sum_{i=1}^M \dfrac{1}{N_i} \sum_{k=1}^{N_i} \text{I}_{\left\{ g(\mathbf{x}) \leq 0 \right\}} (\mathbf{x}_k | A_i) \mathcal{P} \left( \Omega_i \right) = \sum_{i=1}^M \dfrac{w_i}{N_i} \sum_{k=1}^{N_i} \text{I}_{\left\{ g(\mathbf{x}) \leq 0 \right\}} (\mathbf{x}_k | A_i)
\end{equation}
where $ w_i = \mathcal{P} \left( \Omega_i \right) $ 
and $N_i$ is the number of samples drawn from stratum $\Omega_i$ such that $\sum_{i=1}^M N_i = N$ is the total number of Monte Carlo samples. (Note that $ \mathbf{x}_k | A_i $ implies that $ \mathbf{x}_k \in A_i $ and follows the relevant conditional distribution.) The variance of this estimator is
\begin{equation}
    \label{eqn:general_stratified_sampling_variance}
    \operatorname{\mathbb{V}ar} \left[ \hat{P}^{SS}_{\mathcal{F}} \right] = \sum_{i=1}^m \dfrac{w_i^2}{N_i} \sigma_i^2
\end{equation}
where $\sigma_i^2 = \operatorname{\mathbb{V}ar} \left[ \text{I}_{\left\{ g(\mathbf{x}) \leq 0 \right\}} (\mathbf{X} | A_i) \right] $ is the conditional variance within stratum $A_i$.

Historically, stratified sampling has not been particularly effective for reliability analysis. This is due to the inability to define strata that concentrate samples around the failure domain $\Omega_{\mathcal{F}}$, which requires an exhaustive search of the sample space to identify failure domains and methods to stratify the space around these domains. Past attempts have relied on strata that are defined on a cartesian grid, which makes them very highly dependent on dimension~\cite{shields2015targeted}. While these stratified sampling estimates are guaranteed to reduce variance over classical Monte Carlo, they become very ineffective for high-dimensional problems.

However, if the stratification can be efficiently defined, stratified sampling offers the potential for huge variance reduction. One recent strategy stratifies using auxiliary system outputs that are correlated with the performance function $ g (\mathbf{x}) $~\cite{arunachalamspence}. In this work, however, we propose a stratification procedure wherein each stratum has known probability and is bounded by level sets of the joint probability density function $f_{\mathbf{X}}(\mathbf{x})$ of the input random variables. 
Defining the strata in this manner is equivalent to stratifying the one-dimensional output of the density function into disjoint intervals of the form $p_{i-1} \geq f_{\mathbf{X}}(\mathbf{x})\geq p_i$; thus,
the resulting stratified sampling scheme is not explicitly dependent on dimension. Moreover, the level sets are defined to progressively probe deeper into the tails of the joint distribution to produce well-defined bounds on the probability of failure estimate and provably small variance. This requires us to first define the tails of distribution. In the following section, we explain why the tails are hard to define and establish two sufficient criteria that will allow us to provide a formal definition. 

\subsection{The tails are hard to define}
\label{section:defining_the_tails_introduction}

The proposed Tail Stratified Sampling method requires a clear and consistent definition of the tails of a distribution. Perhaps surprisingly, a formal definition of the tails of a distribution is not readily available. Instead, the tails are often loosely defined as the ``extreme regions'' of the distribution, requiring some interpretation on the part of the analyst as to what this means. Moreover, it is usually implied that probability diminishes in the tails such that these ``extreme regions'' of the distribution become increasingly ``rare.'' Not unexpectedly, these definitions give rise to a number of ambiguous cases. One such case is the uniform distribution, which is generally interpreted to have no tail because the probability is constant over the entire domain and drops precipitously at the bounds. However, even the uniform distribution has ``extreme regions" corresponding to values near the bounds. But these extreme regions have the same probability of occurrence as more centrally located regions and they are therefore not considered to lie in the tails. Another example is a simple bimodal mixture of two Gaussian distributions that are located far apart from one another. In this case, there is a valley of low-probability between the modes. According to the vague definition above, it's not clear whether this valley is considered a tail of the distribution. 

Informally, we propose the following definition of the tails from the considerations above; a formal definition will follow later. The tails of a distribution are the regions of the domain that lie outside regions of high probability concentration. Here, we specify a threshold on the probability density function, $u$, such that the region of high probability concentration, $S$, is defined as the region over which $f_{\mathbf{X}}(\mathbf{x})\ge u$, which integrates to contain a total probability $P$. For distributions with no regions of constant probability density, the threshold $u$ uniquely defines the total probability $P$. However, for distributions with regions of constant probability density, $u_c$, regions where $ f_{\mathbf{X}}(\mathbf{x})\ge u_c $ can have a range of total probabilities $P\in [P_{lo}, P_{hi}]$. In such cases, it is further necessary to specify the distance, $ l $, from some central point in the distribution. The mode of the distribution (or the nearest mode for multimodal distributions) is the primary choice for this central point, because it is the point of highest probability density and, therefore, is the limit as $u \to \max{\left( f_{\mathbf{X}}(\mathbf{x}) \right)}$. In the absence of a clearly defined mode (or a set of distinct modes), the mean of the distribution becomes the secondary choice. Together, the threshold $u$ and the distance $ l $ define a unique integrated probability $P$. Following this definition, the tails of the distribution correspond to those regions that lie outside the set having a total probability $P$ where $f_{\mathbf{X}}(\mathbf{x})\geq u$. This definition will be detailed in Section~\ref{section:TSS_theory_formulation_conception} and theoretically formalized in Appendix~\ref{section:TSS_S_uv_theory}. 

Furthermore, it is especially challenging to define meaningful tails for high-dimensional distributions. Applying our threshold-based definition to high-dimensional standard normal variables in Cartesian space, one recognizes that the regions of highest probability concentration occur near the origin and that the pdf diminishes as each component increases. However, expressing the same distribution in spherical coordinates, we recognize that the mode of the radial component ($R\sim \chi(n)$ is a chi random variable) lies at $r = \sqrt{n-1}$ and the $n-1$ angular random variables together specify a uniform density over the surface of the unit hypersphere (arguably having no tail). In high dimensions, this results in the so-called important ring such that most of the probability is concentrated in a thin ring centered at the origin with radius $r = \sqrt{n-1}$~\cite{katafygiotis2008geometric}. This, of course, gives rise to significant ambiguity in how the tail is defined in high dimensions. Does one define the tail based on constant probability contours of the joint distribution in Cartesian space, or does one define the tail based on a threshold of the radial marginal distribution in spherical coordinates? If the latter, how does one treat the angular coordinates? Numerous works in reliability analysis rely on the latter representation, which facilitates methods like importance sampling in high dimension~\cite{WANGSong201642}. In cases that use spherical coordinates, there are also different approaches to handling the angular coordinates~\cite{song2023adaptive, grooteman2011adaptive}. Here, we argue that in high dimensions, the definition of the tail based on a pdf threshold is equally valid regardless of the coordinate transformation, but the choice of coordinate system (and thus the pdf that is stratified) can dramatically influence the usefulness of the tail for reliability estimation purposes in a way that is highly problem-specific. In other words, for some problems it is appropriate to stratify the tails in Cartesian coordinates, for others it is appropriate to stratify in spherical coordinates, and some problems may require stratification in some other coordinate system altogether. 








\subsection{Paper Structure and Important Contributions}
\label{section:paper_structure_and_contributions}

This work is structured as follows. We first introduce the concept of Tail Stratified Sampling, beginning with an intuitive description and then presenting the mathematical formulation and its statistical properties in Sections~\ref{section:TSS_stratification}--\ref{section:TSS_estimator_properties}. Some important statistical considerations and important special cases are presented in Sections~\ref{section:sample_Allocation} and \ref{section:special_cases}, respectively. Section~\ref{section:method} provides the TSS algorithm and additional practical insights to implement TSS for probability of failure estimation. Sections~\ref{section:examples} and \ref{section:Hi-D} present numerous examples, with Section~\ref{section:examples} focusing on lower dimensional but complex performance functions and Section~\ref{section:Hi-D} dedicated to illuminating TSS performance in high dimensions. Section~\ref{section:discussion} discusses the relationship between TSS and widely used existing methods and presents some opportunities for future research at their intersection. Finally, some concluding remarks are made in Section~\ref{section:conclusions}.

The TSS represents, in some ways, a new perspective on statistical methods for reliability analysis. With this in mind, the fundamental contributions of this work are the following:
\begin{itemize}
    \item The paper attempts to formally define the tails of a distribution in a manner that is rigorous and useful for the purposes of reliability analysis;
    \item The paper uses this tail definition to gain insights into failure events and their probabilities. 
    \item The paper introduces a scheme, TSS, to stratify the tails of the distribution to facilitate efficient reliability analysis; 
\end{itemize}
Meanwhile, the specific benefits of the TSS methodology include:
\begin{itemize}
    \item The method is intuitive, generalizable, and straightforward to implement;
    \item The algorithm only requires the failure indicator function to be evaluated, and thus can be applied for cases where the performance function is stochastic or unavailable;
    \item The proposed framework allows for very precise quantification of errors (bias and variance) in probability of failure estimates;
    \item The proposed methodology results in failure probability estimates with provably small variance when failures can be isolated in appropriately defined tails of the distribution;
    \item Safe regions, where it is known that failure will not occur, can be identified and sampling in these regions can be avoided altogether; 
    \item The method has no explicit dependence on dimension, but indirectly depends on dimension through the need to identify an appropriate tail for high-dimensional problems. 
\end{itemize}

\section{Tail Stratified Sampling (TSS): Concept \& Theory}
\label{section:TSS_theory_formulation_conception}

The proposed Tail Stratified Sampling (TSS) approach derives from a simple premise: because failure usually occurs when input random variables take values in the tails of their distribution, samples for probability of failure analysis should often be concentrated in the tails. TSS achieves this through a scheme that efficiently stratifies the tails, 
thus allowing for direct exploration of regions that are hard to reach through simple Monte Carlo sampling, while still maintaining the robustness of independent and identically distributed (i.i.d.) random samples for estimation.


\subsection{Tail Stratification: An Intuitive Explanation}
\label{section:TSS_stratification}

In short, the TSS stratification scheme involves defining a sequence of strata that probe deeper and deeper into the tails of the distribution (we formally define the tail in Appendix \ref{section:TSS_S_uv_theory}). To do this, we begin by defining a sequence of nested subsets $ S_{u_1 v_1} \subset S_{u_2 v_2} \subset \dots \subset \Omega $, such that each successive set incorporates increasingly more of the tails of the input distribution. That is, the innermost set $ S_{u_1 v_1} $ contains only the peak(s) of the distribution, and each subsequent set contains more of the tail than the previous set. The subsets $ S_{uv} $ are defined by a probability density threshold $u$ and an auxiliary parameter $v\in \left[ 0, 1 \right]$, described next.

\begin{figure}[!htbp]
\centering
\begin{subfigure}{.48\textwidth}
  \centering
  \includegraphics[width=\linewidth]{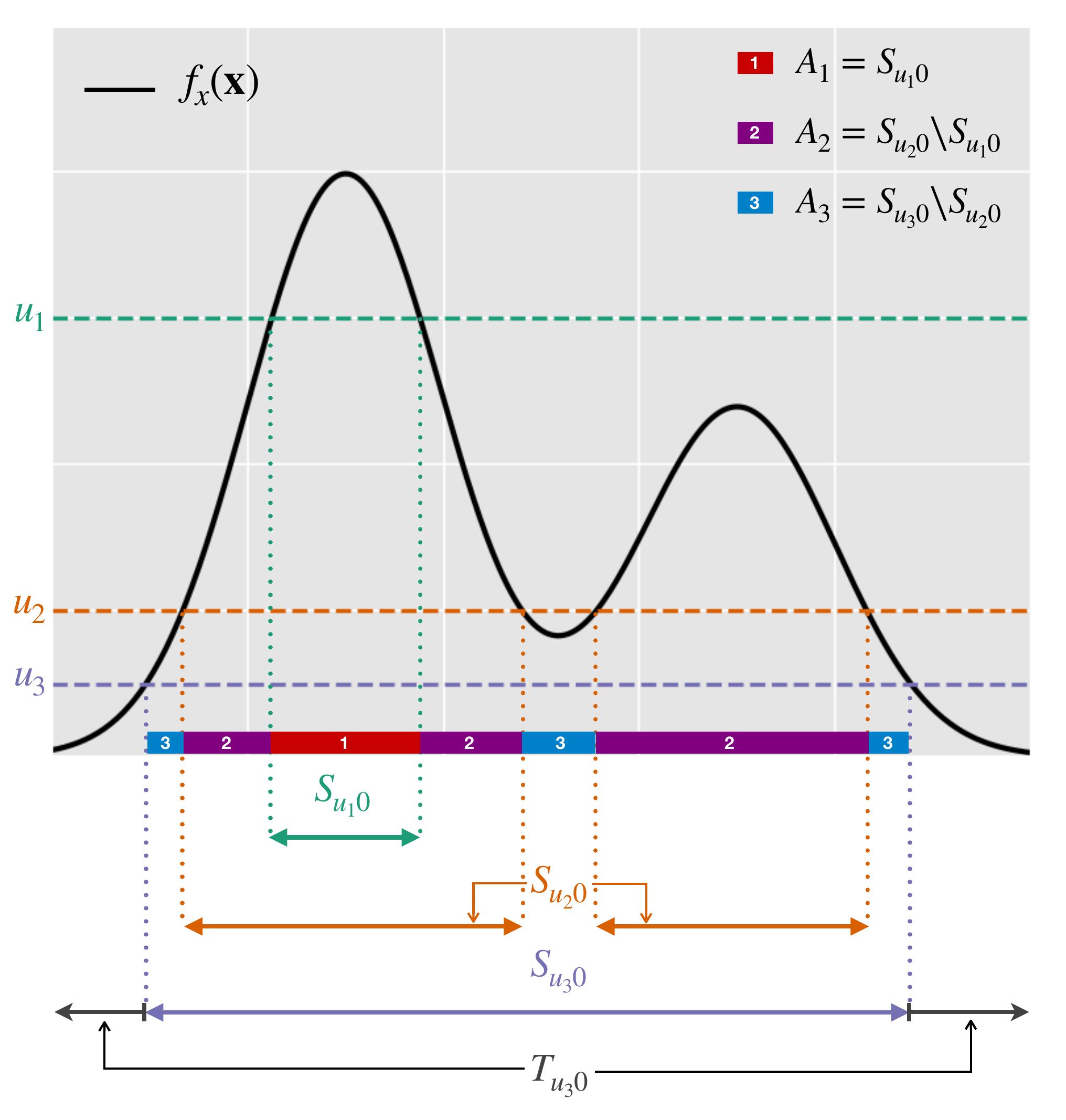}
  \caption{}
  \label{fig:bimodal_stratification}
\end{subfigure}%
\begin{subfigure}{.48\textwidth}
  \centering
  \includegraphics[width=\linewidth]{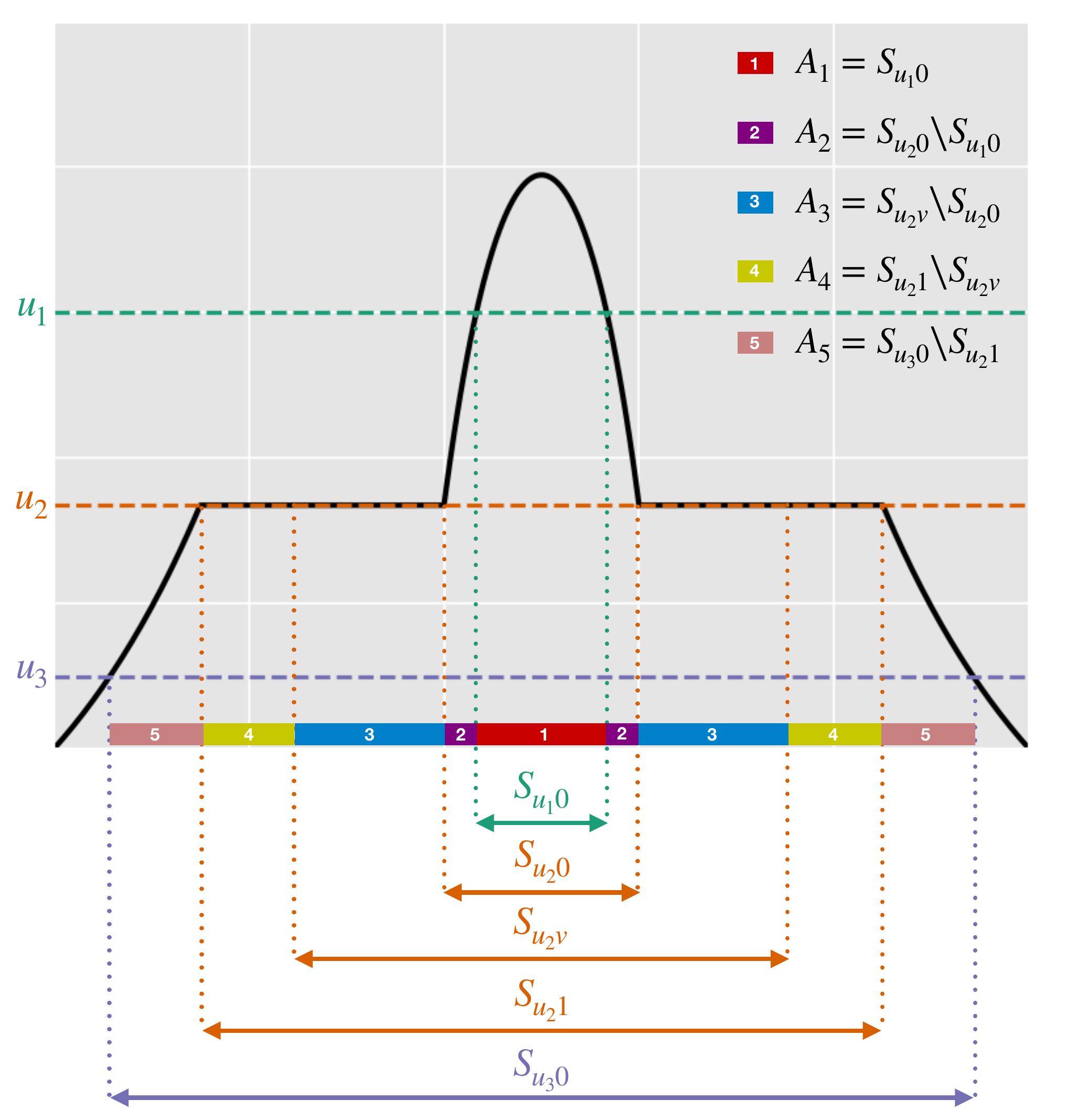}
  \caption{}
  \label{fig:plateau_stratification}
\end{subfigure}
\caption{Visualization of $ S_{uv} $ for different values of $ u $ and $ v $ for (a) a continuously changing density function (i.e., it has no plateaus), and (b) a density function which has regions of uniformity (i.e., it has plateaus, or regions in the domain for which $ f_{\mathbf{X}} (\mathbf{x}) $ is constant). Changing values of $ u $ can naturally deal with multiple modes of the distribution; $ S_{u_2 0} $ in case (a) is composed of disjoint sets corresponding to the two modes of the distribution. When the density function is constant in some region for a value of $ u $, $ v $ changes from $ 0 $ to $ 1 $ and corresponds to different sets within the domain, as shown for case (b) when $ f_{\mathbf{X}} (\mathbf{x}) = u_2 $.}
\label{fig:stratification_visualization}
\end{figure}

To better understand the sets $ S_{uv} $, consider first a 1-D bimodal probability density function, $ f_{\mathbf{X}} (\mathbf{x}) $, as illustrated in Figure~\ref{fig:bimodal_stratification}.  
When $ f_{\mathbf{X}} (\mathbf{x}) $ changes continuously (has no plateaus), $ S_{uv} $ is the subset of the domain for which the density function lies above the threshold $ u $, i.e., $ S_{uv} = S_{u0} = \left\{ \mathbf{x} : f_{\mathbf{X}} (\mathbf{x}) \geq u \right\} $. In this case, the auxiliary parameter $ v $ is always equal to $ 0 $. The threshold $ u $ informally defines the tail of the distribution as the complement of $S_{u0}$, $T_{u0} =  S_{u0}^C = \left\{ \mathbf{x} : f_{\mathbf{X}} (\mathbf{x}) < u \right\} $.
Figure~\ref{fig:bimodal_stratification} shows three sets $ S_{u0} $ for different values of $ u $. It is easy to see how decreasing $ u $ results in an increase in the size of $ S_{u0} $. We can further see that the subsets are nested such that 
$ S_{u_1 0} $ is completely contained within $ S_{u_2 0} $ if $ u_1 > u_2 $. However, each subset does not need to be continuous, as depicted by $ S_{u_2 0} $, which includes two distinct regions. 

The auxiliary parameter, $ v $, becomes important in cases where $ f_{\mathbf{X}} (\mathbf{x}) $ is constant over some region of the domain (i.e. it has plateaus), such as shown in Figure~\ref{fig:plateau_stratification}. In this case, the sets $ \left\{ \mathbf{x} : f_{\mathbf{X}} (\mathbf{x}) \geq u \right\}$ and $ \left\{ \mathbf{x} : f_{\mathbf{X}} (\mathbf{x}) > u \right\} $ have significantly different volumes. The auxiliary parameter is defined such that $ S_{u0} = \left\{ \mathbf{x} : f_{\mathbf{X}} (\mathbf{x}) > u \right\} $ and $ S_{u1} = \left\{ \mathbf{x} : f_{\mathbf{X}} (\mathbf{x}) \geq u \right\} $, and the values of $ v \in \left( 0, 1 \right) $ define a sequence of intermediate sets in the domain, as illustrated in Figure~\ref{fig:plateau_stratification}. In particular, for $ v_1 < v_2 $ and $ v_1, v_2 \in \left( 0, 1 \right) $, $ S_{u0} \subset S_{u v_1} \subset S_{u v_2} \subset S_{u1} $. Unfortunately, since $ f_{\mathbf{X}} (\mathbf{x}) $ does not change for the region $ S_{u1} \setminus S_{u0} $, there is no obvious probabilistic way to construct the intermediate sets $ S_{uv}, v\in (0,1) $. For our purposes, we use a distance measure from a central point, $\mathbf{x}_c$, in $ f_{\mathbf{X}} (\mathbf{x}) $ to define these intermediate sets. Both the measure of distance and the central point can be selected at the user's discretion. Some considerations and corresponding formulations are presented in Section~\ref{section:TSS_S_uv_theory}.

Establishing $ S_{uv} $ in this way allows the tail to be defined as $ T_{uv}=S_{uv}^C $ regardless of the behavior of the density function. Consequently, tails with arbitrary probability can always be found, i.e., one may specify a target value for $ \mathcal{P} \left( T_{uv} \right) $ and find the tail set $ T_{uv} $ 
in accordance with the discussion in Section~\ref{section:defining_the_tails_introduction}.

Given the sequence of nested subsets $ S_{u_1 v_1} \subset S_{u_2 v_2} \subset \dots \subset \Omega $, the strata of $ \Omega $ for TSS, $ \left\{ A_i \right\}_{i \in \left\{1, 2, \dots \right\}} $, are then defined as follows:
\begin{align}
    A_1 &= S_{u_1 v_1} \label{eqn:intuitive_TSS_stratification_scheme_first_set_definition} \\
    A_i &= S_{u_i v_i} \setminus S_{u_{i-1} v_{i-1}} \quad \forall i = 2, 3, \dots \label{eqn:intuitive_TSS_stratification_scheme_definition}
\end{align}
For illustration, Figure~\ref{fig:stratification_visualization} shows the corresponding strata defined by the subsets for each distribution. Here we clearly see that each stratum lies deeper into the tail of the distribution than the previous strata.

Finally, as with any Stratified Sampling scheme (see Section~\ref{section:stratified_sampling})
, Tail Stratified Sampling applies the Law of Total Probability to represent the probability of failure, $P_{\mathcal{F}} $, as
\begin{equation}
\label{eqn:general_TSS_decomposition}
    P_\mathcal{F}^{\text{TSS}} = \sum_{i=1}^m \mathcal{P} \left( \Omega_{\mathcal{F}} | A_i \right) \mathcal{P} \left( A_i \right) = \sum_{i=1}^m \mathcal{P} \left( \mathcal{F}_i \right) \mathcal{P} \left( A_i \right)
\end{equation}
where $ \mathcal{P} \left( \mathcal{F}_i \right) = \mathcal{P} \left( \Omega_{\mathcal{F}} | A_i \right) $. In general, Eq.~\eqref{eqn:general_TSS_decomposition} can be evaluated using estimators for both $ \mathcal{P} \left( \mathcal{F}_i \right) $ and $ \mathcal{P} \left( A_i \right) $. But in this paper, we present a scheme where $ \mathcal{P} \left( A_i \right) $ is known exactly a priori, and thus only consider estimators of the form
\begin{equation}
\label{eqn:general_TSS_estimator}
    \hat{P}_\mathcal{F}^{\text{TSS}} = \sum_{i=1}^m \hat{\mathcal{P}} \left( \mathcal{F}_i \right) \mathcal{P} \left( A_i \right)
\end{equation}



\subsection{Selecting a Sequence of Tails}
\label{section:selecting_sequence_of_subsets_tails_for_TSS}

To begin defining this sequence of tail sets, let us first highlight that strata defined over regions that are known to be safe, which have zero conditional failure probability, can be removed from the estimator. That is, they do not need to be sampled. If the set $ A_0 \subset \Omega $, having complement $A_*=A_0^C\subset \Omega $, is such that $ g(\mathbf{x}) > 0 $ $ \forall \mathbf{x} \in A_0 $ (i.e., $ A_0 \cap \Omega_{\mathcal{F}} = \emptyset $), it follows that $\Omega_\mathcal{F}\subset A_*$ and
\begin{equation}
    \label{eqn:failure_probability_in_reduced_region}
    P_\mathcal{F} = \mathcal{P} \left( \Omega_{\mathcal{F}} \right) = \mathcal{P} \left( A_* \cap \Omega_{\mathcal{F}} \right) = \mathcal{P} \left( \Omega_{\mathcal{F}} | A_* \right) \mathcal{P} \left( A_* \right).
\end{equation}
Thus, the region $A_0$ defines a null stratum within which the conditional probability of failure $ \mathcal{P} \left( \Omega_{\mathcal{F}} | A_0 \right)$ is known to be zero. Failure probability estimation then requires further stratifying (and sampling) only the region $ A_* = \Omega \setminus A_0 $. 
Any computational effort associated with sampling from $ A_0 $
can be conserved. Section~\ref{section:A0_first_safe_region} discusses how to find the region $ A_0 $ using well-established practices. If there is no prior knowledge of safe regions in the domain, then $ A_0 = \emptyset \Rightarrow A_* = \Omega $, and the entire sample space is stratified according to the TSS procedure discussed below.

To construct the stratification $ \left\{ A_i \right\}_{i = \left\{ 1, 2, \dots\right\}} $ of $ A_* $, a sequence of nested subsets $ S_{u_1 v_1} \subset S_{u_2 v_2} \subset \dots $ (or equivalently a sequence of nested tails $T_{u_1v_1}\supset T_{u_2v_2}\supset \dots$) is needed. Since $ S_{uv} $ (and hence $T_{uv})$ can be constructed for arbitrary values of $ u $ and $ v $, constructing the sequence of subsets is equivalent to selecting an ordered sequence of tuples $ \left( u_1, v_1 \right) \prec \left( u_2, v_2 \right) \prec \dots $, where the order is implied by the fact that
\begin{equation}
    \label{eqn:ordering_of_u_v_tuples}
    \begin{aligned}
        u_1 > u_2 &\Rightarrow S_{u_1 v_1} \subset S_{u_2 v_2} \quad \text{and} \\
        (u_1 = u_2) \wedge (v_1 < v_2) &\Rightarrow S_{u_1 v_1} \subset S_{u_2 v_2}
    \end{aligned}
\end{equation}

To select the sequence of tuples, we propose to define these sets such that each successive tail set includes a fixed portion, $p_0$, (by probability where $ 0 < p_0 < 1 $) of the previous tail set (while only considering those regions of the tail sets that lie within $ A_* $).
That is,
\begin{align}
    \mathcal{P}(T_{u_1 v_1} \cap A_*) = \mathcal{P} \left( A_* \setminus S_{u_1 v_1} \right) &= p_0 \mathcal{P} \left( A_* \right) \label{eqn:first_tail_in_sequence_probabiltiy}\\
    \mathcal{P}(T_{u_i v_i} \cap A_*) = \mathcal{P} \left( A_* \setminus S_{u_i v_i} \right) &= p_0 \mathcal{P} \left( A_* \setminus S_{u_{i-1} v_{i-1}} \right) = p_0 \mathcal{P}(T_{u_{i-1} v_{i-1}} \cap A_*) =  p_0^i \mathcal{P} \left( A_* \right) \quad , i \geq 2 \label{eqn:sequence_of_tails_probability}
\end{align}
The above conditions can be simplified as
\begin{align}
    \mathcal{P} \left(S_{u_1 v_1}  \cap A_* \right) &= (1 - p_0) \mathcal{P} \left( A_* \right) \label{eqn:first_subset_in_sequence_probability}\\
    \mathcal{P} \left( S_{u_i v_i}  \cap A_* \right) &= (1 - p_0) \mathcal{P} \left( A_* \right) + p_0 \mathcal{P} \left( S_{u_{i-1} v_{i-1}}  \cap A_* \right) = \left( 1 - p_0^i\right) \mathcal{P} \left( A_* \right) \quad , i \geq 2 \label{eqn:sequence_of_subsets_probability}
\end{align}
and the resultant condition on the stratification $ \left\{ A_i \right\}_{i = \left\{ 1, 2, \dots\right\}} $ is
\begin{align}
    A_1 &= S_{u_1 v_1} \cap A_* &\quad \mathcal{P} \left(A_1 \right) &= 
    (1 - p_0) \mathcal{P} \left( A_* \right) \label{eqn:general_TSS_stratification_scheme_first_set_definition}\\
    A_i &=\left( S_{u_i v_i} \setminus S_{u_{i-1} v_{i-1}} \right) \cap A_* &\quad \mathcal{P} \left(A_i \right) &= 
    p_0^{i-1} \left( 1 - p_0 \right) \mathcal{P} \left( A_* \right) \quad , i \geq 2 \label{eqn:general_TSS_stratification_scheme_definition}
\end{align}
This implies the following sequence of strata probabilities
\begin{equation}
    \label{eqn:geometric_progression_of_partition_sets}
    \mathcal{P} \left(A_i \right) = p_0 \mathcal{P} \left(A_{i-1} \right) \quad , i \geq 2
\end{equation}
For example, setting $p_0=0.1$, $A_* = \Omega$, and $\mathcal{P}(A_0)=0.9$, we have 
\begin{equation*}
    \mathcal{P}(A_1)=0.09, \;\; \mathcal{P}(A_2)=0.009, \;\; \mathcal{P}(A_3)=0.0009, \dots
\end{equation*}
with corresponding set probabilities 
\begin{equation*}
    \mathcal{P}(S_{u_0v_0})=0.9, \;\; \mathcal{P}(S_{u_1v_1})=0.99, \;\; \mathcal{P}(S_{u_2v_2})=0.999, \dots
\end{equation*} 
and tail probabilities 
\begin{equation*}
    \mathcal{P}(T_{u_0v_0})=0.1, \;\; \mathcal{P}(T_{u_1v_1})=0.01, \;\; \mathcal{P}(T_{u_2v_2})=0.001, \dots
\end{equation*}
We discuss how this can be achieved in practice in Section~\ref{section:method}.



\subsection{TSS Estimator \& Its Properties}
\label{section:TSS_estimator_properties}

From the description in Section~\ref{section:selecting_sequence_of_subsets_tails_for_TSS}, we note that $ \left\{ S_{u_i v_i} \right\}_{i = \left\{ 1, 2, \dots\right\}} $ is an infinite sequence (likewise $ \left\{ T_{u_i v_i} \right\}_{i = \left\{ 1, 2, \dots\right\}} $ is an infinite sequence), and therefore the stratification of $ A_* $ defined by Eqs.~\eqref{eqn:general_TSS_stratification_scheme_first_set_definition} and~\eqref{eqn:general_TSS_stratification_scheme_definition} consists of an infinite number of sets. Thus, constructing the TSS estimator requires a truncation to a finite number of tail strata, $m$, which can be achieved in two ways. To form an unbiased estimator, we can assign the proportion for stratum $m$, $p_{0,m}=0$, such that the final stratum $A_m$ extends to include the full tail (equivalently, the tail set $T_{u_{m}v_{m}} \cap A_* =\emptyset$). This is discussed further in Appendix~\ref{appendix:unbiased_TSS}.

Alternatively, if we simply truncate after $m$ tail strata, the TSS estimator, as presented in Eq.~\eqref{eqn:general_TSS_estimator}, is biased. This bias has an upper bound of
\begin{equation}
    \label{eqn:bias_bound_TSS_estimator}
    \begin{aligned}
        \operatorname{Bias} \left( \hat{P}_{\mathcal{F}}^{\text{TSS}} , P_{\mathcal{F}} \right) &\leq \mathcal{P} \left( A_* \setminus S_{u_m v_m} \right) = \mathcal{P} (T_{u_{m} v_{m}} \cap A_*) \\
        \Rightarrow \operatorname{Bias} \left( \hat{P}_{\mathcal{F}}^{\text{TSS}} , P_{\mathcal{F}} \right) &\leq p_0^{m} \mathcal{P} \left( A_* \right)
    \end{aligned}
\end{equation}
It is straightforward to see that the bias decreases exponentially as the number of strata increases, and the simplicity of this formulation allows the user to carefully control the precision of $ \hat{P}^{\text{TSS}}_\mathcal{F}$. Notably, by setting $p_0=0.1$, we can precisely specify the order of magnitude to be resolved, and lower orders of magnitude can be effectively ignored. For example, with six tail strata, the estimator will have a bias of less than $10^{-6}$, even when $ A_* = \Omega $. Precision beyond the prescribed truncation level is deemed inconsequential. For these reasons, we suggest applying the biased estimator in practice, and only discuss the biased estimator in the main body of this manuscript.




Eq.~\eqref{eqn:general_TSS_decomposition} can then be simplified by substituting in Eqs.~\eqref{eqn:general_TSS_stratification_scheme_first_set_definition} and~\eqref{eqn:general_TSS_stratification_scheme_definition} as
\begin{equation}
    P_\mathcal{F}^{\text{TSS}} = \left( 1 - p_0 \right) \mathcal{P} \left( A_* \right) \sum_{i=1}^m p_0^{i-1} \mathcal{P} \left( \mathcal{F}_i \right) \label{eqn:biased_TSS_simplified}
\end{equation}
Drawing $N_i$ samples independently from each stratum $A_i$ and applying $\hat{\mathcal{P}}(\mathcal{F}_i) = \dfrac{1}{N_i} \sum_{k=1}^{N_i}  \text{I}_{\left\{g(\mathbf{x}) \le 0 \right\}} \left(\mathbf{x}_k | A_i \right)$ leads to the following statistical TSS estimator
\begin{equation}
    \hat{P}_\mathcal{F}^{\text{TSS}} = \left( 1 - p_0 \right) \mathcal{P} \left( A_* \right) \sum_{i=1}^m p_0^{i-1} \dfrac{1}{N_i} \sum_{k=1}^{N_i}  \text{I}_{\left\{ g(\mathbf{x}) \leq 0 \right\}} \left(\mathbf{x}_k| A_i \right)\label{eqn:biased_TSS_estimator_simplified}
\end{equation}

The exact variance of this estimator is given by
\begin{equation}
    \operatorname{\mathbb{V}ar} \left[ \hat{P}_{\mathcal{F}}^{\text{TSS}} \right] = \left[ \left( 1 - p_0 \right) \mathcal{P} \left( A_* \right) \right]^2 \sum_{i=1}^m p_0^{2 (i-1)} \frac{ \mathcal{P} \left( \mathcal{F}_i \right) \left( 1 - \mathcal{P} \left( \mathcal{F}_i \right) \right) }{N_i}. \label{eqn:biased_TSS_estimator_variance}
\end{equation}
Inspection of Eq.~\eqref{eqn:biased_TSS_estimator_variance} shows that this variance follows the form of the classical stratified sampling variance relation from Eq.~\eqref{eqn:general_stratified_sampling_variance}. However, the stratum weights, $w_i^2 \propto p_0^{2(i-1)}$, in the TSS estimator vanish exponentially with the selected tail factor $p_0$ as we move deeper into the tails, while the stratum variances satisify $ \sigma_i^2 = \mathcal{P} \left( \mathcal{F}_i \right) \left( 1 - \mathcal{P} \left( \mathcal{F}_i \right) \right) \leq 1 $, such that variance in the estimator is dominated by the large central strata.
For example, letting $A_*=\Omega$ and setting $p_0=0.1$, we see that the variance factors for the first four strata are $w_1^2=0.81$, $w_2^2=0.0081$, $w_3^2=0.000081$, $w_4^2=8.1\times 10^{-7}$. This is important because it implies that variance in the failure probability estimator is dominated by high probability density regions of the distribution where failure does not usually occur in engineering applications. In other words, if failure occurs deep in the tails (as is typical in engineering), then $\mathcal{P}(\mathcal{F}_i)=0$ for the initial strata and all variance in the estimator derives from the later strata, yielding an estimator with very small variance. However, if failure occurs in regions of high probability density, the TSS estimator will have high variance and require a large sample size in the early strata. 

The preceding implies that, for TSS to be effective, the conditional failure probabilities in the tails $P(\mathcal{F}_i)$ should be large compared to the total probability of failure $P_\mathcal{F}$. If the conditional failure probabilities are of comparable magnitude to $P_\mathcal{F}$, then the appropriate tail, $A_*$, (or equivalently the null stratum $A_0$) has not been identified and the variance of the TSS estimator will not be provably small. 


For similar reasons, two clear benefits of stratifying and sampling from $ A_* $ instead of $ \Omega $ become apparent. First, comparing Eq.~\eqref{eqn:failure_probability_in_reduced_region} with Eq.~\eqref{eqn:biased_TSS_simplified} implies
\begin{equation}
\label{eqn:stratified_estimator_of_failure_given_A0_complement}
    \mathcal{P} \left( \Omega_{\mathcal{F}} | A_* \right) = \sum_{i=1}^m p_0^{i-1} \left( 1 - p_0 \right) \mathcal{P} \left( \mathcal{F}_i \right)
\end{equation}
where the right-hand side can readily be interpreted as a TSS Estimator with the $ i $-th strata having probability $ p_0^{i-1} \left( 1 - p_0 \right) $. 
This makes intuitive sense as well since the stratification was constructed as a partition on $ A_* $ from the beginning. Thus, the general TSS decomposition can be written as
\begin{equation}
    \label{eqn:failure_porbability_conditional_decomp_stratified_region}
    P_{\mathcal{F}}^{\text{TSS}} = \mathcal{P} \left( A_* \right) P^{\text{TSS}}_{\mathcal{F}|A_*} 
\end{equation}
where $ P^{\text{TSS}}_{\mathcal{F}|A_*} $ is a tail stratified representation of $ \mathcal{P} \left( \Omega_{\mathcal{F}} | A_* \right) $, and $ \mathcal{P} \left( A_* \right) $ is computed separately. Importantly, 
Eq.~\eqref{eqn:stratified_estimator_of_failure_given_A0_complement} shows that $ \mathcal{P} \left( \Omega_{\mathcal{F}} | A_* \right) $ is independent of $\mathcal{P}(A_*)$, which we later show can be readily calculated using conventional methods. Thus, the problem of estimating the small failure probability of a system where failure is known to occur in the tails (far from regions of high probability density) 
can be converted into the comparatively simpler task of estimating $ \mathcal{P} \left( \Omega_{\mathcal{F}} | A_* \right) $, which can be much larger. Moreover, the performance of TSS \textit{improves} for smaller failure probabilities that occur deeper in the tails of the distribution. This is because $\mathcal{P}(A_*)$ reduces (i.e. the ``safe'' region $A_0$ gets larger), which reduces the variance of the TSS estimator in Eq.~\eqref{eqn:biased_TSS_estimator_variance}
In other words, unlike most reliability methods, TSS can work better for smaller failure probabilities, provided that they occur deep in readily identifiable tails.


The second benefit is the computational savings gained by not sampling from $ A_0 $. To highlight this, let us consider an alternative stratified sampling estimator $ P_{\mathcal{F}, \text{full}}^{\text{TSS}} $ which uses $ \left\{ A_0, A_1, A_2, \dots \right\} $ as the partition of $ \Omega $, where $ A_0 $ is as defined in Section~\ref{section:TSS_theory_formulation_conception} and $ A_1, A_2, \dots $ are as defined in Eqs.~\eqref{eqn:general_TSS_stratification_scheme_first_set_definition} and~\eqref{eqn:general_TSS_stratification_scheme_definition}. We can write
\begin{equation}
    \label{eqn:TSS_estimator_without_safe_region}
    P_{\mathcal{F}, \text{full}}^{\text{TSS}} = \mathcal{P} \left( \mathcal{F}_0 \right) \mathcal{P} \left( A_0 \right) + \sum_{i=1}^m \mathcal{P} \left( \mathcal{F}_i \right) \mathcal{P} \left( A_i \right)
\end{equation}
where $ \mathcal{P} \left( \mathcal{F}_0 \right) = \mathcal{P} \left( \Omega_{\mathcal{F}} | A_0 \right) $. Therefore, 
using i.i.d. samples from each stratum, the variance can be expressed as
\begin{equation}
    \operatorname{\mathbb{V}ar} \left[ \hat{P}_{\mathcal{F}, \text{full}}^{\text{TSS}} \right] = \left[ \mathcal{P} \left( A_0 \right) \right]^2 \frac{{\mathcal{P}} \left( \mathcal{F}_0 \right) \left( 1 - {\mathcal{P}} \left( \mathcal{F}_0 \right) \right)}{N_0} + \left[ \left( 1 - p_0 \right) \mathcal{P} \left( A_* \right) \right]^2 \sum_{i=1}^m p_0^{2 (i-1)} \frac{ {\mathcal{P}} \left( \mathcal{F}_i \right) \left( 1 - {\mathcal{P}} \left( \mathcal{F}_i \right) \right) }{N_i} \label{eqn:variance_TSS_estimator_without_safe_region} 
\end{equation}
It follows that the difference between Eq.~\eqref{eqn:variance_TSS_estimator_without_safe_region} and Eq.~\eqref{eqn:biased_TSS_estimator_variance} is
\begin{equation}
    \operatorname{\mathbb{V}ar} \left[ \hat{P}_{\mathcal{F}, \text{full}}^{\text{TSS}} \right] - \operatorname{\mathbb{V}ar} \left[ \hat{P}_{\mathcal{F}}^{\text{TSS}} \right] = \left[ \mathcal{P} \left( A_0 \right) \right]^2 \frac{{\mathcal{P}} \left( \mathcal{F}_0 \right) \left( 1 - {\mathcal{P}} \left( \mathcal{F}_0 \right) \right)}{N_0} = 0 \quad \left[ \because \mathcal{P} \left( \mathcal{F}_0 \right) := 0 \right]
\end{equation}
where the same number of samples, $ N_i $, are drawn from each stratum $ i \in \left\{ 1, 2, \dots \right\} $ for both cases $ \hat{P}_{\mathcal{F}, \text{full}}^{\text{TSS}} $ and $ \hat{P}_{\mathcal{F}}^{\text{TSS}} $, and $ N_0 $ additional samples are drawn from $ A_0 $ for the former. Here we see that both estimators have the same variance despite $ \hat{P}_{\mathcal{F}, \text{full}}^{\text{TSS}} $ requiring an additional $ N_0 $ model evaluations. This implies that within the proposed TSS scheme, it is more efficient to not sample from $ A_0 $. If the safe stratum $A_0$ can be identified, which we discuss in Section~\ref{section:A0_first_safe_region}, it is wasteful to sample from it.

A final benefit of the proposed TSS estimator is that there is no \textit{explicit} dependency on the dimension, $d$, of the random vector $\mathbf{X}$. While conventional stratified sampling estimators unconditionally reduce variance regardless of dimension, it is well known that the variance reduction diminishes rapidly with increasing dimension. This is due to the challenges of stratifying high-dimensional spaces. However, the TSS stratification is effectively one-dimensional, regardless of the dimension $d$, because stratification occurs according to the scalar-valued joint probability density. Defining the strata in this way creates $d$-dimensional strata whose probability and distance from the center of the distribution are controlled and sequentially probe deeper into the tails \textit{equally} in all dimensions. Therefore, as long as a meaningful distance metric can be defined in $d$ dimensions to measure the tails of the distribution and an appropriate null stratum $A_0$ can be defined, the TSS stratification is effectively one-dimensional and variance improvements should scale to high dimensions. This can, however, implicitly introduce a dimensional dependence by necessitating an appropriate coordinate transformation for certain high-dimensional problems. 



\subsection{Sample Allocation Strategies}
\label{section:sample_Allocation}

An important consideration for achieving the best results from any stratified sampling estimator is the sample allocation among different strata. Let the total number of samples $ N = \sum_{i=1}^m N_i$, where $ m $ is the total number of strata.
When each sample evaluation has the same cost, 
the Neyman sample allocation is optimal for any stratified sampling scheme~\cite{tocher1967art}. For the biased TSS estimator, this allocation is given by
\begin{equation}
    \label{eqn:neyman_alloc_optimal}
    N_i = N \frac{\mathcal{P} \left( A_i \right) \sqrt{\operatorname{\mathbb{V}ar} \left[ \hat{\mathcal{P}} \left(\mathcal{F}_i \right) \right]}}{\displaystyle \sum_{i=1}^m \mathcal{P} \left( A_i \right) \sqrt{\operatorname{\mathbb{V}ar} \left[ \hat{\mathcal{P}} \left(\mathcal{F}_i \right) \right]}} 
    = N \frac{p_0^{i-1} \sqrt{\mathcal{P} \left(\mathcal{F}_i \right) \left( 1 - \mathcal{P} \left(\mathcal{F}_i \right) \right)}}{\displaystyle \sum_{i=1}^m p_0^{i-1} \sqrt{ \mathcal{P} \left(\mathcal{F}_i \right) \left( 1 - \mathcal{P} \left(\mathcal{F}_i \right) \right)}}.
\end{equation}
The optimal allocation can be similarly defined for the unbiased TSS estimator, but is not shown here for brevity. 
The variance of the TSS estimator under the Neyman allocation is
\begin{equation}
    \label{eqn:neyman_optimal_alloc_variance}
    \operatorname{\mathbb{V}ar} \left[ \hat{P}_{\mathcal{F}, \text{Opt}}^{\text{TSS}} \right] = \frac{1}{N} \left[ \sum_{i=1}^m \mathcal{P} \left( A_i \right) \sqrt{ \operatorname{\mathbb{V}ar} \left[ \hat{\mathcal{P}} \left(\mathcal{F}_i \right) \right]} \right]^2 = \frac{\left[ \left( 1 - p_0 \right) \mathcal{P} \left( A_* \right) \right]^2}{N} \left[ \sum_{i=1}^m p_0^{i-1} \sqrt{ \mathcal{P} \left(\mathcal{F}_i \right) \left( 1 - \mathcal{P} \left(\mathcal{F}_i \right) \right) } \right]^2
\end{equation}
Unfortunately, the Neyman allocation is not achievable in practice because it depends on $ {\mathcal{P}} \left(\mathcal{F}_i \right) $, which must be estimated. 

Another popular sample allocation strategy is proportional allocation, where each stratum receives a number of samples proportional to its probability content,
i.e.,
\begin{equation}
    \label{eqn:proportional_alloc}
    N_i = N \frac{\mathcal{P} \left( A_i \right)}{\mathcal{P} \left( A_* \right)}
\end{equation}
This allocation is identical to the Neyman allocation when the variance in each stratum $ \operatorname{\mathbb{V}ar} \left[ \hat{\mathcal{P}} \left(\mathcal{F}_i \right) \right] $ is equal, and the variance of the overall estimator is given by
\begin{equation}
    \label{eqn:proportional_alloc_variance}
    \operatorname{\mathbb{V}ar} \left[ \hat{P}_{\mathcal{F}, \text{Prop}}^{\text{TSS}} \right] = \frac{1}{N} \sum_{i=1}^m \mathcal{P} \left( A_i \right) \mathcal{P} \left( A_* \right) \operatorname{\mathbb{V}ar} \left[ \hat{\mathcal{P}} \left(\mathcal{F}_i \right) \right] = \frac{1}{N} \sum_{i=1}^m p_0^{i-1} \left( 1 - p_0 \right) \left( \mathcal{P} \left( A_* \right) \right)^2 \mathcal{P} \left(\mathcal{F}_i \right) \left( 1 - \mathcal{P} \left(\mathcal{F}_i \right) \right)
\end{equation}
The benefit of proportional allocation is that it can be approximately achieved simply by sampling from $ f_{\mathbf{X} | A_*} \left( \mathbf{x} | A_* \right) $ and sorting the resultant samples by stratum. 
This makes the sample generation procedure equivalent to direct Monte Carlo sampling, 
but with a guarantee of lower (or at worst equal) variance compared to a conventional Monte Carlo estimator.

Another easily implementable allocation assigns an equal number of samples to each stratum, such that
\begin{equation}
    \label{eqn:equal_alloc}
    N_i = N / m
\end{equation}
which results in an estimator having variance given by
\begin{equation}
    \label{eqn:equal_alloc_variance}
    \operatorname{\mathbb{V}ar} \left[ \hat{P}_{\mathcal{F}, \text{Eq}}^{\text{TSS}} \right] = \frac{1}{N} \sum_{i=1}^m m \left(\mathcal{P} \left( A_i \right) \right)^2 \operatorname{\mathbb{V}ar} \left[ \hat{\mathcal{P}} \left(\mathcal{F}_i \right) \right] = \frac{1}{N} \sum_{i=1}^m m p_0^{2(i-1)} \left( 1 - p_0 \right)^2 \left( \mathcal{P} \left( A_* \right) \right)^2 \mathcal{P} \left(\mathcal{F}_i \right) \left( 1 - \mathcal{P} \left(\mathcal{F}_i \right) \right)
\end{equation}
This sample allocation strategy can outperform proportional allocation if
\begin{gather}
    \operatorname{\mathbb{V}ar} \left[ \hat{P}_{\mathcal{F}, \text{Eq}}^{\text{TSS}} \right] - \operatorname{\mathbb{V}ar} \left[ \hat{P}_{\mathcal{F}, \text{Prop}}^{\text{TSS}} \right] < 0 \label{eqn:comparing_equal_and_proportional_alloc_generic} \\
    \Rightarrow \frac{1}{N} \sum_{i=1}^m \mathcal{P} \left( A_i \right) \operatorname{\mathbb{V}ar} \left[ \hat{\mathcal{P}} \left(\mathcal{F}_i \right) \right] \left\{ m \mathcal{P} \left( A_i \right) - \mathcal{P} \left( A_* \right) \right\} < 0 \label{eqn:comparing_equal_and_proportional_alloc} \\
    \Rightarrow \sum_{i=1}^m p_0^{i-1} \mathcal{P} \left(\mathcal{F}_i \right) \left( 1 - \mathcal{P} \left(\mathcal{F}_i \right) \right) \left[ m p_0^{i-1} \left( 1 - p_0 \right) - 1 \right] < 0
\end{gather}
However, this condition is rarely met in practice.

Finally, considering the general sample allocation strategy with
\begin{equation}
    \label{eqn:sample_allocation_general}
    N_i = \gamma_i N \quad \text{where } \sum_{i=1}^m \gamma_i = 1
\end{equation}
we get an estimator with the following variance
\begin{equation}
    \label{eqn:general_alloc_variance}
    \operatorname{\mathbb{V}ar} \left[ \hat{P}_{\mathcal{F}, \text{Gen}}^{\text{TSS}} \right] = \frac{1}{N} \sum_{i=1}^m \frac{\left(\mathcal{P} \left( A_i \right) \right)^2 \operatorname{\mathbb{V}ar} \left[ \hat{\mathcal{P}} \left(\mathcal{F}_i \right) \right]}{\gamma_i} = \frac{1}{N} \sum_{i=1}^m \frac{p_0^{2(i-1)} \left( 1 - p_0 \right)^2 \left( \mathcal{P} \left( A_* \right) \right)^2 \mathcal{P} \left(\mathcal{F}_i \right) \left( 1 - \mathcal{P} \left(\mathcal{F}_i \right) \right)}{\gamma_i}
\end{equation}
Again, we see that any general allocation will outperform proportional allocation if
\begin{gather}
    \frac{1}{N} \sum_{i=1}^m \mathcal{P} \left( A_i \right) \operatorname{\mathbb{V}ar} \left[ \hat{\mathcal{P}} \left(\mathcal{F}_i \right) \right] \left\{ \frac{\mathcal{P} \left( A_i \right)}{\gamma_i} - \mathcal{P} \left( A_* \right) \right\} < 0 \\
    \Rightarrow \frac{1}{N} \sum_{i=1}^m p_0^{i-1} \mathcal{P} \left(\mathcal{F}_i \right) \left( 1 - \mathcal{P} \left(\mathcal{F}_i \right) \right) \left\{ \frac{p_0^{i-1} \left( 1 - p_0 \right)}{\gamma_i} - 1 \right\} < 0 \label{eqn:condition_general_allocation_better}
\end{gather}
A sufficient condition for Eq.~\eqref{eqn:condition_general_allocation_better} to hold is
\begin{equation}
    \label{eqn:condition_general_allocation_better_simplified}
    \gamma_i > \frac{\mathcal{P} \left( A_i \right)}{\mathcal{P} \left( A_* \right)} \Rightarrow \gamma_i > p_0^{i-1} \left( 1 - p_0 \right) \quad \forall i = 1, \dots m
\end{equation}
which can only be met if the biased TSS estimator is used.

\subsection{Important Special Cases}
\label{section:special_cases}

\subsubsection{TSS for Independent Standard Normal Random Variables}
\label{section:gaussian_formulation}

In the field of reliability analysis, special attention is paid to analyzing systems whose inputs follow independent standard normal distributions. This is due to the existence of isoprobabilistic mappings (e.g., the Nataf transformation~\cite{lebrun2009innovating}) that allow known non-Gaussian random variables to be transformed to standard normal and the many simplifications that result from the normal distribution. In fact, a variety of efficient failure estimators have been designed solely for standard normal random variables (e.g., \cite{papaioannou2015mcmc}), including variance reduction techniques such as directional simulation \cite{bjerager1988probability}. 

While TSS is a general algorithm applicable to random variables with arbitrary distributions, it simplifies considerably when applied to uncorrelated normal random variables. In fact, 
while all formulations in this manuscript were developed independently, the TSS formulation for standard normal random variables has notable similarities to the stratified ``beta-sphere" sampling approach proposed by Hong et al.~\cite{HONG2025102546}. However, these methods differ in important ways. Most notably, the beta-sphere sampling applies only to normal random variables, and it does not take advantage of the large variance reduction that can be achieved by (a) identifying, quantifying, and deliberately \textit{not sampling} from the known safe region, $A_0$, and (b) exploiting the rapidly diminishing variance in the tails. Consequently, it requires coupling with specialized directional sampling techniques (e.g.\ importance sampling) to achieve maximum statistical benefit, while TSS does not. That said, additional variance reduction may be possible by coupling TSS with advanced sampling methods, but this is not explored here. 

Consider a $d$-dimensional standard normal random vector, $\mathbf{Z}$, having joint probability density function
\begin{equation}
    \phi_d(\mathbf{z}) \equiv \phi_d(r_z) = \frac{1}{\left(\sqrt{2 \pi}\right)^d} \exp{\left( - \frac{r_z^2}{2} \right)}
\end{equation}
where $r_z=\lVert \mathbf{z}\rVert_2$. 
Note that $\phi_d(r_z)$ is an invertible function that utilizes spherical symmetry to express the density function of a point $ \mathbf{z} $ in terms of the scalar Chi random variable $ r_z \sim \chi \left( d \right)$ with $d$ degrees of freedom. 

We begin by assuming that the initial safe region, $A_0$, can be defined as follows
\begin{equation}
    \label{eqn:gaussian_TSS_safe_region}
    A_0 = \left\{ \mathbf{z} \in \Omega : \phi_{d} (\mathbf{z}) > u_0 \right\}
\end{equation}
This is the region around the mode having probability density greater than some threshold $ u_0 $. Leveraging the geometry of the standard normal distribution, we determine the stratum probability as
\begin{equation}
    \mathcal{P} \left( A_0 \right) =F_{\chi} \left( \phi_d^{-1} (u_0); d \right) 
\end{equation}
where $F_{\chi}(\cdot; d)$  is the cumulative distribution function of $ \chi \left( d \right) $. It follows that,
\begin{equation}
    \mathcal{P} \left( A_* \right) = 1 - F_{\chi} \left( \phi_d^{-1} (u_0); d \right) 
\end{equation}

\begin{figure}[!htbp]
\centering
\includegraphics[width=\linewidth]{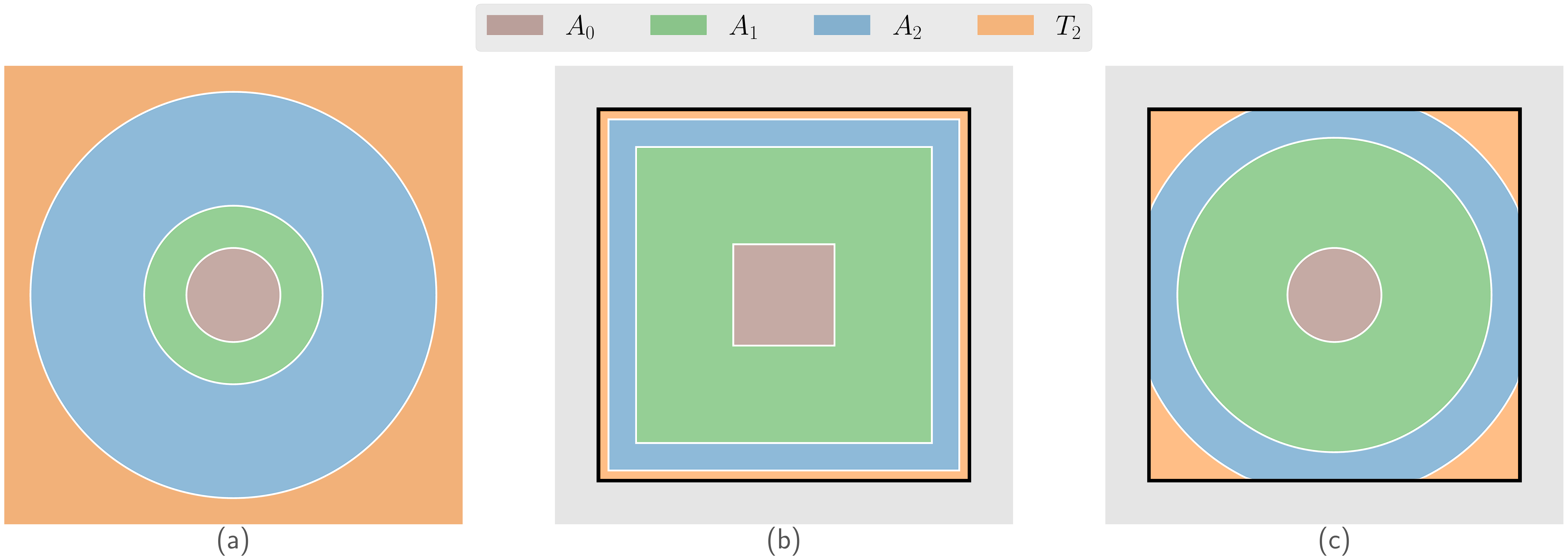}
\caption{Visualization of tail stratification for (a) independent standard Gaussian inputs
(b) independent uniform inputs when an $ l_{1} $ norm is used, and (c) independent uniform inputs when an $ l_{2} $ norm is used. In (b) and (c) the bounds of the domain are visualized by the solid black border.
}
\label{fig:special_cases_stratification_visualization}
\end{figure}

The remainder of the TSS stratification can be defined directly from Eqs.
\eqref{eqn:first_subset_in_sequence_probability} --
\eqref{eqn:general_TSS_stratification_scheme_definition} as
\begin{align}
    r_i &= \phi_d^{-1} \left( \phi_d \left( r_0 \right) + \left( 1 - p_0^i \right) \left[ 1 - F_{\chi} \left( \phi_d^{-1} (u_0); d \right) \right] \right) &\quad i &\geq 1 \label{eqn:gaussian_radial_strata_boundary} \\
    A_1 &= \left\{ \mathbf{x} \in \Omega : r_0 \leq \lVert \mathbf{x} \rVert_2 \leq r_1 \right\} \\
    A_i &= \left\{ \mathbf{z} \in \Omega : r_{i-1} < \lVert \mathbf{z} \rVert_2 \leq r_i \right\} &\quad i &> 1 \label{eqn:gaussian_strata_definition} \\
    \mathcal{P} \left( A_i \right) &= p_0^{i-1} \left( 1 - p_0 \right) \left[ 1 - F_{\chi} \left( \phi_d^{-1} (u_0); d \right) \right] &\quad i &\geq 1 \label{eqn:gaussian_strata_probability}
\end{align}
where $r_0 = \phi_d^{-1} \left( u_0 \right)$, and the resultant failure probability estimator can be computed using Eq.~\eqref{eqn:biased_TSS_estimator_simplified}.
The resulting stratification is visualized for a two-dimensional domain in Figure~\ref{fig:special_cases_stratification_visualization}(a).

Note that we desire $A_0$ to define an initial safe region. This can be achieved by determining the design point (i.e., the point along the limit surface $g(\mathbf{z})=0$ closest to the origin) as $r_0$ and setting $u_0 = \phi_d(r_0)$.

\subsubsection{TSS for Independent Uniform Random Variables}
\label{section:uniform_formulation}

Anther case of practical importance involves problems having independent Uniform inputs. Let $ \mathbf{X} = \begin{bmatrix}  X_1 & \dots & X_d  \end{bmatrix} $ with independent components $ X_j \sim \mathcal{U} \left[ -1, 1 \right] $ ($ j =  1, 2, \dots, d $).
Since the joint uniform distribution is constant everywhere in $ \Omega $, 
its tails must be defined using some distance measure as discussed in Section~\ref{section:TSS_theory_formulation_conception} and detailed in
Appendix~\ref{section:TSS_S_uv_theory}.
Further, because the uniform distribution does not have any distinct peaks or disjoint regions, the mean value is a natural choice around which to define the spread of the distribution (here $ \mathbf{x}_c $ is the origin), and the 
$ l_\mathfrak{p} $ norm is a reasonable choice for the distance measure, i.e., $ l_\mathfrak{p} \left( \mathbf{x}, \mathbf{x}_c \right) = \lVert \mathbf{x} \rVert_\mathfrak{p} $. 

Let us consider the initial safe domain $ A_0 $ to be given by,
\begin{equation}
    \label{eqn:uniform_case_safe_region}
    A_0 = \left\{ \mathbf{x} \in \Omega: \lVert \mathbf{x} \rVert_\mathfrak{p} < \lambda_0 \right\}
\end{equation}
such that all points within a certain distance $ \lambda_0 $ (under the chosen measure) from the origin are known to be safe. Additionally, since $ f_{\mathbf{X}} (\mathbf{x}) = 1/2 $ $ \forall \mathbf{x} \in \Omega $, $\mathcal{P} \left( A_0 \right)$ can be easily computed.

We can then define the sets $ S_{10} $ and $ S_{11} $ 
as follows, 
\begin{gather}
    S_{10} = \left\{ \mathbf{x} \in \Omega: \lVert \mathbf{x} \rVert_\mathfrak{p} = \lambda_0 \right\} \label{eqn:uniform_TSS_smallest_set} \\
    S_{11} = \left\{ \mathbf{x} \in \Omega: \lambda_0 \leq \lVert \mathbf{x} \rVert_\mathfrak{p} \leq \lambda_\text{max} \right\} = A_* \label{eqn:uniform_TSS_largest_set}
\end{gather}
and the general intermediate sets $S_{1 v}$ can be defined by
\begin{equation}
    S_{1 v} = \left\{ \mathbf{x} \in \Omega: \lambda_0 \leq \lVert \mathbf{x} \rVert_\mathfrak{p} \leq \lambda_0 + v \left( \lambda_\text{max} - \lambda_0 \right) \right\}, \quad v \in \left[ 0, 1 \right]
\end{equation}
where $ \lambda_\text{max} = \displaystyle \max_{\mathbf{x} \in \Omega} \left( \lVert \mathbf{x} \rVert_\mathfrak{p} \right) $. The stratification is then performed by selecting a sequence of values $ v_1 < v_2 < \dots $ using 
Eqs.~\eqref{eqn:first_subset_in_sequence_probability} --
\begin{align}
    A_1 &= \left\{ \mathbf{x} \in \Omega: \lambda_0 \leq \lVert \mathbf{x} \rVert_\mathfrak{p} \leq \lambda_0 + v_i \left( \lambda_\text{max} - \lambda_0 \right) \right\} \label{eqn:uniform_TSS_partition1}\\
    A_i &= \left\{ \mathbf{x} \in \Omega: \lambda_0 + v_{i-1} \left( \lambda_\text{max} - \lambda_0 \right) < \lVert \mathbf{x} \rVert_\mathfrak{p} \leq \lambda_0 + v_i \left( \lambda_\text{max} - \lambda_0 \right) \right\}  \quad , i > 1 \label{eqn:uniform_TSS_partition2}
\end{align}
where each $ v_i \in \left[ 0, 1 \right] $ is computed using the equation
\begin{equation}
\label{eqn:uniform_TSS_v_equation}
    \mathcal{V} \left( S_{1 v_i} \right) = \left( 1 - p_0^i \right) \left[ 1 - \mathcal{V} \left( A_0 \right) \right] \quad , i \geq 1
\end{equation}
where $\mathcal{V}$ is the standard Lebesgue measure of volume.
The failure probability can then be estimated using  Eq.~\eqref{eqn:biased_TSS_estimator_simplified} 
with partition defined by Eqs.~\eqref{eqn:uniform_TSS_partition1} and~\eqref{eqn:uniform_TSS_partition2}. Stratified uniform domains using the $l_1$ and $l_2$ norms are visualized in Figures~\ref{fig:special_cases_stratification_visualization}(b) and (c), respectively.

\section{Methodology \& Implementation}
\label{section:method}

Algorithm~\ref{Algo:TSS_algo} presents the steps for applying TSS to estimate failure probabilities. 
The following subsections discuss various practical considerations to guide the implemention of TSS in practice.

\begin{algorithm}[!ht]
\caption{Failure Probability Estimation using (biased) TSS}
\label{Algo:TSS_algo}
\begin{codefont}
\begin{algorithmic}[1]
\Require $ f_{\mathbf{X}} (\mathbf{x}) $, $ p_0 $, $ m $, $ N $     \Comment{where $ m $ is the number of strata and $ N $ is the total number of samples.}
\State Determine $ A_* $ and $ \mathcal{P} (A_*) $      \Comment{More details in Section~\ref{section:A0_first_safe_region}}
\State Define $ A_i $ $ \forall i = 1, 2, \dots, m $ using Eqs.~\eqref{eqn:general_TSS_stratification_scheme_first_set_definition} and~\eqref{eqn:general_TSS_stratification_scheme_definition}
\State $ \mathcal{P} (A_1) = \left( 1 - p_0 \right) \mathcal{P} \left( A_* \right) $ \& $ \mathcal{P} (A_i) = p_0^{i-1} \left( 1 - p_0 \right) \mathcal{P} \left( A_* \right) $ $ \forall i = 2, \dots, m $ \Comment{Eqs.~\eqref{eqn:general_TSS_stratification_scheme_first_set_definition},~\eqref{eqn:general_TSS_stratification_scheme_definition}}
\State $ N_i = N \mathcal{P} (A_i) $    \Comment{Proportional sample allocation (Section~\ref{section:sample_Allocation})}
\For{i = 1:$ m $}
\For{k = 1:$ N_i $}
\State Draw $ \mathbf{x}_{ik} \sim f_{\mathbf{X} | A_i } \left( \mathbf{x} | A_i \right) $   \Comment{More details in Section~\ref{section:strata_sampling_strategies}}
\EndFor
\State $ \hat{\mathcal{P}} (\mathcal{F}_i) = \frac{1}{N_i} \sum_{k = 1}^{N_i} \text{I}_{\left\{ g(\mathbf{x}) \leq 0 \right\}} \left( \mathbf{x}_{ik} | A_i \right) $ \Comment{??}
\EndFor
\State $ \hat{P}_\mathcal{F}^{\text{TSS}} = \left( 1 - p_0 \right) \mathcal{P} \left( A_* \right) \sum_{i=1}^m p_0^{i-1} \hat{\mathcal{P}} (\mathcal{F}_i) $   \Comment{Eq.~\eqref{eqn:biased_TSS_estimator_simplified}}
\State Estimate upper bound on $ \operatorname{Bias} \left( \hat{P}_{\mathcal{F}}^{\text{TSS}} , P_{\mathcal{F}} \right) $ as $ \operatorname{Bias} \left( \hat{P}_{\mathcal{F}}^{\text{TSS}} , P_{\mathcal{F}} \right) \leq p_0^{m} \mathcal{P} \left( A_* \right) $     \Comment{Eq.~\eqref{eqn:bias_bound_TSS_estimator}}
\State $ \hat{\operatorname{\mathbb{V}ar}} \left[ \hat{P}_{\mathcal{F}}^{\text{TSS}} \right] = \left[ \left( 1 - p_0 \right) \mathcal{P} \left( A_* \right) \right]^2 \sum_{i=1}^m p_0^{2 (i-1)} \frac{ \hat{\mathcal{P}} \left( \mathcal{F}_i \right) \left( 1 - \hat{\mathcal{P}} \left( \mathcal{F}_i \right) \right) }{N_i} $     \Comment{Eq.~\eqref{eqn:biased_TSS_estimator_variance}}
\end{algorithmic}
\end{codefont}
\end{algorithm}

\subsection{Finding a Known Safe Region}
\label{section:A0_first_safe_region}

A major benefit of the TSS formulation is the ability to remove a known safe region $ A_0 $ from consideration and only sample from $ A_* = \Omega \setminus A_0 $. Thus, confidently identifying this ``safe set'' 
is an important step in implementing TSS. 
In many cases, knowledge of the physics or engineering behavior of the system may provide the analyst with strong justification for defining $A_0$.
Alternatively, a surrogate model for the response function $ g(\mathbf{x}) $ can then be used to define $A_0$.

A classical approach to define $ A_0 $ (and the approach used here) is to locate the design point $ \mathbf{x}_* $ having reliability index $\lVert \mathbf{x}_* \rVert_2=\beta$, also known as the most probable point of failure, defined by 
\begin{equation}
    \label{eqn:design_point_definition}
    \mathbf{x}_* = \arg \max_{\mathbf{x} \in \Omega_{\mathcal{F}}} \left( f_{\mathbf{X}} (\mathbf{x}) \right) = \arg \max_{\mathbf{x} \in \Omega} \left( f_{\mathbf{X}} (\mathbf{x}) \; | \; g(\mathbf{x}) \leq 0 \right)
\end{equation}
This, in turn, ensures that all points in $ \Omega $ with a higher density function value are safe. 
It is easy to see that $ \forall \mathbf{x} \in \Omega $ such that $ f_{\mathbf{X}} (\mathbf{x}) > f_{\mathbf{X}} (\mathbf{x}_*) $, $ g(\mathbf{x}) > 0 \Rightarrow \mathbf{x} \notin \Omega_{\mathcal{F}} $. Thus, we can define $ A_0 $ and $ A_* $ as
\begin{align}
    A_0 &= \left\{ \mathbf{x} \in \Omega : f_{\mathbf{X}} (\mathbf{x}) > f_{\mathbf{X}} (\mathbf{x}_*) \right\} \label{eqn:design_point_based_safe_set} \\
    A_* &= \left\{ \mathbf{x} \in \Omega : f_{\mathbf{X}} (\mathbf{x}) \leq f_{\mathbf{X}} (\mathbf{x}_*) \right\} \label{eqn:design_point_based_A*}
\end{align}

While this definition can be applied for any parametric distribution, it is the easiest to implement when the inputs follow independent standard Gaussian distributions such that
\begin{align}
    A_0 &= \left\{ \mathbf{x} \in \Omega : \lVert \mathbf{x} \rVert_2 < \beta \right\} \label{eqn:design_point_based_safe_set_gaussian} \\
    A_* &= \left\{ \mathbf{x} \in \Omega : \lVert \mathbf{x} \rVert_2 \geq \beta \right\} \label{eqn:design_point_based_A*_gaussian}
\end{align}

Many robust and efficient optimization algorithms to locate the design point in standard Gaussian space exist~\cite{rackwitz1978structural, liu1991optimization, haukaas2006strategies}, as locating $ \mathbf{x}_* $ is also an important step in the implementation of FORM and SORM. 

A careful observation of Eqs.~\eqref{eqn:design_point_based_safe_set} and~\eqref{eqn:design_point_based_A*} reveals that using the design point to define $ A_* $ is the natural choice for low dimensional problems as it is structurally consistent with the proposed TSS stratification scheme. Further, since defining $ A_* $ in this way excludes the regions of $ \Omega $ with the highest probability density, it is almost assured 
that if $P_{\mathcal{F}}$ is small and failure occurs in the tails then $ \mathcal{P} \left( A_* \right) $ will be small. 
One can therefore expect that defining $ A_* $ using Eq.~\eqref{eqn:design_point_based_A*} (or Eq.~\eqref{eqn:design_point_based_A*_gaussian} for the standard Gaussian case) will lead to appreciable computational benefits for low-dimensional problems. This is, indeed, observed to be the case in the applications in Section~\ref{section:examples}.

However, as illustrated by the examples in Section~\ref{section:Hi-D}, defining $A_0$ using the design point is not effective in many high-dimensional problems because a hypersphere with radius $\beta$ often has vanishingly small probability in high-dimensions. Consequently $P(A_0)$ will be vanishingly small, which will result in the conditional probability in the tail being comparable in magnitude to the failure probability, i.e. $P(\mathcal{F}|A_*)\approx P_\mathcal{F}$. The variance reduction from TSS will then be modest.


\subsection{Sampling within each Stratum}
\label{section:strata_sampling_strategies}

TSS requires independent samples from each stratum $ A_i $ drawn from the conditional distribution $ f_{\mathbf{X} | A_i} \left( \mathbf{x} | A_i \right) $. 
In the most general case, MCMC methods (e.g.~\cite{brooks2011handbook,vrugt2009accelerating,goodman2010ensemble,chakroborty2024intrepid}) can be used to sample from an arbitrary distribution. However, i.i.d. samples can be drawn in certain special situations.

Consider cases where the components of the input random vector $ \mathbf{X} $ are independent, 
If, additionally, each stratum $ A_i $ can also be decomposed into a product of one-dimensional intervals (or the union of multiple one-dimensional intervals), i.e., if $ A_i = a_{i1} \times a_{i2} \times \dots \times a_{id} $ where each $ a_{ij} \subseteq \mathbb{R} $ such that $ \mathbf{X} \in A_i \Rightarrow X_j \in a_{ij} $ $ \forall j = 1, \dots, d $, then $ f_{\mathbf{X} | A_i} \left( \mathbf{x} | A_i \right) $ can be constructed in closed form as
\begin{equation}
    f_{\mathbf{X} | A_i} \left( \mathbf{x} | A_i \right) = \prod_{j=1}^d f_{X_j | a_{ij}} (x_j | a_{ij}) \quad \forall i = 1, \dots, m
\end{equation}
Here, the inverse transform method can be used to generate i.i.d. samples from each conditional component distribution $ f_{X_j | a_{ij}} (x_j | a_{ij}) $, and thus i.i.d. samples of $ \mathbf{X}|A_i $ can be drawn. While not immediately obvious, independent standard Gaussian random variables can be sampled in this manner after an appropriate coordinate transformation.

There can also be cases where the strata boundaries do not decompose into independent intervals, but the region of interest $ A_* $ can still be written as a product of one dimensional intervals as before, i.e., $ A_* = a_{*1} \times a_{*2} \times \dots \times a_{*d} $, $ a_{*j} \subseteq \mathbb{R} $ $ \forall j = 1, \dots, d $. If the components of the input random vector are independent, then i.i.d. samples can be drawn from the conditional distribution $ f_{\mathbf{X} | A_*} \left( \mathbf{x} | A_* \right) $ using the inverse transform method as discussed above. More generally, and as discussed in Section~\ref{section:sample_Allocation}, a proportional allocation of samples (see Section~\ref{section:sample_Allocation}) following the conditional distribution $ f_{\mathbf{X} | A_i} \left( \mathbf{x} | A_i \right) $ in each stratum can be achieved by drawing $ N $ i.i.d. samples from $ f_{\mathbf{X} | A_*} \left( \mathbf{x} | A_* \right) $ (using any appropriate method, e.g. inverse transform or rejection sampling) and simply sorting them by strata. 

\subsection{Empirical Strata Boundaries for Arbitrary Distributions}
\label{section:strata_boundaries_in_practice}

The previous discussions assumed that the boundaries of each strata, $ A_i $, are known prior to sampling. This will be true for simple input distributions, such as independent standard Gaussian or Uniform random variables. 
However, defining the strata 
using the framework presented in Section~\ref{section:selecting_sequence_of_subsets_tails_for_TSS} is not a trivial task for arbitrary input distributions, as it requires solving a complex constrained optimization problem. Instead, we propose a simple scheme in which samples are first drawn from $ f_{\mathbf{X} | A_*} \left(\mathbf{x} | A_* \right) $ and then sorted into appropriate strata by evaluating $ f_{\mathbf{X}} (\mathbf{x}) $ (and other functions) at the sampled points. 

Consider we draw a set of $ N $ samples $ \left\{ \mathbf{x}_k \right\}_{k=\left\{ 1, \dots, N \right\}} $ from $ f_{\mathbf{X} | A_*} \left(\mathbf{x} | A_* \right) $. 
These samples can be sorted into ascending order based on the following:
\begin{itemize}
    \item If $ f_{\mathbf{X}} (\mathbf{x}_i) > f_{\mathbf{X}} (\mathbf{x}_j) \Rightarrow \mathbf{x}_i \prec \mathbf{x}_j $.
    \item If $ f_{\mathbf{X}} (\mathbf{x}_i) = f_{\mathbf{X}} (\mathbf{x}_j) $, then $ \mathbf{x}_i \prec \mathbf{x}_j $ if $ l (\mathbf{x}_i, \mathbf{x}_c) < l (\mathbf{x}_j, \mathbf{x}_c) $, where $ l $ is some measure of distance. This can be generalized to condition on any general measure of sample dispersion as discussed in Appendix~\ref{section:TSS_S_uv_theory}. 
\end{itemize}

Let $ \mathfrak{X} $ be a tuple that denotes the $ N $ samples re-indexed to be in increasing order, i.e., $ \mathfrak{X} = \left( \mathbf{x}_{k} \right)_{k=\left\{ 1, \dots, N \right\}} $ where $ \mathbf{x}_{i} \prec \mathbf{x}_{j} $ if $ 1 \leq i < j \leq N $. Empirically, the strata can be constructed from subsets of these ordered samples. 
Further, recognizing that
\begin{align}
    \mathcal{P} (A_i) &= \mathcal{P} \left( A_* \right) \int_{A_*} \text{I}_{A_i} \left( \mathbf{x} | A_* \right) f_{\mathbf{X} | A_*} \left(\mathbf{x} | A_* \right) d \mathbf{x} = \mathcal{P} \left( A_* \right) \mathbb{E}_{f|A_*} \left[ \text{I}_{A_i} \left( \mathbf{X} | A_* \right) \right] 
\end{align}
we can estimate $\mathcal{P} (A_i)$ by
\begin{align}
    \hat{\mathcal{P}} (A_i) &= \frac{\mathcal{P} \left( A_* \right)}{N} \sum_{k=1}^{N} \text{I}_{A_i} \left( \mathbf{x}_k \right) \quad \text{where } \mathbf{x}_k \sim f_{\mathbf{X} | A_*} \left(\mathbf{x} | A_* \right) \label{eqn:estimating_strata_probability} 
\end{align}
and use this estimator to adaptively define empirical strata $ \tilde{A}_i $ using  the procedure in Algorithm~\ref{Algo:TSS_strata_construction} to achieve a proportional allocation of samples.

\begin{algorithm}[!ht]
\caption{Adaptive Empirical Stratification with Proportional Sample Allocation}
\label{Algo:TSS_strata_construction}
\begin{codefont}
\begin{algorithmic}[1]
\Require $ f_{\mathbf{X}} \left(\mathbf{x} \right) $, $ A_* $, $ \mathcal{P} (A_*) $, $ f_{\mathbf{X}|A_*} \left(\mathbf{x} | A_* \right) $, $ p_0 $, $ m $, $ N $
\For{k = 1:$N$}
\State Draw $ \mathbf{x}_k \sim f_{\mathbf{X}|A_*} \left(\mathbf{x} | A_* \right) $
\State Evaluate $ f_{\mathbf{X}} \left( \mathbf{x}_k \right) $ and $ l \left( \mathbf{x}_k , \mathbf{x}_c \right) $
\EndFor
\State Re-index the samples $ \mathbf{x}_k $ to be in increasing order such that for all $ 1 \leq i < j \leq N $, $ f_{\mathbf{X}} \left(\mathbf{x}_i \right) > f_{\mathbf{X}} \left(\mathbf{x}_j \right) $ or $ l (\mathbf{x}_i, \mathbf{x}_c) < l (\mathbf{x}_j, \mathbf{x}_c) $ if $ f_{\mathbf{X}} \left(\mathbf{x}_i \right) = f_{\mathbf{X}} \left(\mathbf{x}_j \right) $
\State Set $ k = 0 $
\For{i=1:$m$}
\State Initialize $ \tilde{A}_i \gets \emptyset $
\While{$ n \left( \tilde{A}_i \right) < N \left( 1 - p_0 \right) p_0^{i-1} $} \Comment{$ n \left( \tilde{A}_i \right) $ denotes the number of elements in $ \tilde{A}_i $} 
\State $ \tilde{A}_i \gets \tilde{A}_i \cup \left\{ \mathbf{x}_k \right\} $
\State $ k \gets k+1 $
\EndWhile
\EndFor
\end{algorithmic}
\end{codefont}
\end{algorithm}

As discussed in Section~\ref{section:sample_Allocation}, there are alternative sample allocation strategies that achieve better variance reduction than proportional sample allocation. While it is difficult to achieve a specific sample allocation without knowing the strata boundaries in advance, it is possible to obtain a sample allocation with lower variance than proportional allocation while also discovering the strata bounds adaptively using concepts of importance sampling. This is beyond the scope of the current work, but a preliminary discussion can be found in Appendix~\ref{app:is_allocation}.



\section{Challenging Benchmark Problems}
\label{section:examples}

The proposed TSS is fully generalizable to problems with arbitrary input distributions, with theoretical guarantees as detailed above. However, for proof of concept, it is not necessary to explore this generality. Rather, we choose to consider problems with standard normal inputs (recognizing that these can be determined from a transformation of non-normal inputs~\cite{lebrun2009innovating}) and focus on problems with challenging failure domains. In all cases, the null stratum is defined using the design point as in Eq.~\eqref{eqn:design_point_based_safe_set_gaussian}, which is effective in low dimensions. Where possible, we compare TSS with the state-of-the-art subset simulation method, demonstrating its efficiency and robustness.

\subsection{2-D Analytical Case Studies}
\label{section:analytical_2d_examples}

\begin{figure}[!ht]
\centering
\begin{subfigure}{.35\textwidth}
  \centering
  \includegraphics[width=\textwidth]{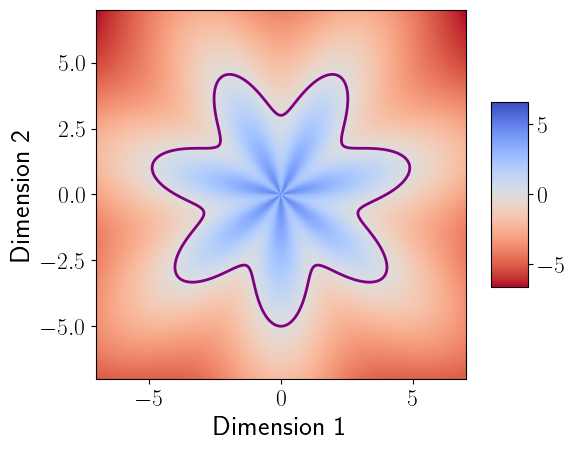}
  \caption{}
  \label{fig:wavy_circle_performance_function}
\end{subfigure}
\begin{subfigure}{.35\textwidth}
  \centering
  \includegraphics[width=\textwidth]{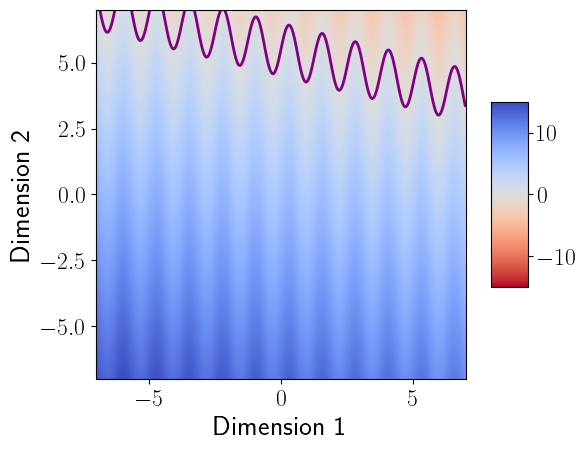}
  \caption{}
  \label{fig:wavy_line_performance_function}
\end{subfigure}
\begin{subfigure}{.35\textwidth}
  \centering
  \includegraphics[width=\textwidth]{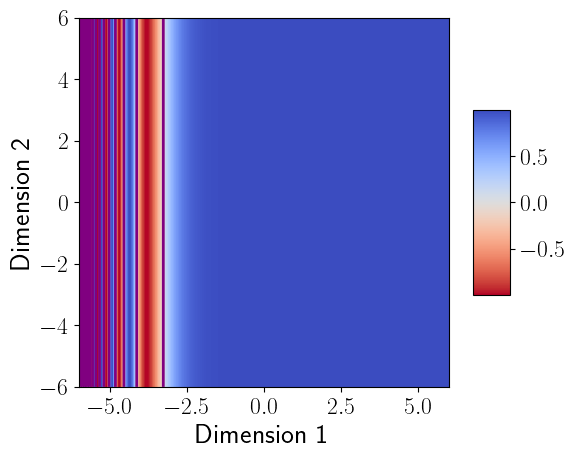}
  \caption{}
  \label{fig:alternating_domains_performance_function}
\end{subfigure}
\begin{subfigure}{.35\textwidth}
  \centering
  \includegraphics[width=\textwidth]{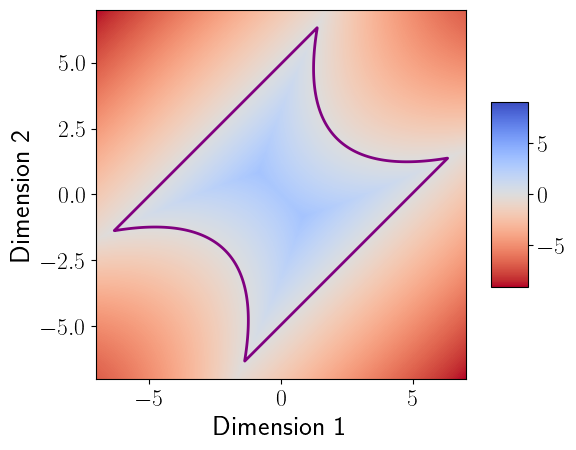}
  \caption{}
  \label{fig:four_branch_performance_function}
\end{subfigure}
\begin{subfigure}{.35\textwidth}
  \centering
  \includegraphics[width=\linewidth]{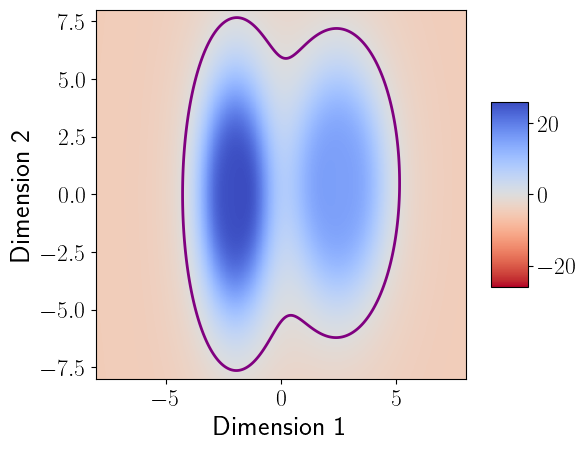}
  \caption{}
  \label{fig:metaball_performance_function}
\end{subfigure}
\begin{subfigure}{.35\textwidth}
  \centering
  \includegraphics[width=\linewidth]{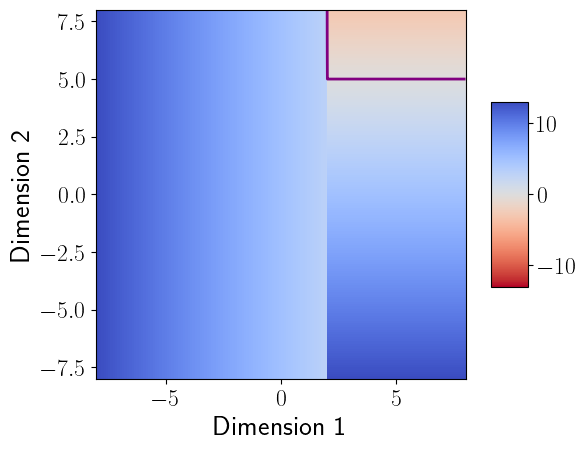}
  \caption{}
  \label{fig:black_swan_performance_function}
\end{subfigure}
\begin{subfigure}{.35\textwidth}
  \centering
  \includegraphics[width=\linewidth]{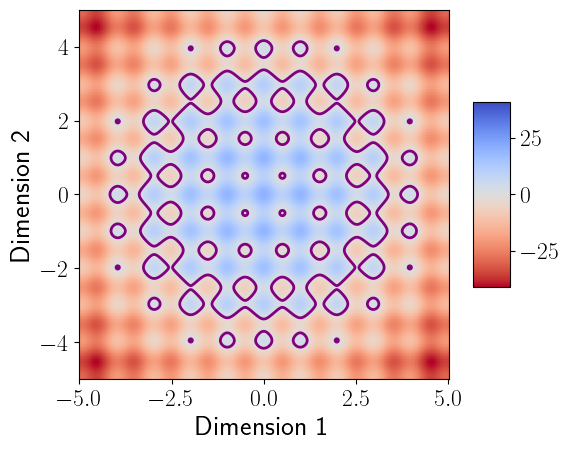}
  \caption{}
  \label{fig:modified_rastrigin_performance_function}
\end{subfigure}
\caption{2D Analytical Problems: Performance function heatmaps for the (a) Wavy Circle, (b) Wavy Line, (c) Alternating Domains, (d) Four-Branch Function, (e) Metaball Function, (f) Black Swan, and (g) Modified Rastrigin. In all cases, the purple line represents the limit surface.}
\label{fig:response_function_visualization_for_2d_analytical_examples}
\end{figure}

First, we apply TSS to seven two-dimensional benchmark reliability problems that illustrate different challenges in failure probability estimation. These problems have been carefully studied in past works~\cite{vovrechovsky2022reliability,breitung2019geometry}, with each shown to pose unique challenges for existing methods. 
Each performance function and the associated true failure probability is provided in Table~\ref{tab:analytical_response_functions}, and plotted in Figure~\ref{fig:response_function_visualization_for_2d_analytical_examples}. 
In each case, $ \mathbf{X} = \begin{bmatrix} X_1 & X_2 \end{bmatrix}^T $, where $ X_1 , X_2 \sim \mathcal{N} \left( 0, 1 \right) $.

\begin{table}[!ht]
    \centering
    \resizebox{\columnwidth}{!}{%
    \begin{tabular}{c|c|c|c}
        Name & Performance Function $ g (\mathbf{x}) $ & $ P_\mathcal{F} $ & $ \beta $ \\
        \hline \hline
        Wavy Circle & $ 4 + \sin(7 \atantwo{\left( x_2, x_1 \right)}) - \lVert \mathbf{x} \rVert_2 $ & $ 2.582 \times 10^{-3} $ & $ 3.0 $ \\
        Wavy Line & $ \displaystyle 5.5 + \sin(5 x_1) - \frac{1}{4} x_1 - x_2 $ & $ 1.217 \times 10^{-6} $ & $ 4.36 $\\
        Alternating Domains & $ \cos \left( x_1 \exp \left( -x_1 - 4 \right) \right) $ & $ 5.266 \times 10^{-4} $ & $ 3.26 $ \\
        Four-Branch Function & $ \displaystyle \min \left( \begin{cases}
            3 + 0.1 \left( x_1 - x_2 \right)^2 - 0.5 \sqrt{2} \left( x_1 + x_2 \right) \\
            3 + 0.1 \left( x_1 - x_2 \right)^2 + 0.5 \sqrt{2} \left( x_1 + x_2 \right) \\
            x_1 - x_2 + 3.5 \sqrt{2} \\
            x_2 - x_1 + 3.5 \sqrt{2}
        \end{cases} \right) $ & $ 2.222 \times 10^{-3} $ & $ 3.0 $ \\
        Metaball Function & $ \displaystyle \frac{30}{\left( \frac{4 \left( x_1 + 2 \right)^2 }{9} + \frac{x_2^2}{25} \right)^2 + 1} + \frac{20}{\left( \frac{\left( x_1 - 2.5 \right)^2}{4} + \frac{\left( x_2 - 0.5 \right)^2}{25} \right)^2 + 1} - 5 $ & $ 1.129 \times 10^{-5} $ & $ 4.26 $ \\
        Black Swan & $ \displaystyle \begin{cases}
            5 - x_1 & \text{for } x_1 \leq 2 \\
            5 - x_2 & \text{for } x_1 > 2
        \end{cases} $ & $ 6.521 \times 10^{-9} $ & $ 5.38 $ \\
        Modified Rastrigin & $ \displaystyle 10 - \sum_{j = 1}^2 \left( x_{j}^2 - 5 \cos \left( 2 \pi x_{j} \right) \right) $ & $ 7.299 \times 10^{-2} $ & $ 0.64 $ \\
        \hline
    \end{tabular}
    }
    \caption{2D analytical performance functions along with their known analytical failure probabilities ($ P_\mathcal{F} $) and reliability index ($ \beta $).}
    \label{tab:analytical_response_functions}
\end{table}

\begin{figure}[!ht]
  \centering
  \includegraphics[width=0.85\linewidth]{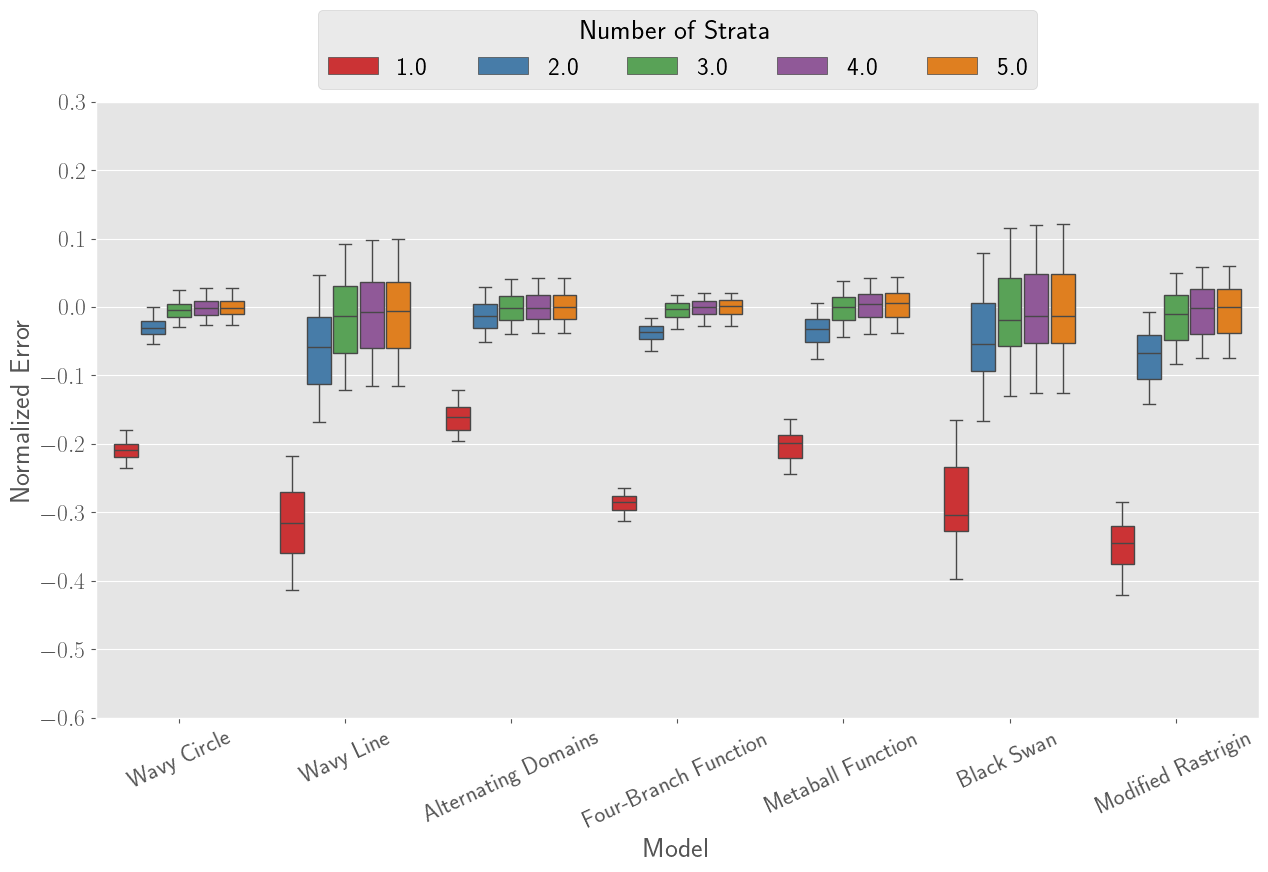}
  \label{fig:tss_lhs_3000_samples_2d_results}
\caption{2D Analytical Problems: Normalized error of the TSS predicted failure probability (i.e., $ \left( P_{\mathcal{F}}^{\text{TSS}} - P_{\mathcal{F}} \right) / P_{\mathcal{F}} $) for all examples with increasing number of strata $ m $ and $ 3000 $ samples per stratum. Boxplots show 95 percentile values from 100 independent trials.
}
\label{fig:tss_lhs_2d_results_normalized_error}
\end{figure}

We begin this study by evaluating the bias in the TSS estimator for increasing number of strata, $m$.
Figure~\ref{fig:tss_lhs_2d_results_normalized_error} shows 95 percentile boxplots 
of the normalized error in the TSS-predicted failure probability for $ m = \left\{1, 2, 3, 4, 5\right\} $ on each example computed from $ 100 $ independent trials of TSS with equal sample allocation of $ N_i = 3000 \; \forall i $ samples per stratum. 
It is clear that the bias becomes indistinguishably small after $m = 4$ in all examples.
Beyond this, increasing to $ m = 10 $ reveals no discernable change in the asymptotic behavior, despite a much larger number of total samples. This is consistent the discussion in Section~\ref{section:TSS_estimator_properties}, which ensures that the bias for $m=4$ in $ P_{\mathcal{F}}^{\text{TSS}} $ is $ \approx 4 $ orders of magnitude less than the target failure probability 
(for $ p_0 = 0.1 $). We therefore suggest using $ p_0 = 0.1 $ and $ m = 4 $ as standard practice.

Figure~\ref{fig:tss_lhs_2d_results_normalized_error} 
further highlights similar behavior across the seven examples, despite the vast difference in the failure probabilities (ranging from $ \mathcal{O}(10^{-2}) $ to $ \mathcal{O}(10^{-9}) $). All seven cases show a rapid
decrease in the bias as $ m $ increases, asymptoting around $ m = 4 $ and comparable bias magnitude at each $ m $. Notice that the variation in error does not decrease as $m$ increases. This is because increasing $m$ places the added samples in regions of much smaller probability that have rapidly diminishing influence on the variance of the estimator, as described in Section~\ref{section:TSS_estimator_properties}.

Next, we estimate the failure probability for each of the seven cases using proportional sample allocation and equal sample allocation with $ p_0 = 0.1 $, $ m = 4 $ strata, and a fixed number of $ N = 4000 $ total samples. For each sample allocation scheme, the samples in each stratum were generated using simple random i.i.d. sampling (MCS) and Latin Hypercube Sampling (LHS). 
Table~\ref{tab:analytical_2d_example_tss_results} shows the mean probability of failure estimate and coefficient of variation from 100 independent trials for each case. We see that in all cases, TSS provides precise and accurate estimates of the failure probability, with very low coefficient of variation, regardless of its magnitude. Proportional allocation generally performs better than equal allocation and LHS provides some additional variance reduction compared to MCS. This is expected because proportional allocation places more samples in the larger strata that contribute significantly more to the variance, as discussed in Section~\ref{section:TSS_estimator_properties}.


\begin{table}[!ht]
    \centering
    \resizebox{\columnwidth}{!}{%
    \begin{tabular}{c|c|c|c|c}
        \multirow{2}{*}{Name} & \multicolumn{2}{c|}{Proportional Allocation} & \multicolumn{2}{c}{Equal Allocation} \\
         & MCS & LHS & MCS & LHS \\
        \hline \hline
        Wavy Circle & $ 2.58 \times 10^{-3} $ ($ 2.8 \% $) & $ 2.58 \times 10^{-3} $ ($ 1.5 \% $) & $ 2.58 \times 10^{-3} $ ($ 4.5 \% $) & $ 2.58 \times 10^{-3} $ ($ 2.1 \% $) \\
        Wavy Line & $ 1.23 \times 10^{-6} $ ($ 12.2 \% $) & $ 1.21 \times 10^{-6} $ ($ 6.4 \% $) & $ 1.21 \times 10^{-6} $ ($ 21.2 \% $) & $ 1.24 \times 10^{-6} $ ($ 12.3 \% $) \\
        Alternating Domains & $ 5.28 \times 10^{-4} $ ($ 4.4 \% $) & $ 5.28 \times 10^{-4} $ ($ 2.2 \% $) & $ 5.19 \times 10^{-4} $ ($ 7.7 \% $) & $ 5.29 \times 10^{-4} $ ($ 4.3 \% $) \\
        Four-Branch Function & $ 2.22 \times 10^{-3} $ ($ 2.8 \% $) & $ 2.22 \times 10^{-3} $ ($ 1.6 \% $) & $ 2.22 \times 10^{-3} $ ($ 5.1 \% $) & $ 2.22 \times 10^{-3} $ ($ 2.4 \% $) \\
        Metaball Function & $ 1.13 \times 10^{-5} $ ($ 5.1 \% $) & $ 1.13 \times 10^{-5} $ ($ 2.3 \% $) & $ 1.11 \times 10^{-5} $ ($ 7.8 \% $) & $ 1.13 \times 10^{-5} $ ($ 4.5 \% $) \\
        Black Swan & $ 6.58 \times 10^{-9} $ ($ 12.8 \% $) & $ 6.52 \times 10^{-9} $ ($ 7.9 \% $) & $ 6.50 \times 10^{-9} $ ($ 20.3 \% $) & $ 6.65 \times 10^{-9} $ ($ 15.7 \% $) \\
        Modified Rastrigin & $ 7.27 \times 10^{-2} $ ($ 4.7 \% $) & $ 7.28 \times 10^{-2} $ ($ 3.9 \% $) & $ 7.22 \times 10^{-2} $ ($ 8.2 \% $) & $ 7.14 \times 10^{-2} $ ($ 6.4 \% $) \\
        \hline
    \end{tabular}
    }
    \caption{2D Analytical Problems: TSS failure probability estimates using $ p_0 = 0.1 $, $ m = 4 $, and $ N = 4000 $ for proportional allocation and equal allocation with simple Monte Carlo sampling (MCS) and Latin hypercube sampling (LHS). Estimates are the mean value from 100 independent trials and the coefficient of variation is presented in parentheses. The true failure probabilities are provided in Table~\ref{tab:analytical_response_functions}.}
    \label{tab:analytical_2d_example_tss_results}
\end{table}

To further demonstrate the efficiency and robustness of TSS, we compare it against SuS for all the examples except the Black Swan problem, where SuS does not converge.
The results are shown in Table~\ref{tab:analytical_2d_example_tss_sus_comparison_equal_total_sample_size} with mean probability of failure and coefficient of variation from 100 independent trials. 
In all cases, SuS is run with $ 1000 $ and $ 10,000 $ samples per subset. In both cases, the total number of samples for TSS, $N$ is taken as the total number of samples from SuS averaged over $ 100 $ trials. 
TSS is applied with $ p_0 = 0.1 $, $ m = 4 $, and proportional sample allocation, using both LHS and MCS to sample within each stratum. 
In all cases, TSS significantly outperforms SuS in terms of both accuracy and precision, with the CoV of the TSS estimate being $ 1-2 $ orders of magnitude lower than that of SuS. Even for cases where SuS is not stable, viz. the Wavy Line, Metaball Function, and Black Swan, TSS has no difficulty estimating $ P_{\mathcal{F}} $ accurately.

\begin{table}[!ht]
    \centering
    \begin{tabular}{c|c|c|c|c}
        Name & Total Samples & Subset Simulation & \multicolumn{2}{c}{Tail Stratified Sampling} \\
         &  &  & LHS & MCS \\
        \hline \hline
        Wavy Circle & $ 2809 $ & $ 2.22 \times 10^{-3} $ ($ 33.2 \% $) & $ 2.58 \times 10^{-3} $ ($ 1.6 \% $) & $ 2.58 \times 10^{-3} $ ($ 3.5 \% $) \\
         & $ 28000 $ &  $ 2.65 \times 10^{-3} $ ($ 8.2 \% $) & $ 2.58 \times 10^{-3} $ ($ 0.6 \% $) & $ 2.58 \times 10^{-3} $ ($ 1.2 \% $) \\
        Wavy Line & $ 6076 $ &  $ 1.16 \times 10^{-6} $ ($ 101.5 \% $) & $ 1.23 \times 10^{-6} $ ($ 5.5 \% $) & $ 1.25 \times 10^{-6} $ ($ 9.2 \% $) \\
         & $ 58150 $ &  $ 1.36 \times 10^{-6} $ ($ 50.4 \% $) & $ 1.22 \times 10^{-6} $ ($ 2.1 \% $) & $ 1.22 \times 10^{-6} $ ($ 2.9 \% $) \\
        Alternating Domains & $ 3646 $ &  $ 4.95 \times 10^{-4} $ ($ 52.8 \% $) & $ 5.26 \times 10^{-4} $ ($ 2.2 \% $) & $ 5.25 \times 10^{-4} $ ($ 4.3 \% $) \\
         & $ 37000 $ & $ 4.97 \times 10^{-4} $ ($ 11.2 \% $) & $ 5.27 \times 10^{-4} $ ($ 0.7 \% $) & $ 5.28 \times 10^{-4} $ ($ 1.4 \% $) \\
        Four-Branch Function & $ 2809 $ & $ 2.12 \times 10^{-3} $ ($ 25.0 \% $) & $ 2.22 \times 10^{-3} $ ($ 1.9 \% $) & $ 2.21 \times 10^{-3} $ ($ 3.6 \% $) \\
         & $ 28000 $ & $ 2.17 \times 10^{-3} $ ($ 8.6 \% $) & $ 2.22 \times 10^{-3} $ ($ 0.6 \% $) & $ 2.22 \times 10^{-3} $ ($ 1.0 \% $) \\
        Metaball Function & $ 6985 $ & $ 1.34 \times 10^{-5} $ ($ 238.9 \% $) & $ 1.13 \times 10^{-5} $ ($ 1.6 \% $) & $ 1.13 \times 10^{-5} $ ($ 3.0 \% $) \\
         & $ 55450 $ & $ 1.25 \times 10^{-5} $ ($ 88.7 \% $) & $ 1.13 \times 10^{-5} $ ($ 0.6 \% $) & $ 1.13 \times 10^{-5} $ ($ 1.4 \% $) \\
        Black Swan & $ 4000 $ & Does not converge & $ 6.52 \times 10^{-9} $ ($ 7.9 \% $) & $ 6.58 \times 10^{-9} $ ($ 12.8 \% $) \\
        Modified Rastrigin & $ 1900 $ & $ 7.49 \times 10^{-2} $ ($ 12.7 \% $) & $ 7.23 \times 10^{-2} $ ($ 6.4 \% $) & $ 7.26 \times 10^{-2} $ ($ 7.5 \% $) \\
         & $ 19000 $ & $ 7.13 \times 10^{-2} $ ($ 2.3 \% $) & $ 7.27 \times 10^{-2} $ ($ 1.9 \% $) & $ 7.27 \times 10^{-2} $ ($ 2.2 \% $) \\
        \hline
    \end{tabular}
    \caption{2D Analytical Problems: Comparison of failure probabilities estimated by SuS and TSS ($ p_0 = 0.1 $, $ m = 4 $) with proportional sample allocation. Samples for TSS were generated using both LHS and MCS. Presented results are the mean failure probability of 100 independent trials and the coefficient of variation is presented in parenthesis. The true failure probabilities for all examples are provided in Table~\ref{tab:analytical_response_functions}. 
    }
    \label{tab:analytical_2d_example_tss_sus_comparison_equal_total_sample_size}
\end{table}

\begin{figure}[!ht]
\centering
\begin{subfigure}{.32\textwidth}
  \centering
  \includegraphics[width=\linewidth]{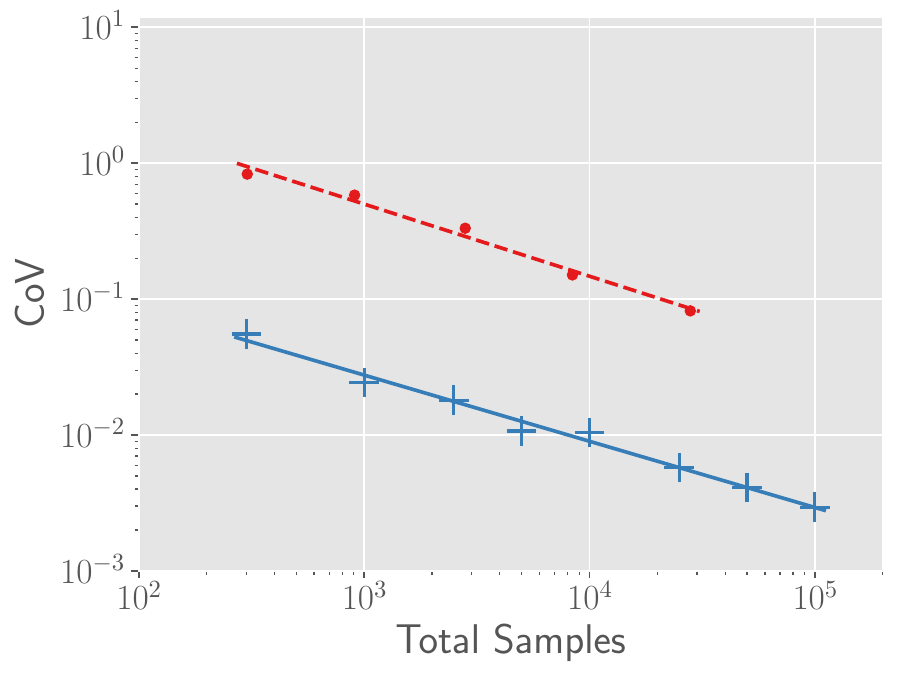}
  \caption{}
  \label{fig:wavy_circle_cov_convergence}
\end{subfigure}%
\begin{subfigure}{.32\textwidth}
  \centering
  \includegraphics[width=\linewidth]{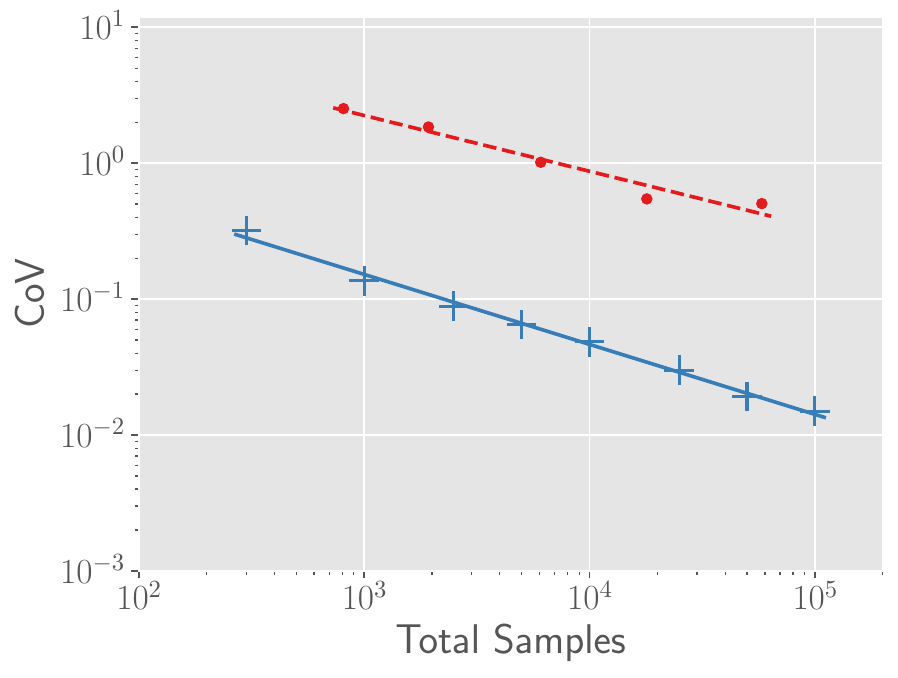}
  \caption{}
  \label{fig:wavy_line_cov_convergence}
\end{subfigure}%
\begin{subfigure}{.32\textwidth}
  \centering
  \includegraphics[width=\linewidth]{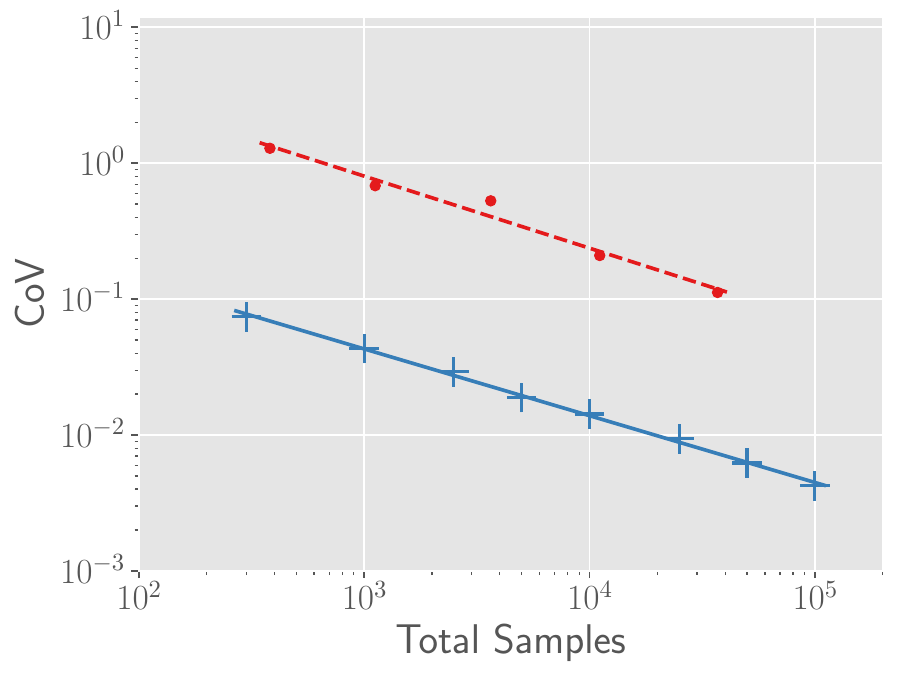}
  \caption{}
  \label{fig:alternating_domains_cov_convergence}
\end{subfigure}
\begin{subfigure}{.32\textwidth}
  \centering
  \includegraphics[width=\linewidth]{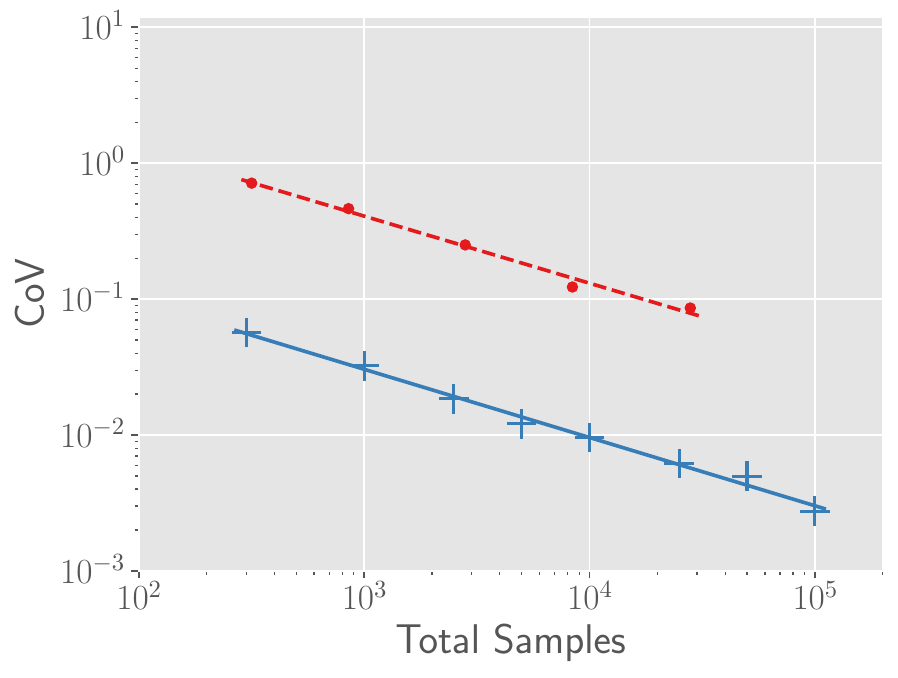}
  \caption{}
  \label{fig:four_branch_cov_convergence}
\end{subfigure}%
\begin{subfigure}{.32\textwidth}
  \centering
  \includegraphics[width=\linewidth]{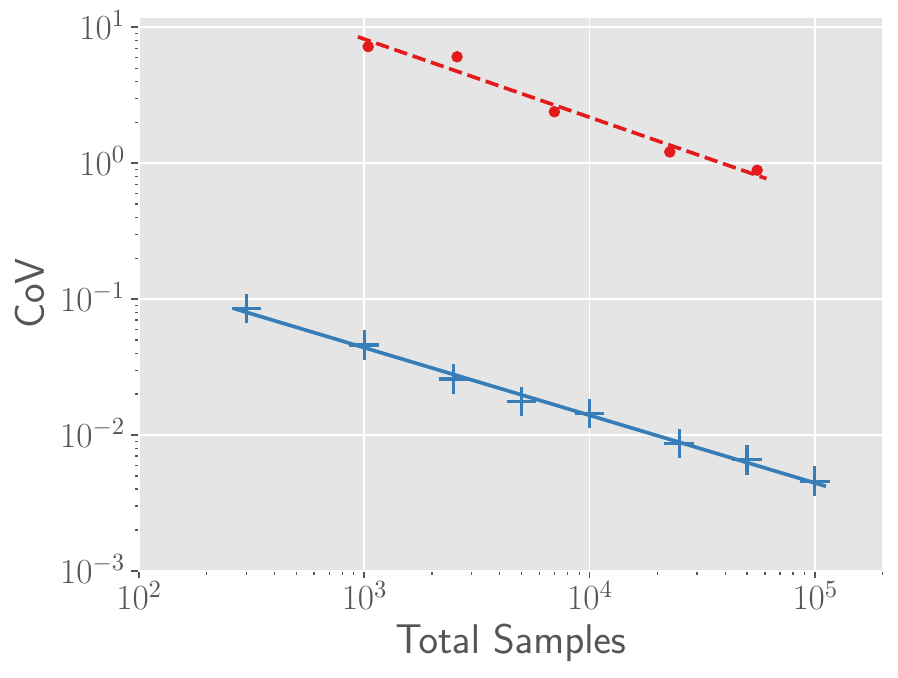}
  \caption{}
  \label{fig:metaball_cov_convergence}
\end{subfigure}%
\begin{subfigure}{.32\textwidth}
  \centering
  \includegraphics[width=\linewidth]{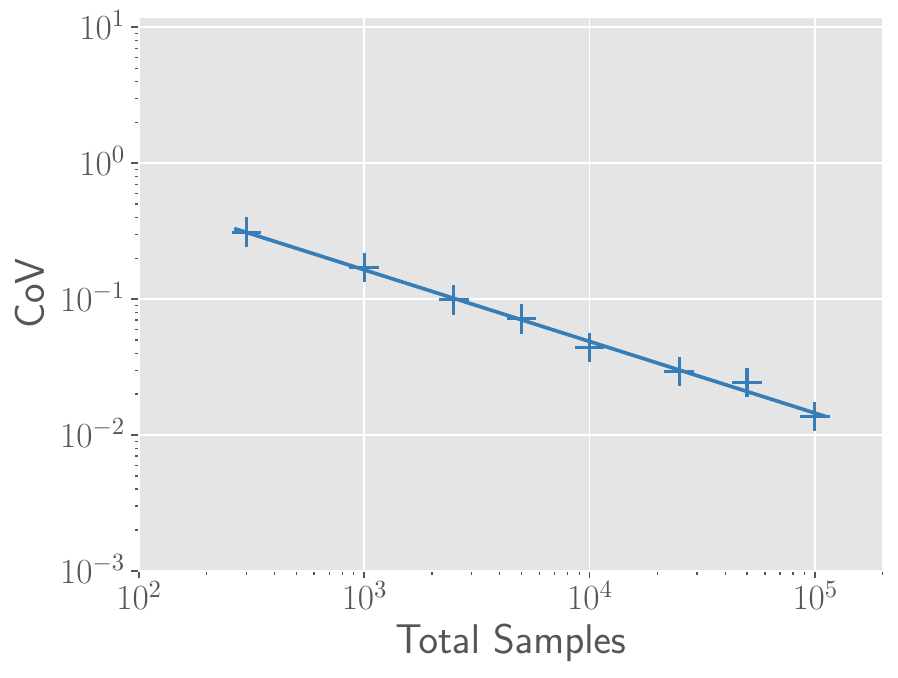}
  \caption{}
  \label{fig:black_swan_cov_convergence}
\end{subfigure}
\begin{subfigure}{.32\textwidth}
  \centering
  \includegraphics[width=\linewidth]{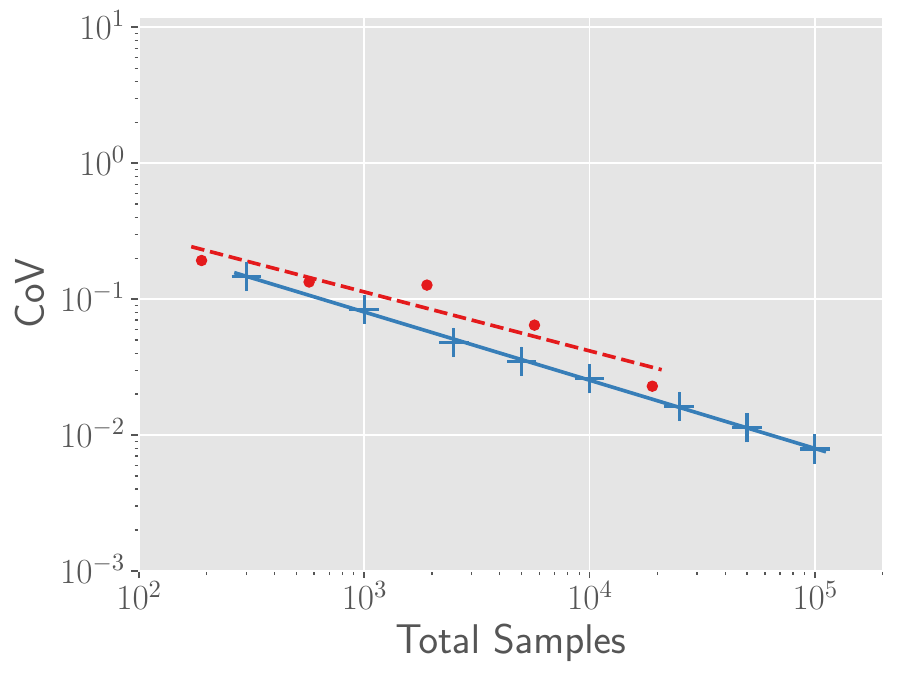}
  \caption{}
  \label{fig:modified_rastrigin_cov_convergence}
\end{subfigure}
\caption{2D Analytical Problems: Convergence of the CoV for TSS (blue lines) and SuS (red lines) for increasing sample sizes for (a) Wavy Circle, (b) Wavy Line, (c) Alternating Domains, (d) Four-Branch Function, (e) Metaball Function, (f) Black Swan, and (g) Modified Rastrigin. The solid blue line indicates the best linear regression fit for the estimated CoV values (blue crosses) for TSS ($p_0 = 0.1$, $ m = 4 $, proportional sample allocation with LHS), while the dashed red line indicates the same for SuS (with estimated CoV values shown as red dots). All CoV values are estimated from 100 independent trials. The red dotted line is absent for figure (f) since SuS does not converge for the Black Swan example.}
\label{fig:CoV_convergence_for_analytical_2d_examples}
\end{figure}

Finally, to underscore the performance of TSS, Figure~\ref{fig:CoV_convergence_for_analytical_2d_examples} compares the convergence of the CoV for TSS with that of SuS for increasing sample size, for all seven examples. As before, TSS was implemented with $p_0=0.1$, $ m = 4 $, and proportional sample allocation, using LHS to sample each stratum. The plots show CoV values for TSS for $ N =  300, 1000, 2500, 5000, 10000, 25000, 50000, 100000  $. In parallel, SuS was implemented with $ 100 $, $ 300 $, $ 1000 $, $ 3000$, and $ 10000 $ samples per subset. The CoV for both methods was computed from $ 100 $ independent trials. In all cases, we see the expected exponential decrease in the CoV with increasing total sample size, with TSS and SuS having the same rate of decrease. 
The TSS estimates, however, are consistently 1-2 orders of magnitude lower than the SuS estimates for a fixed sample size. The CoV magnitudes are only comparable for the Modified Rastrigin problem, which is a priori expected to show relatively poor performance for TSS because failure is not concentrated deep in the tails. Nonetheless, TSS performs marginally better than SuS. In the case of the Black Swan, SuS does not converge but TSS nonetheless performs well. 
Further, with just $ N = 1000 $ samples, TSS achieves around $ 10 \% $ CoV in all cases and even lower in several cases. SuS struggles to achieve this benchmark even with $ 10,000 $ samples or more. In short, for this wide variety of 2-D geometries and magnitudes of target failure probability, TSS proves to be extremely effective.

\subsection{Stable Buckling of Rigid Bars}
\label{section:buckling_problem}

\begin{figure}[!ht]
\centering
\includegraphics[width=0.4\linewidth]{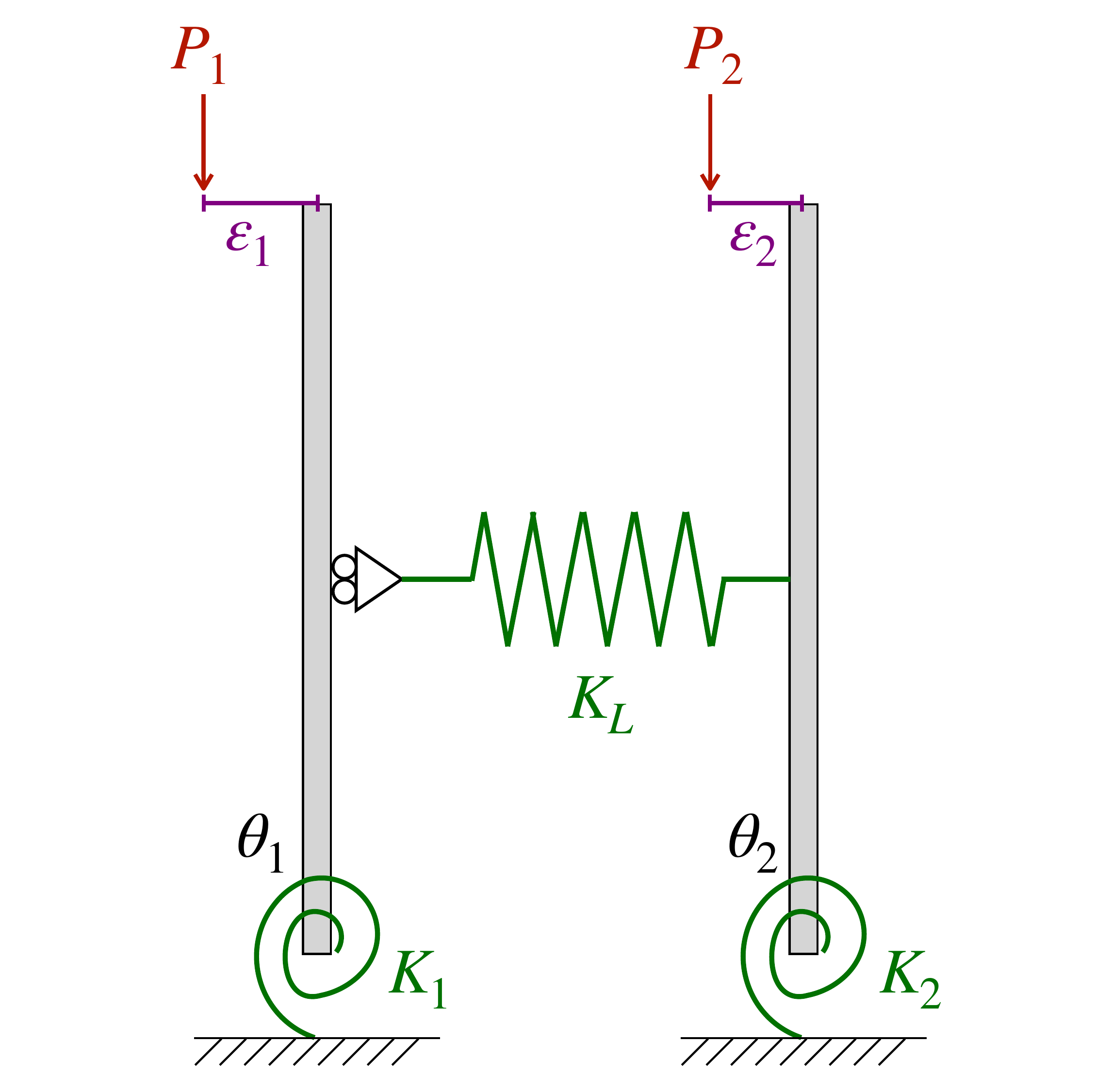}
\caption{System of rigid vertical bars prone to stable buckling behavior}
\label{fig:buckling_example_visualization}
\end{figure}

Next, we consider the case of two connected rigid bars subjected to eccentric axial loads, as depicted in Figure~\ref{fig:buckling_example_visualization}. The bars are each of length $ \SI{1}{m} $, attached vertically to the ground with torsional springs (stiffnesses $ K_1 $ and $ K_2 $), and connected to each other by a linear spring (stiffness $ K_L $) which is rigidly affixed to the midpoint of the right bar and attached via roller to the left bar, ensuring the spring stays horizontal. Loads $ P_1 $ and $ P_2 $ are applied vertically downwards to the left and right bars, respectively, at an eccentricity of $ \varepsilon_1 $ and $ \varepsilon_2 $ from their respective axial centers, causing the bars to rotate by angles $ \theta_1 $ and $ \theta_2 $. The load-deflection behavior is assumed to be quasi-static.

This system was previously studied by Sharma and Manohar~\cite{sharma2023modified}, who considered stochastic eccentricities and defined failure as the tip of the left bar experiencing vertical displacement beyond an acceptable threshold, leading to two disjoint failure regions in the $ 2 $-D input space. We consider the applied loads $ P_1 $ and $ P_2 $, the eccentricities $ \varepsilon_1 $ and $ \varepsilon_2 $, and all spring stiffnesses $ K_1 $, $ K_2 $ and $ K_L $ to be stochastic, leading to a $ 7 $-dimensional input space. Letting $ \mathbf{X} = \begin{bmatrix} X_1 & X_2 & \dots & X_7 \end{bmatrix}^T $ denote the uncorrelated standard normal input vector, the stochastic parameters are defined through the following transformations
\begin{equation}
    \begin{gathered}
        \begin{aligned}
            P_1 &= 1 + 0.1 \exp \left( 1 + 0.25 X_1 \right) &\quad P_2 &= 1 + 0.1 \exp \left( 1 + 0.25 X_2 \right) \\
            \varepsilon_1 &= 0.1 \left( \Phi (X_3) - 0.5 \right) &\quad \varepsilon_2 &= 0.1 \left( \Phi (X_4) - 0.5 \right) \\
            K_1 &= 2 + 0.6 \left( F_{\left(\beta, 5, 5 \right)}^{-1} \left( \Phi (X_5) \right) - 0.5 \right) &\quad K_2 &= 2 + 0.6 \left( F_{\left(\beta, 5, 5 \right)}^{-1} \left( \Phi (X_6) \right) - 0.5 \right)
        \end{aligned} \\
        K_L = 1 + 0.4 \left( F_{\left(\beta, 5, 5 \right)}^{-1} \left( \Phi (X_7) \right) - 0.5 \right)
    \end{gathered}
\end{equation}
where $ P_1 $ and $ P_2 $ are in \SI{}{GN}, $ \varepsilon_1 $ and $ \varepsilon_2 $ are in $ \SI{}{m} $, $ K_1 $ and $ K_2 $ are in $ \SI{}{GN/m} $, and $ K_L $ is in $ \SI{}{GNm/rad}$. $ \Phi (\cdot) $ represents the standard normal cumulative distribution function (cdf) and $ F_{\left(\beta, 5, 5 \right)}^{-1} (\cdot) $ represents the inverse cdf of the $ \text{Beta} \left(\alpha=5, \beta=5 \right) $ distribution.

For a given realization of the input vector $ \mathbf{X} = \mathbf{x} $, a stable deflection state, denoted by $ \bm{\theta}_{S|\mathbf{x}} = \begin{bmatrix} \theta_{1s|\mathbf{x}} & \theta_{2s|\mathbf{x}} \end{bmatrix}^T $, occurs when the bars are in static equilibrium for $ \theta_1 = \theta_{1s|\mathbf{x}} $ and $ \theta_2 = \theta_{2s|\mathbf{x}} $. In accordance with Thompson and Hunt~\cite{ThompsonHuntElasticStability}, stable solutions for the deflection are found for values of $ \bm{\theta} = \begin{bmatrix} \theta_{1} & \theta_{2} \end{bmatrix}^T $ such that
\begin{enumerate}
    \item $ \nabla_{\bm{\theta}} V = \mathbf{0} $, where $ \nabla_{\bm{\theta}} V $ is the gradient of the potential energy of the system $ V $ w.r.t. $ \bm{\theta} $, and
    \item $ \mathbf{H}_V $ (the Hessian of the potential energy) is positive definite, which is satisfied if $ \det \left( \mathbf{H}_V \right) > 0 $ and $ \frac{\partial^2 V}{\partial \theta_1^2} > 0$.
\end{enumerate}
For our system, the potential energy can be written as
\begin{equation}
    V = \frac{1}{2} K_1 \theta_1^2 + \frac{1}{2} K_2 \theta_2^2 + \frac{1}{2} K_L \left[ 0.5 \cos \theta_2 \left( \tan \theta_2 - \tan \theta_1 \right) \right]^2 - P_1 \left[ 1 - \cos \theta_1 + \varepsilon_1 \sin \theta_1 \right] - P_2 \left[ 1 - \cos \theta_2 + \varepsilon_2 \sin \theta_2 \right]
\end{equation}
and stable deflection solutions can be found by numerically solving for the gradient and Hessian for a given $ \mathbf{x} $.

Next, we define the set $ \Theta_{S| \mathbf{x}} $, which contains every stable deflection state $ \bm{\theta}_{S|\mathbf{x}} $ of the system for a given realization $ \mathbf{x} $. Using this, the performance function of the system can be represented as follows.
\begin{equation}
    g(\mathbf{x}) = 0.2 - \max_{\bm{\theta} \in \Theta_{S| \mathbf{x}}} \left( \lVert \bm{\theta} \rVert_\infty \right)
\end{equation}
That is, failure occurs when any of the stable solutions for a given $ \mathbf{x} $ cause either of the bars to deflect by more than $ \SI{0.2}{rad} $. Using simple Monte Carlo with $ 10^8 $ samples, the system was found to have a failure probability of $ 2.424 \times 10^{-5} $. 

\begin{figure}[!ht]
\centering
\begin{subfigure}{.49\textwidth}
  \centering
  \includegraphics[width=\linewidth]{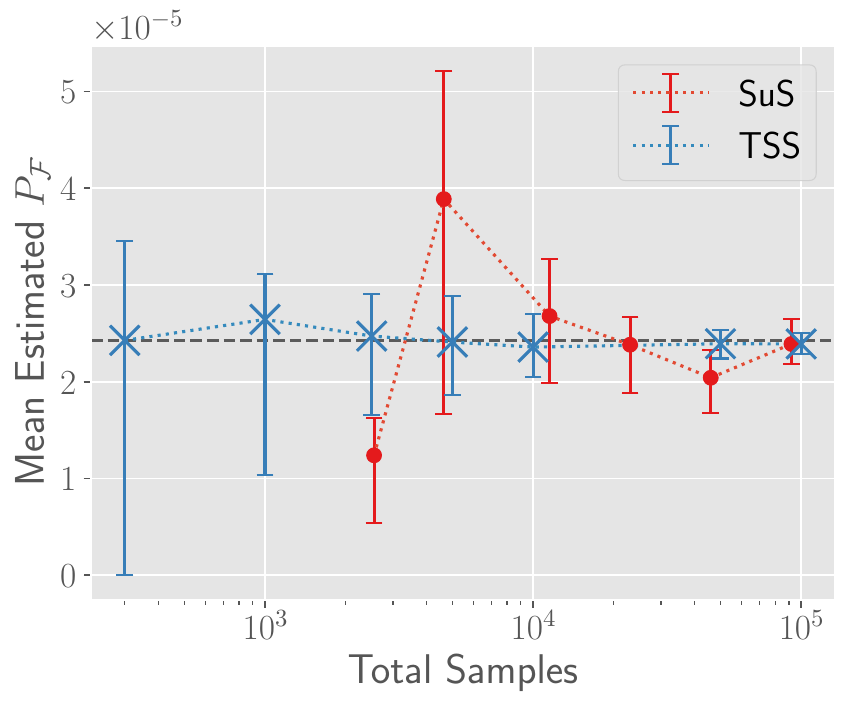}
  \caption{}
  \label{fig:buckling_problem_means}
\end{subfigure}%
\begin{subfigure}{.49\textwidth}
  \centering
  \includegraphics[width=\linewidth]{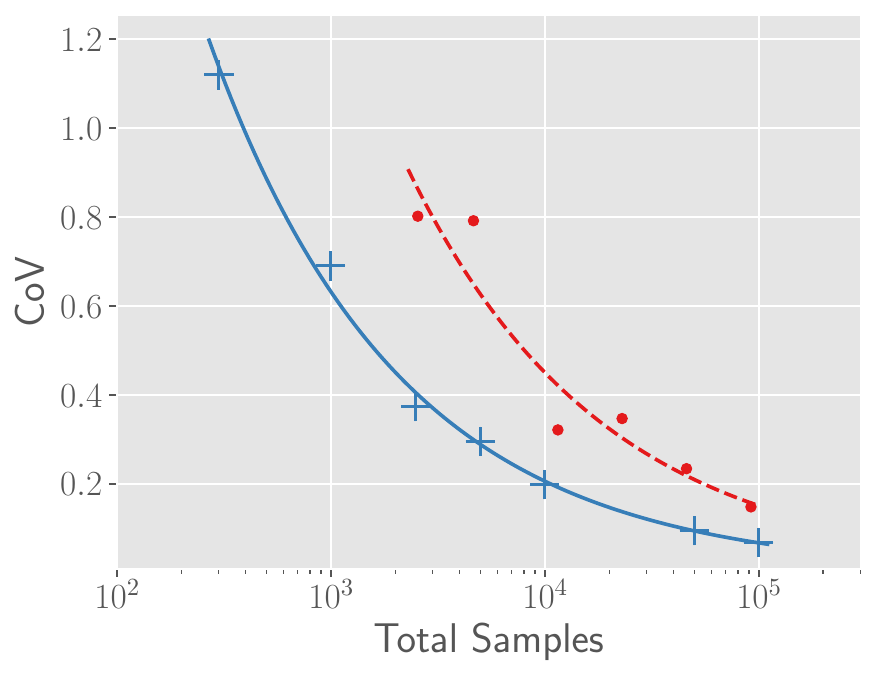}
  \caption{}
  \label{fig:buckling_problem_cov}
\end{subfigure}
\caption{Stable Buckling Problem: Performance comparison between SuS and TSS. (a) Mean predicted failure probability computed from $ 100 $ independent trials with error bars depicting the $ 25 $ - $ 75 $ percentile range for increasing sample size. (b) Coefficient of Variation of the predicted failure probability computed from $ 100 $ independent trials for increasing sample size. The dashed black line in (a) represents the true failure probability.}
\label{fig:buckling_problem_results}
\end{figure}

The performance of TSS for this problem is compared with SuS in Figure~\ref{fig:buckling_problem_results} for 100 independent trials. TSS was applied using $ p_0 = 0.1 $, $ m = 4 $, and $ N \in \left\{ 300, 1000, 2500, 5000, 10000, 50000, 100000 \right\} $ with proportional sample allocation and simple Monte Carlo sampling in each stratum, while SuS was implemented with $ 500 $, $ 1000 $, $ 2500 $, $ 5000 $, $ 10000 $, and $ 20000 $ samples per subset. TSS consistently produces estimates with smaller coefficients of variation than SuS by a factor of $ \approx 2 $ across all sample sizes. Furthermore, SuS produces noticeably biased results, particularly for small sample sizes, while TSS does not. 

\section{High-Dimensional Problems}
\label{section:Hi-D}

As mentioned above, the TSS method does not explicitly depend on the dimension of the problem. But, it requires identifying an effective stratification that can isolate the distribution tails (i.e. identifying a good $A_0$), which can be challenging in high dimensions. In this section, we explore three high-dimensional examples and show how its performance varies for cases where the tails can be effectively isolated and cases where they cannot.

\subsection{Cantilever Beam Tip Deflection}
\label{section:cantilever_tip_deflection}

\begin{figure}[!ht]
\centering
\includegraphics[width=0.7\linewidth]{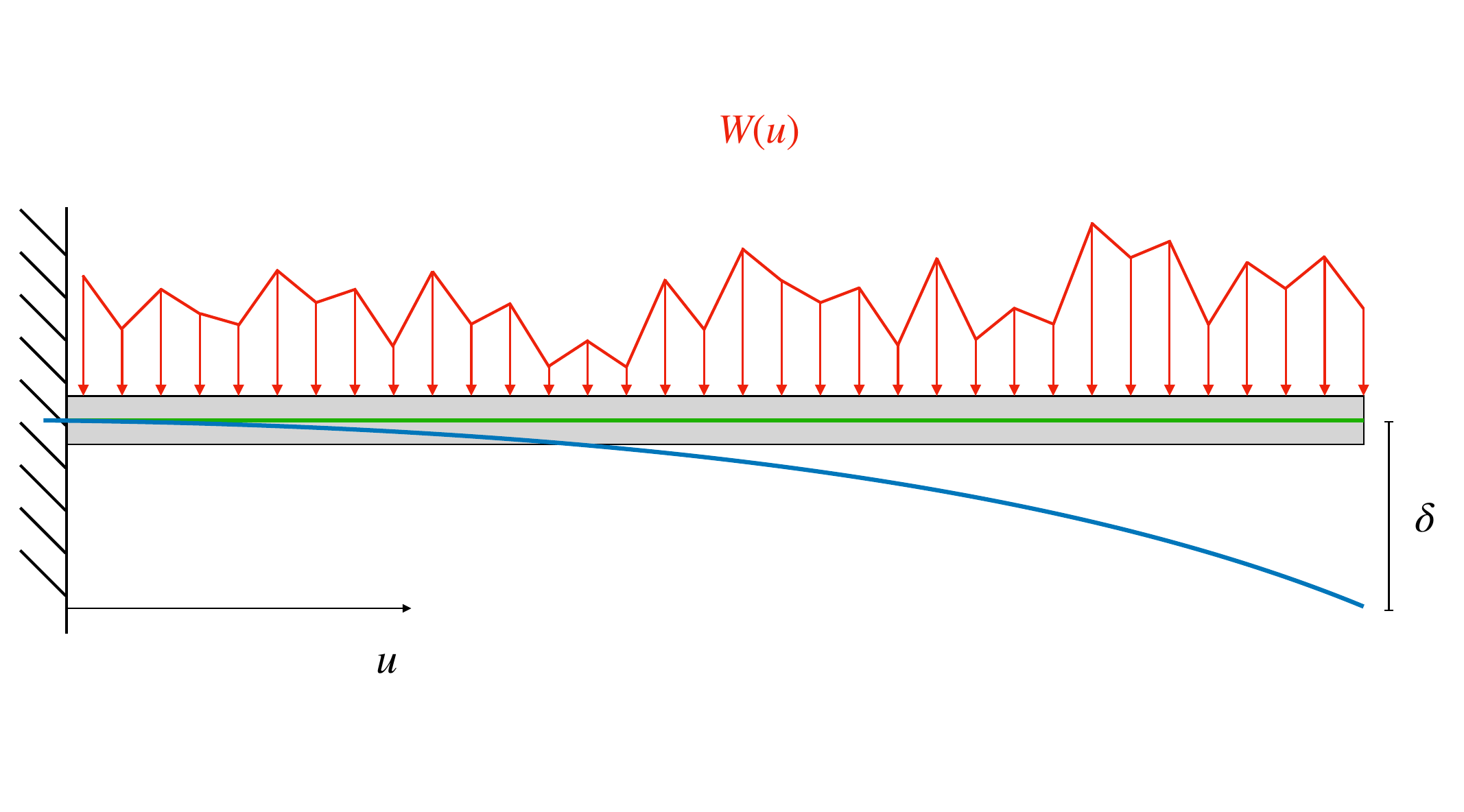}
\caption{Cantilever beam subjected to a stochastically distributed load $ W(u) $, leading to a tip deflection of $ \delta $. 
}
\label{fig:deflection_example_visualization}
\end{figure}

For the first high-dimensional case, we consider the tip deflection of a steel cantilever beam subjected to a stochastically distributed load, as depicted in Figure~\ref{fig:deflection_example_visualization}. The beam has a rectangular cross section with a width of $ \SI{160}{mm} $, a depth of $ \SI{240}{mm} $, and is $ \SI{10}{m} $ long. The applied load $ W(u) $ is nominally uniformly distributed across the length of the beam with a random error at each point defined as follows:
\begin{equation}
    \label{eqn:beam_loading}
    W(u_j) = w_0 + \frac{w_0}{2} \left( \exp \left( X_j \right) - 1 \right) \quad j = 1, \dots, d
\end{equation}
where $ u_j = j \Delta u $ is the location of the $j^{th}$ point load, $ d $ is the number of discretizations, $ w_0 = \nicefrac{20}{d} $ (in $ \SI{}{kN} $) is the integrated contribution of the nominal uniform distributed load at each point $ u_j $, and 
$ X_j \sim \mathcal{N} (0, 1) $. 
We take $ \Delta u = \SI{0.01}{m} \Rightarrow d = 1000 $, leading to a $ 1000 $-D input vector.

Failure occurs when the total tip deflection $ \delta \geq \SI{10}{cm} $, 
where the tip deflection can be determined by superposition as:
\begin{align}
    \delta &= \sum_{j=1}^d \delta_j \label{eqn:total_beam_tip_deflection} \\
    \delta_j &= W(u_j) \frac{u_j^2 \left( 3L - u_j \right)}{6 E I} \label{eqn:beam_tip_deflection_component}
\end{align}
where $ L $ is the length of the beam, $ E = \SI{200}{GPa} $ is the Young's modulus, and $ I $ is the moment of inertia of the beam.
Using $ 10^8 $ samples, simple Monte Carlo was used to compute the true failure probability as $ 1.782 \times 10^{-3} $. 

We again compare the mean predicted failure probability and the sample CoV from $ 100 $ independent trials of TSS and SuS. For TSS, $ p_0 = 0.1 $, $ m = 4 $, and $ N = 1000, 2500, 5000, 10000, 30000 $ with simple Monte Carlo sampling in each stratum and proportional sample allocation, while $500$, $1000$, $2500$, $5000$, and $10000$ samples per subset were used for SuS. In TSS, we define the null stratum $A_0$ as a hypersphere whose radius equals the reliability index, $\beta $. As we'll see, this is not a good choice in high dimensions; past works~\cite{katafygiotis2008geometric} have highlighted the statistical insignificance of the design point in high dimensions. Correspondingly, in our case, the probability of the null stratum is vanishingly small ($\mathcal{P}(A_0) \approx 0.002 $) in 1000 dimensions, and does not effectively isolate the tails of the distribution. Nonetheless, TSS provides a robust estimator.

\begin{figure}[!ht]
\centering
\begin{subfigure}{.49\textwidth}
  \centering
  \includegraphics[width=\linewidth]{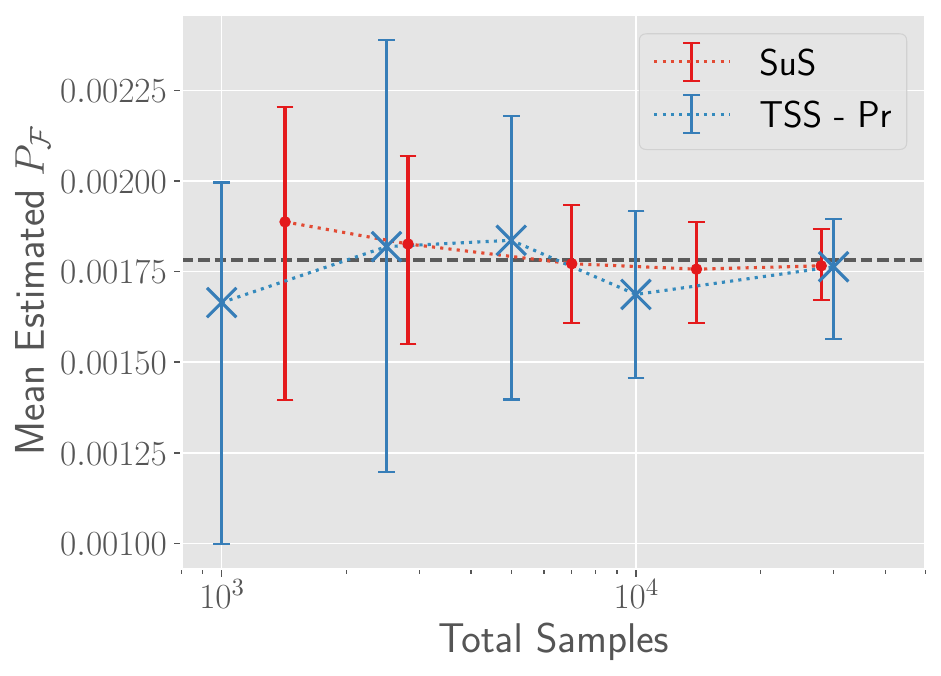}
  \caption{}
  \label{fig:deflection_problem_means}
\end{subfigure}%
\begin{subfigure}{.49\textwidth}
  \centering
  \includegraphics[width=\linewidth]{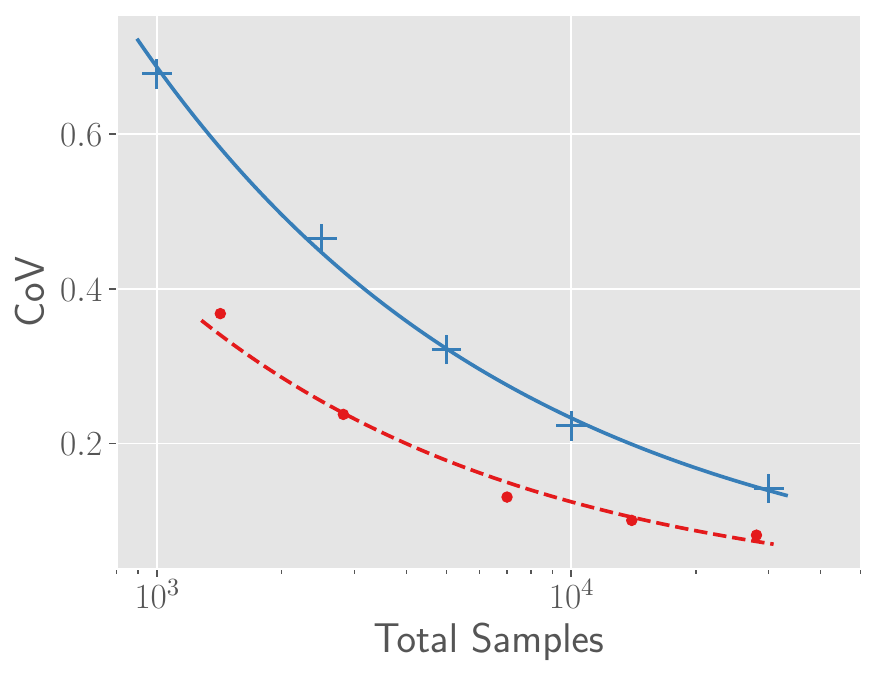}
  \caption{}
  \label{fig:deflection_problem_cov}
\end{subfigure}
\caption{Cantilever beam problem: Performance comparison between SuS and TSS. (a) Mean predicted failure probability computed from $ 100 $ independent trials with error bars depicting the $ 25 $ - $ 75 $ percentile range for increasing sample size. (b) Coefficient of Variation of the predicted failure probability computed from $ 100 $ independent trials for increasing sample size. The dashed black line in (a) represents the true failure probability.}
\label{fig:deflection_problem_results}
\end{figure}

Figure~\ref{fig:deflection_problem_results} shows the results where we see that SuS outperforms TSS, despite both methods producing stable estimates with low bias for moderate-to-large sample sizes. This happens because we have not identified a low-probability tail for TSS over which failures can be isolated. Although failures certainly occur in the tails of this distribution, they happen in the tails of a subset of dimensions. Additional insights would be necessary to identify a low probability tail $A_*$. Meanwhile, SuS uses the performance function to guide samples toward the failure domain in the tails and is therefore effective here.  



Importantly, because TSS is guaranteed to provide a variance reduction and give an unbiased estimate, it will always provide improvements over Monte Carlo sampling. Moreover, it provides a robust estimator with convergence guarantees that most contemporary reliability analysis algorithms, SuS included, do not provide.

\subsection{SDOF Exceedance Probability} 
\label{section:first_passage_problem}

The second example studies an adapted version of the single-degree-of-freedom system from Au and Beck~\cite{au2001estimation}. A linear SDOF oscillator is governed by the following equation
\begin{equation}
    \label{eqn:sdof_gov_eq}
    \Ddot{U} (t) + 2 \zeta \Dot{U} (t) + \omega^2 U (t) = W (t)
\end{equation}
where the natural frequency $ \omega = \SI{7.85}{rad/s} $ ($ \SI{1.25}{Hz} $) and the damping ratio $ \zeta = 0.02 $. The forcing function is a Gaussian white noise excitation, which is generated through the following time discretization
\begin{equation}
    \label{eqn:forcing_function_defn}
    W (t_j) = X_j \quad \text{where } t_j = \left( j \Delta t : j = 1, 2, \dots, d \right)
\end{equation}
where $ \Delta t = 0.1$ s and $ d = 1000$ such that we solve the system for a total of $T = 10$ s.  

The performance function is defined by the probability of exceedance of the threshold $b$ within the 10 s time duration as
\begin{equation}
    \label{eqn:passage_problem_performance_function}
    g (\mathbf{x}) = b - \max_{j \in \left\{ 1, \dots, d \right\}} \left( | U (t_j) | \right)
\end{equation}
Three values of the failure threshold are considered as $ b = 0.20 $, $ 0.26 $, and $ 0.32 $. Using simple MCS with $ 10^7 $ samples, the corresponding failure probabilities were found to be $ 5.569 \times 10^{-2} $, $ 5.283 \times 10^{-3} $, and $ 2.727 \times 10^{-4} $.

Once again, using the design point to define $A_0$ as a hypersphere results in vanishingly small $  \mathcal{P} (A_0) $,
meaning that the design point does not isolate the tail of the distribution. We therefore expect that TSS will lose efficiency.

\begin{figure}[!ht]
\centering
\begin{subfigure}{.45\textwidth}
  \centering
  \includegraphics[width=\linewidth]{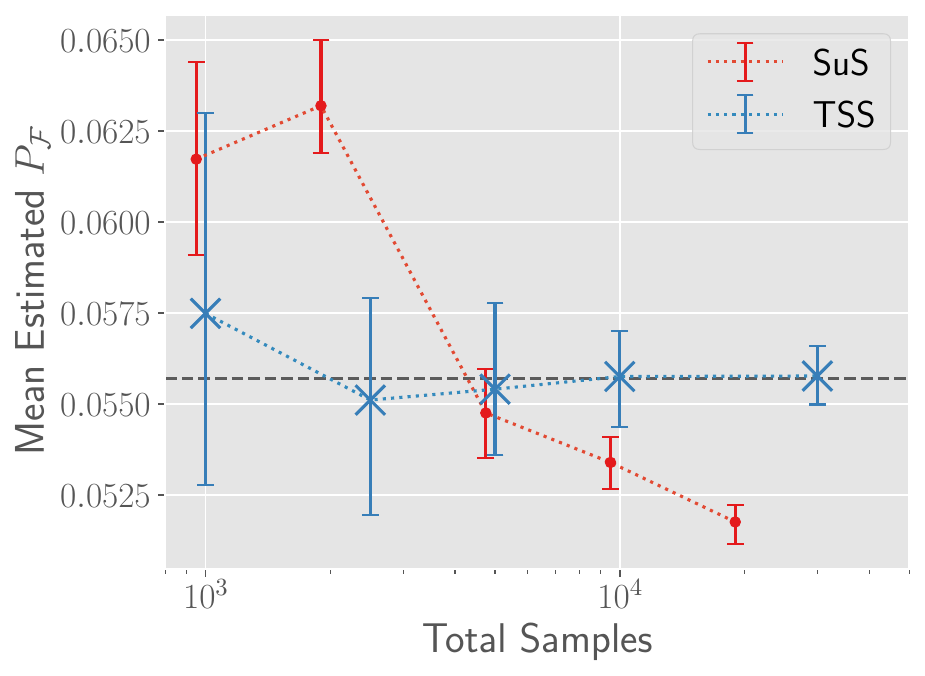}
  \caption{}
  \label{fig:passage_problem_means_b1}
\end{subfigure}%
\begin{subfigure}{.45\textwidth}
  \centering
  \includegraphics[width=\linewidth]{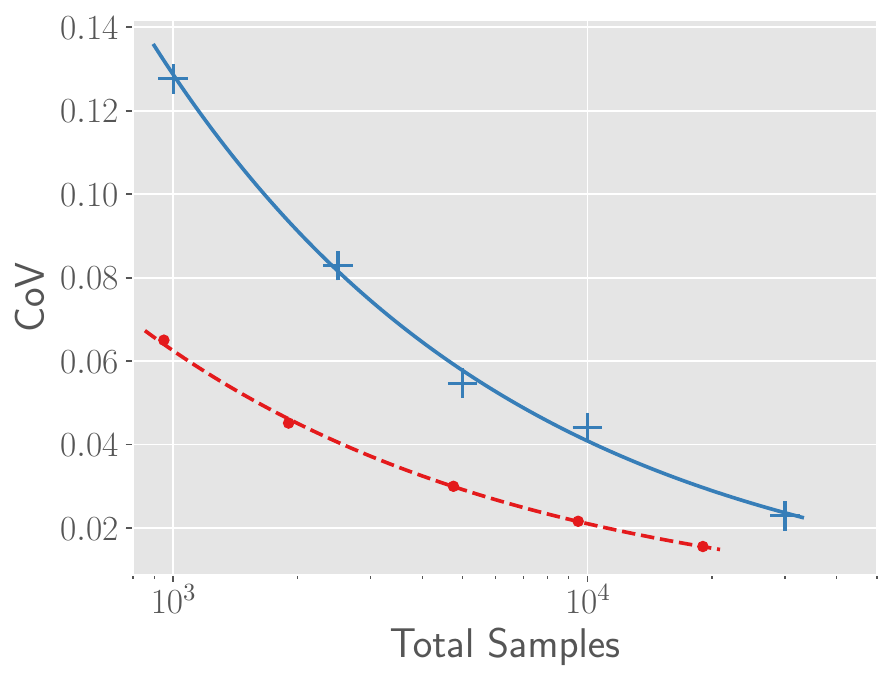}
  \caption{}
  \label{fig:passage_problem_cov_b1}
\end{subfigure}
\begin{subfigure}{.45\textwidth}
  \centering
  \includegraphics[width=\linewidth]{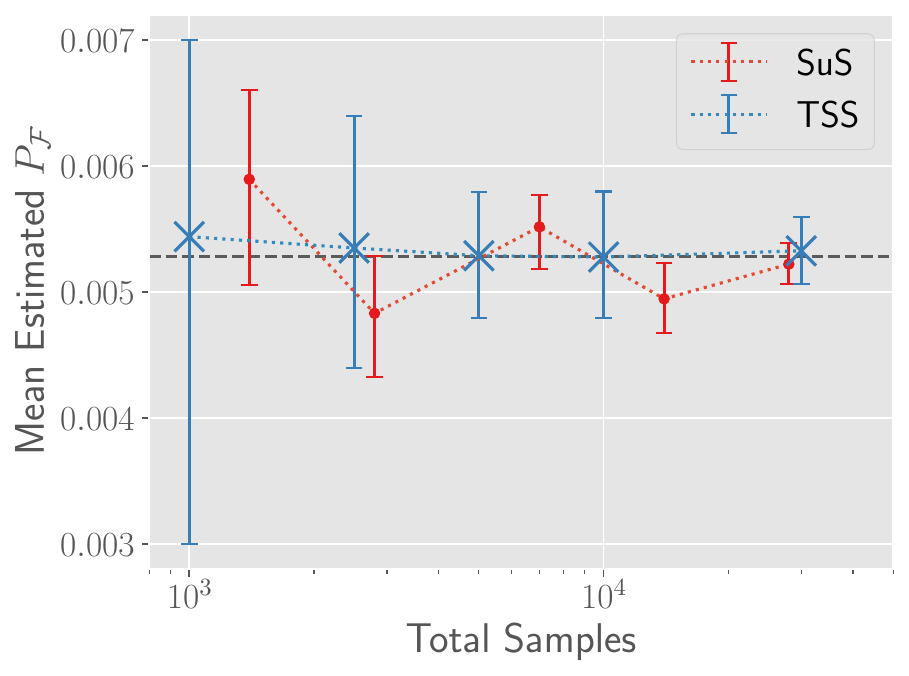}
  \caption{}
  \label{fig:passage_problem_means_b2}
\end{subfigure}%
\begin{subfigure}{.45\textwidth}
  \centering
  \includegraphics[width=\linewidth]{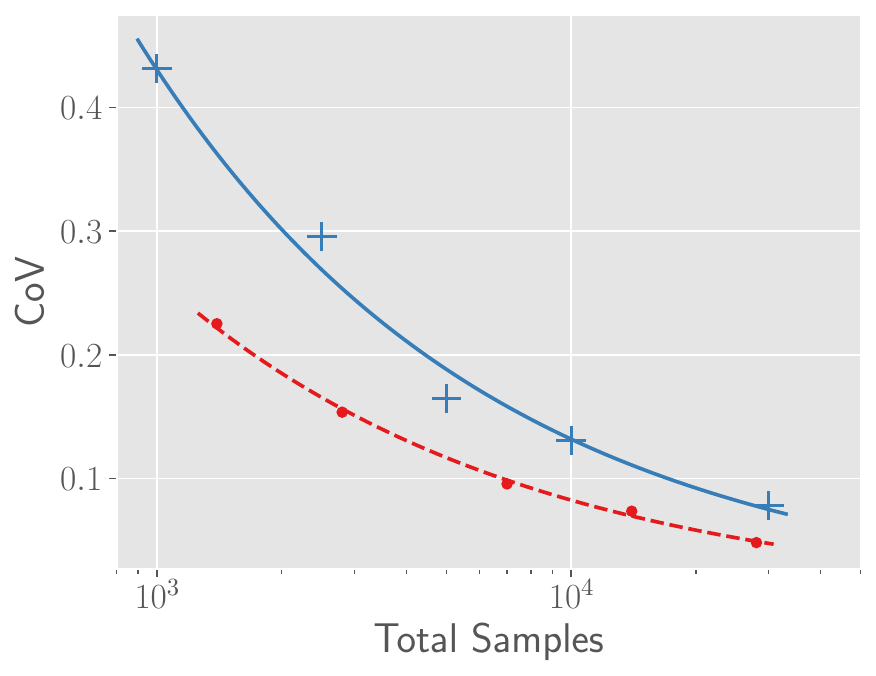}
  \caption{}
  \label{fig:passage_problem_cov_b2}
\end{subfigure}
\begin{subfigure}{.45\textwidth}
  \centering
  \includegraphics[width=\linewidth]{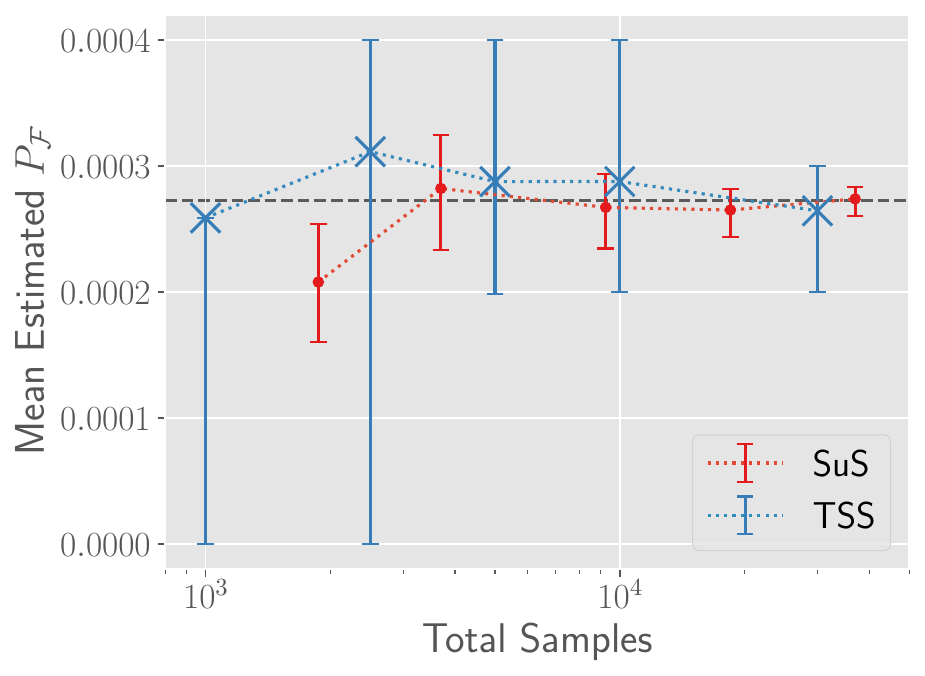}
  \caption{}
  \label{fig:passage_problem_means_b3}
\end{subfigure}%
\begin{subfigure}{.45\textwidth}
  \centering
  \includegraphics[width=\linewidth]{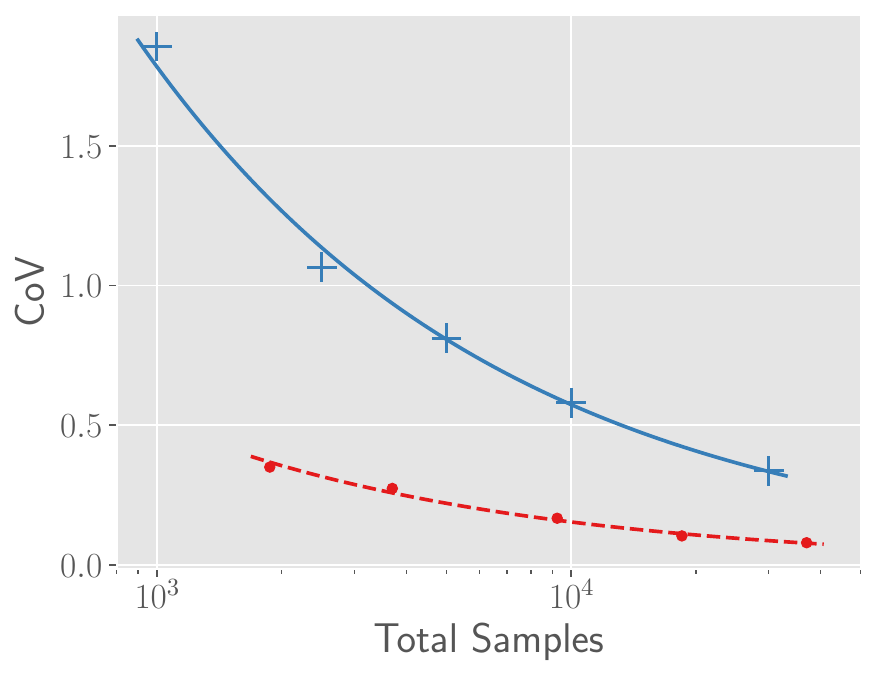}
  \caption{}
  \label{fig:passage_problem_cov_b3}
\end{subfigure}
\caption{SDOF Oscillator: Performance comparison between SuS and TSS. (a, c, e) Mean predicted failure probability computed from $ 100 $ independent trials with error bars depicting the $ 25 $ - $ 75 $ percentile range for increasing sample size and $ b = 0.20, 0.26, 0.32 $. (b, d, f) Coefficient of Variation of the predicted failure probability computed from $ 100 $ independent trials for increasing sample size and $ b = 0.20, 0.26, 0.32 $. The dashed black line in  (a, c, e) represents the true failure probability.}

\label{fig:passage_problem_results}
\end{figure}

As before, we compare $ 100 $ independent trials of TSS and SuS. TSS uses $ p_0 = 0.1 $, $ m = 4 $, and $ N \in \left\{1000, 2500, 5000, 10000, 30000 \right\} $ with simple MCS sampling within each stratum, while SuS is evaluated using $ 500 $, $ 1000 $, $ 2500 $, $ 5000 $, and $ 10000 $ samples per subset. The mean estimated failure probability and sample CoV are plotted in Figure~\ref{fig:passage_problem_results} for all cases.

In this case, TSS mostly gives accurate failure probability predictions and usually has lower bias than SuS, but it has consistently higher CoV. Again, SuS outperforms in this regard because the samples are guided to the failure domain by the performance function, while the design point-based implementation of TSS does not isolate the relevant tails of the distribution. 
However, we once again note that TSS does not suffer from the many drawbacks that other variance reduction methods have, such as a lack of convergence guarantees. We provide a further analysis 
in Section~\ref{section:discussion}.

\subsection{Stranded Wire Cable}
\label{section:corded_rope}

The previous examples illustrate the limitation of TSS in high dimensions when the null stratum is defined by the design point, which does not effectively isolate the tails in high dimensions. 
Here, we present an example in which a different null stratum $ A_0 $ is defined to better isolate the tail and improve performance. 

Consider a steel cable composed of $N_w$  thin strands of steel wire, each having diameter $ \phi_j $ ($ j = 1, \dots, w $) that together resist a total tensile load $P$. 
The performance function can be written as
\begin{equation}
    g(\mathbf{x}) = \left( \sum_{j=1}^{N_w} \frac{\pi}{4} \phi_j^2 \sigma_y \right)  - P, \label{eqn:rope_response}
\end{equation}
where 
\begin{equation}
    \phi_j = \left[ 1 + \varepsilon_\phi X_j \right] \phi_0, \label{eqn:rope_response_diameters} 
\end{equation}
\begin{equation}
    P = \left[ 1 + \varepsilon_P \Phi \left( X_0 \right) \right]P_0, \label{eqn:rope_response_force}
\end{equation}
$ \sigma_y = \SI{250}{MPa} $ denotes the yield strength of steel, $ \phi_0 = \SI{1}{mm} $ is the nominal diameter of each strand, $ P_0 = \SI{193.5}{kN} $ is the nominal applied force, and $ \varepsilon_\phi = 0.1 $ and $ \varepsilon_P = 0.05 $ denote the error magnitudes in the diameters and applied load, respectively. The input vector is once again the uncorrelated standard normal random vector $ \mathbf{X} = \begin{bmatrix} X_0 & X_1 & \dots & X_{N_w} \end{bmatrix}^T $, and $ \Phi (\cdot) $ is the standard normal cdf. 
We initially consider the case where the rope has a known fixed number of strands, i.e., $ N_w = 1000 $. This leads to a $1001$-D random vector and a true failure probability of $ 1.709 \times 10^{-4} $.

As before, we compare $ 100 $ independent trials of TSS and SuS. However, we run two variations of TSS. First, we use the design point to define $ A_0 $. Then, a careful analysis of the performance function allows us to define an alternate $ A_0 $ with desirable geometry. Since the maximum load can be defined by $ P_\text{max} = \left[ 1 + \varepsilon_P \right] P_0 $, we can infer the minimum allowable total cross-sectional area that will guarantee the safety of the cable as $\mathsf{A}_{\min}$.  In this way, we can define the null stratum as the set of all $\mathbf{x}$ such that the total cross-sectional area, $\mathsf{A}_T$, is greater than $\mathsf{A}_{\min}$. That is:
\begin{equation}
\label{eqn:area_criterion}
    A_0 = \left\{ \mathbf{x} \in \Omega : \mathsf{A}_T(\mathbf{x}) > \mathsf{A}_{\min} \right\}
\end{equation}
This may seem trivial, but it highlights the important point that defining the tails for a high-dimensional joint distribution requires insight into the nature of the problem. Defining $A_0$ based on the design point does not produce a null stratum with high probability, but providing some physical insight allows us to define a large null stratum from which the tails can be naturally defined. 

\begin{figure}[!ht]
\centering
\begin{subfigure}{.48\textwidth}
  \centering
  \includegraphics[width=\linewidth]{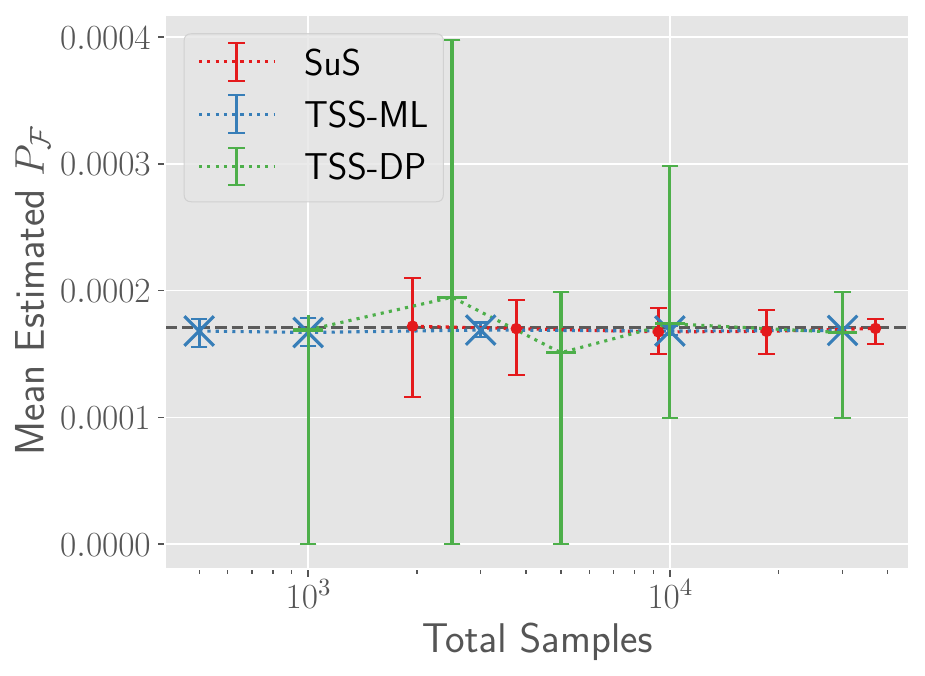}
  \caption{}
  \label{fig:rope_problem_means_all}
\end{subfigure}%
\begin{subfigure}{.48\textwidth}
  \centering
  \includegraphics[width=\linewidth]{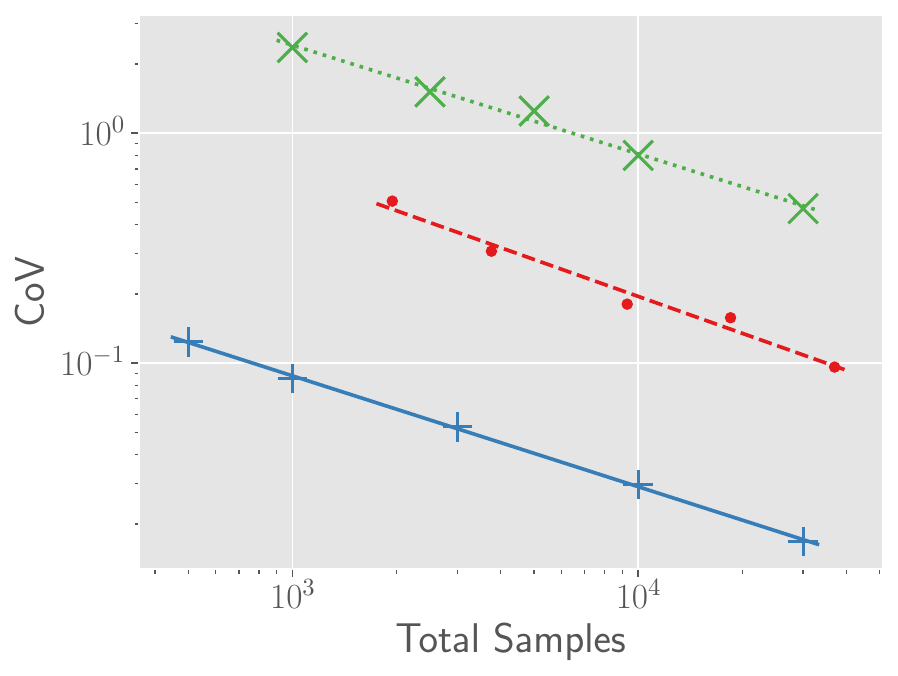}
  \caption{}
  \label{fig:rope_problem_cov_all}
\end{subfigure}
\caption{Stranded Wire Cable: Performance comparison between SuS and TSS. (a) Mean predicted failure probability computed from $ 100 $ independent trials with error bars depicting the $ 25 $ - $ 75 $ percentile range for increasing sample size. (b) Coefficient of Variation of the predicted failure probability computed from $ 100 $ independent trials for increasing sample size. The dashed black line in  (a) represents the true failure probability. TSS-DP denotes the design point based implementation while TSS-ML denotes a maximum load based implementation.}
\label{fig:rope_problem_results}
\end{figure}

The results are plotted in Figure~\ref{fig:rope_problem_results}, with the design point case labeled `TSS-DP' and the maximum load case labeled `TSS-ML'. Aside from the difference in $ A_0 $, both versions of TSS are stratified the same way, using Eqs~\eqref{eqn:general_TSS_stratification_scheme_first_set_definition} and~\eqref{eqn:general_TSS_stratification_scheme_definition}, with $ p_0 = 0.1 $ and $ m = 4 $ with proportional sample allocation. For TSS-DP, the inverse transform method was used in each stratum to generate i.i.d. samples as before. However, since the strata for TSS-ML are hard to define in closed form, Algorithm~\ref{Algo:TSS_strata_construction} was used with rejection sampling to achieve the desired proportional sample allocation with i.i.d. samples. It is important to note here that rejection sampling is possible because only a simple transformation of the input vector is necessary to judge whether it belongs to stratum $ A_i $. Hence, performance function evaluations are not necessary for every candidate state generated by rejection sampling, but only on the final samples retained for TSS.

While TSS-DP remains worse than SuS, similar to the cases in Sections~\ref{section:cantilever_tip_deflection} and~\ref{section:first_passage_problem}, TSS-ML performs significantly better than SuS, and achieves lower than $ 10 \% $ CoV with minimal bias for even tiny sample sizes. Simply by changing the definition of the null stratum, the behavior of TSS is vastly improved in this case, highlighting that even in high dimensions, if it is possible to remove regions of significant probability from the input domain where failure is not expected to occur, such that the conditional probabilities $ P \left( \mathcal{F}_i \right) $ are large, TSS produces highly efficient and accurate estimates of failure. This is in keeping with the discussion from Section~\ref{section:TSS_estimator_properties} where we highlight the lack of explicit dimensional dependence in the closed-form variance of the TSS estimator.

Next, consider a more complicated version of the same problem in which the total number of strands $ N_w $ is not known. This may occur, for example, in a bridge cable where one or more strands have broken due to corrosion~\cite{montoya2015physics}. To further increase the complexity, we also assume that the distribution of $ N_w $ is unknown and is specified only in terms of a range of possible values, i.e. $ N_w \left( \omega \right) \in \left[ N_{w_\text{min}}, N_{w_\text{max}} \right] $. Here, $ \omega $ represents a latent random variable whose distribution can only be inferred from samples. This results in a stochastic performance function
\begin{equation}
\label{eqn:rope_response_stochastic}
    g(\mathbf{x}, \omega) = \left( \sum_{j=1}^{N_w \left( \omega \right)} \frac{\pi}{4} \phi_j^2 \sigma_y \right)  - P, 
\end{equation}
which can be evaluated as a black box; for a given realization $ \mathbf{x} $, having dimension $ N_{w_\text{max}} + 1 $, a sample value of $ N_w (\omega) $ is used to compute the load capacity of the cable. For this example, we take $ N_{w_\text{min}} = 990 $, $ N_{w_\text{max}} = 1000 $, and reduce the nominal load to $ P_0 = \SI{192.25}{kN} $ to maintain a true failure probability of $ 2.587 \times 10^{-4} $. Furthermore, $ g(\mathbf{x}, \omega) $ selects $ N_w \left( \omega \right) $ uniformly at random from the integers between $ 990 $ and $ 1000 $, but this is considered unknown, is stated here for reproducibility purposes only, and not used in any way in the application of TSS. All remaining variables are defined as before. 

\begin{figure}[!ht]
\centering
\begin{subfigure}{.48\textwidth}
  \centering
  \includegraphics[width=\linewidth]{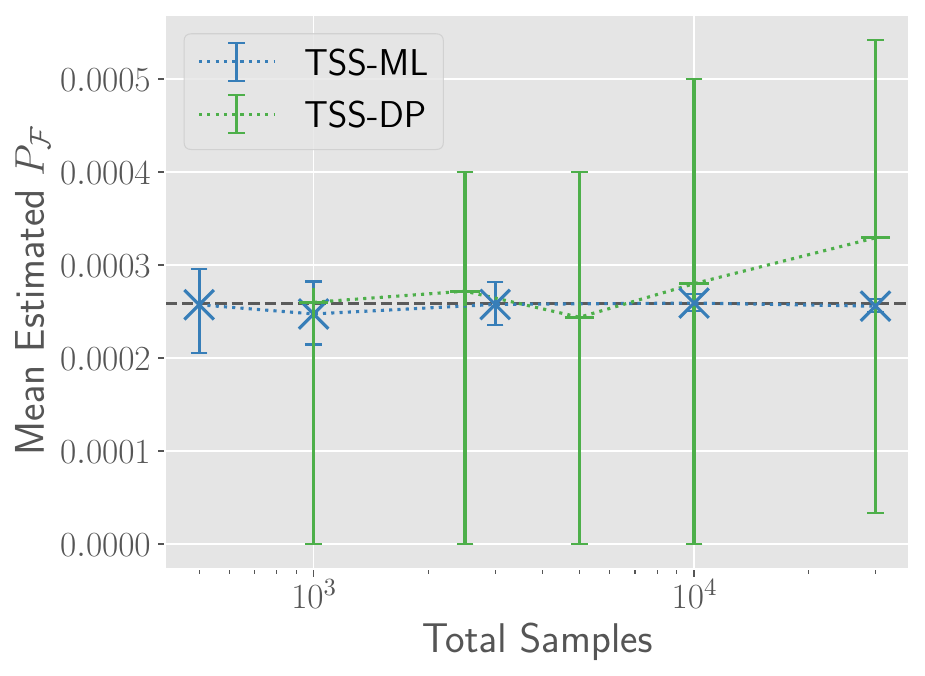}
  \caption{}
  \label{fig:rope_problem_means_all_stochastic}
\end{subfigure}%
\begin{subfigure}{.48\textwidth}
  \centering
  \includegraphics[width=\linewidth]{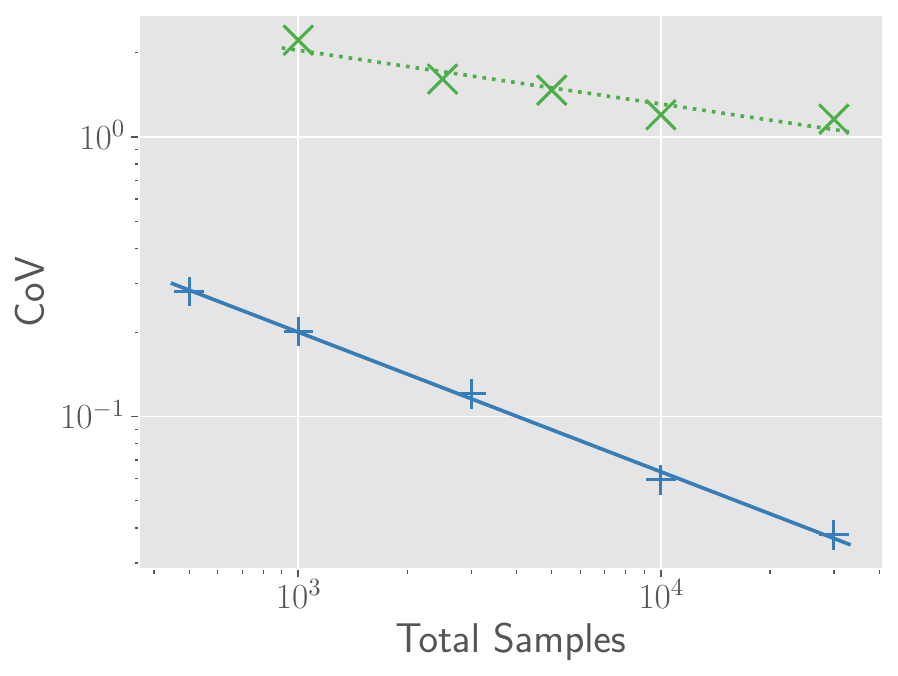}
  \caption{}
  \label{fig:rope_problem_cov_all_stochastic}
\end{subfigure}
\caption{Stochastic Stranded Wire Cable: Performance comparison between TSS-DP and TSS-ML. SuS cannot be used for stochastic performance functions. (a) Mean predicted failure probability computed from $ 100 $ independent trials with error bars depicting the $ 25 $ - $ 75 $ percentile range for increasing sample size. (b) Coefficient of Variation of the predicted failure probability computed from $ 100 $ independent trials for increasing sample size. The dashed black line in  (a) represents the true failure probability. TSS-DP denotes the design point based implementation while TSS-ML denotes a maximum load based implementation.}
\label{fig:rope_problem_stochastic_results}
\end{figure}

Since $ g \left( \mathbf{x}, \omega \right) $ is random even for a given realization of $ \mathbf{x} $, SuS cannot be used for this example. Meanwhile, TSS can be employed without any additional difficulty. Therefore, in Figure~\ref{fig:rope_problem_stochastic_results} we only compare TSS-DP with TSS-ML, to underscore the utility of defining a good $ A_0 $ when possible.  Following the same logic as before, we define $A_0$ using Eq.~\eqref{eqn:area_criterion} such that the total cross-sectional area considers only $N_{w_\text{min}}$ strands.
Here, we see that TSS-ML has at least one order of magnitude lower CoV than TSS-DP, and produces highly accurate failure probability estimates. It is clear that defining the tails based on the design point is not effective (although TSS-DP converges for large sample sizes) and that defining an appropriate tail for the problem can make TSS very efficient.

\section{Discussion, Observations, and Opportunities}
\label{section:discussion}

In Section~\ref{section:examples}, we see the power of TSS to efficiently and accurately estimate small failure probabilities for a variety of problems with complex geometries. 
Moreover, akin to methods like \cite{vovrechovsky2022reliability}, TSS only uses the failure indicator function $ \text{I}_{\Omega_{\mathcal{F}}} (\mathbf{x}) $ and no other information (for example, the specific value of the performance function $ g(\mathbf{x}) $ is not used), which makes it more robust and versatile that many of the other widely-used reliability methods. However, the examples highlight some limitations of TSS as well. In this section, we discuss these benefits and drawbacks of TSS in comparison to the existing foundational reliability algorithms, and provide some potential directions for future improvement.

\subsection{Tail Stratified Sampling (TSS) \& Subset Simulation (SuS)}
\label{section:TSS_Vs_SuS}

Throughout the manuscript, we compare the performance of TSS to that of SuS because SuS is widely considered the benchmark method for estimating failure probability and, like TSS, it also decomposes the small target failure probability into a sequence of larger conditional failure probabilities. Here, we discuss some important takeaways from comparing TSS and SuS.

In SuS, both the variance of the estimator and the total computational budget depend on the failure probability through the number of subsets necessary for estimation. Since this is not known a priori, but rather determined adaptively during the algorithm, it can be difficult to predict the computational cost of SuS. For TSS, the number of strata and the total number of samples can be specified independently of the failure probability. This makes TSS more consistent and easier to implement for a limited computational budget. Furthermore, the conditional distributions in SuS are defined using the performance function $ g (\mathbf{x}) $ (which is usually expensive to evaluate) and can often become highly complex, making them very challenging to sample from. In contrast, the conditional distributions used by TSS are defined purely from the input distribution, which makes them both geometrically interpretable and easier/cheaper to sample because performance function evaluations are not needed for sample generation.
Finally, while SuS requires full performance functions evaluations, TSS uses only the failure indicator function $ \text{I}_{\Omega_{\mathcal{F}}} (\mathbf{x}) $, allowing it to be used when the performance function is unavailable or stochastic,
as demonstrated in Section~\ref{section:corded_rope}.

On the other hand, a benefit of SuS 
is that it is designed to concentrate samples only in regions that are expected to lead to the failure domain. In other words, it is highly exploitative of the performance function. By exploiting the performance function values, SuS sequentially narrows its sampling toward $ \Omega_{\mathcal{F}} $. On the other hand, by only stratifying the tail, TSS instead needs to search each stratum in its entirety for the failure domain, and in its current form has no way to guide itself towards the failure domain within each stratum. Opportunities to integrate SuS-like sequences for conditional failure probability estimates in TSS are potentially promising. As a result, while the conditional failure probabilities in SuS are large by construction (usually chosen to be $ 0.1 $), it is not possible to control $ \mathcal{P} \left( \mathcal{F}_i \right) $ for TSS. This makes selection of the null stratum in TSS crucial because, in the worst pathological cases where $A_0$ is chosen poorly, $ \mathcal{P} \left( \mathcal{F}_i \right) $ might be as difficult to estimate as $ \mathcal{P}_\mathcal{F} $, making TSS inefficient. This is precisely why TSS can become ineffective for high-dimensional problems in which there is little insight into defining a meaningful null stratum. 

In summary, while SuS and TSS use similar geometric decompositions, SuS is a highly exploitative algorithm, while TSS is a highly exploratory algorithm -- with exploration concentrated in the tails. In certain cases, SuS can be too exploitative and fail to sufficiently explore the space to find all failure domains. Meanwhile, TSS can lose efficiency because it does not exploit the structure of the problem except when insights are available to define an appropriate null stratum.  


\subsection{Tail Stratified Sampling (TSS) \& Importance Sampling (IS)}
\label{section:TSS_Vs_IS}

As introduced in Section~\ref{section:Introduction}, IS can be a powerful method for variance reduction, but can be quite difficult to implement because constructing a good importance sampling density (ISD) - an essential step for using IS - can be very challenging, and a poor ISD can lead to unstable and inaccurate estimates of the failure probability. When implemented well, IS nearly eliminates the need to sample from unimportant regions that do not contribute to failure.
In fact, similar to TSS, many IS methods use the design point to inform the construction of the ISD~\cite{au2003important}.

TSS can also be viewed from the perspective of IS. Consider, for example, the case where $ \Omega $ is decomposed only into the null stratum $ A_0 $ and one other stratum $ A_1 = A_* $. In this case, it is easy to see from Eqs.~\eqref{eqn:IS_general_defn} and~\eqref{eqn:IS_monte_carlo_estimator} that the TSS estimator is equivalent to an IS estimator with ISD given by $ q_{\mathbf{X}} \left( \mathbf{x} \right) = f_{\mathbf{X|A_*}} \left( \mathbf{x} | A_* \right) $. Results from Section~\ref{section:examples} suggest that a significant variance reduction occurs due to the removal of $ A_0 $. 
Unfortunately, while the strata for TSS are easy to define in practice (see Sections~\ref{section:TSS_stratification},~\ref{section:strata_boundaries_in_practice}), there is currently no general strategy to construct $ A_* $ that focuses on the failure region. In this way, TSS my suffer similar drawbacks to IS where defining the tail region to sample, $A_*$, is analogous to defining the ISD. This implies a need to develop general strategies for determining an appropriate null stratum, as motivated by the results in Section~\ref{section:corded_rope}.

Additional improvements to TSS may be achieved by employing IS to improve sample allocation, as previously discussed in Section~\ref{section:strata_boundaries_in_practice} and elaborated in Appendix~\ref{app:is_allocation}.

\subsection{Opportunities for Improvements to TSS}
\label{section:TSS_improvements_and_high_d}

The two main drawbacks of TSS are the difficulty of defining meaningful tail strata in high dimensions and the inability to focus samples around the failure domain while sampling within $ A_i $ (i.e., near $ \Omega_{\mathcal{F}} \cap A_i $). 
In many cases, the first limitation can be overcome by recognizing that the failure domain often has a lower-dimensional structure that can be leveraged by stratifying only the relevant marginal tails rather than the full-input distribution. Section~\ref{section:Hi-D} explores this challenge and highlights the need to identify an appropriate and informed null stratum, but does so for a very simple application. In particular, it is emphasized that defining $ A_0 $ (Eq.~\eqref{eqn:design_point_based_safe_set_gaussian}) using the design point is insufficient in high dimensions (at least for independent Gaussian inputs) in keeping with insights about the design point from past works~\cite{katafygiotis2008geometric}. There are abundant opportunities to leverage reduced order models, surrogate models, and data-driven methods to identify transformations and low-dimensional manifolds/subspaces over which the tails can be more effectively defined.


To fully leverage the power of TSS, these methods can be coupled with schemes that focus samples around the conditional failure domains within the tail strata. IS provides a promising path forward where, in the case of independent standard Gaussian inputs, TSS can build upon existing work in Directional Importance Sampling~\cite{moarefzadeh1999directional,grooteman2011adaptive,CHENG2023102291} and Directional Stratified Sampling
~\cite{song2023adaptive,SONG2023102292}. Alternative solutions could also be derived using surrogate modeling approaches. The constructions of Vorechovsky~\cite{vovrechovsky2022reliability} and Hong et al.~\cite{HONG2025102546} may provide guidance for surrogate-driven improvements to TSS, specifically in the case of independent standard Gaussian inputs. Readers may also gain useful insight from~\cite{arunachalamspence} into using alternate system outputs that are correlated with the relevant system performance function to guide the stratification.

\section{Conclusions}
\label{section:conclusions}

In this work, we introduce a stratified sampling approach for efficient rare event reliability analysis that concentrates samples in the tails of the input distribution where failures tend to occur. In doing so, we argue for the need to rigorously define the tails of a distribution as, surprisingly, existing definitions of the tails are notoriously ambiguous. We offer one such definition based on limiting thresholds of the joint density function (see Appendix~\ref{section:TSS_S_uv_theory}). From this definition, the proposed Tail Stratified Sampling (TSS) estimator is derived by defining a sequence of strata that sequentially probe deeper into the tails. 

Key theoretical and algorithmic details that dictate the performance of the TSS method are explored throughout.  
A general TSS stratification scheme is presented, with two special cases highlighted: (1) independent standard Gaussian input variables, and (2) independent uniform input variables. Two versions of the TSS estimator are discussed as well: (1) a truncated estimator with tunable bias that is recommended for practical application, and (2) an unbiased estimator with theoretical guarantees.
Closed-form variance estimators are derived, and various sample allocation schemes are discussed. Of particular note is that TSS allows for the removal of a null stratum where it is known that failure will not occur (through a priori expert knowledge, design point considerations, etc.). Identification of an informative, high-probability null stratum is critical to the efficiency of TSS and is discussed at length.
Practical implementation strategies for finding the initial null stratum, defining the strata boundaries, and sampling from each stratum are also detailed. 

The performance of TSS was compared against Subset Simulation as a well-established benchmark reliability method. For a variety of $ 2 $-D cases with challenging failure domain geometries and varying failure probabilities (ranging from $ 10^{-2} $ to $ 10^{-9} $), TSS was shown to thoroughly outperform SuS in terms of estimator accuracy and stability, achieving at least one order of magnitude lower CoV in all cases. In several of these cases, Subset Simulation either fails altogether or results in unstable predictions while TSS is very accurate and efficient for all of them.
The methods were also compared for a challenging numerical example -- a $ 7 $-D stable buckling problem. We then discuss the application of TSS for high-dimensional problems through three examples: a $ 1000 $-D cantilever beam probem, a $ 1000 $-D SDOF oscillator, and a wire cable composed of 1000 individual wire strands. It is demonstrated through these examples that identifying an informative, high-probability null stratum is essential.  

In the final sections, some observations are made regarding the benefits and drawbacks of TSS in comparison to existing state-of-the-art methods.
Some suggestions on ways to mitigate the recognized limitations are discussed and opportunities for future directions to evolve TSS are made.
Overall, TSS proves to be highly robust and efficient.
With the current landscape of structural reliability methods trending toward more and more complicated, sophisticated, and narrow-scope algorithms, TSS is a reminder of the potential power and near-unparalleled robustness of simple variance-reduction techniques applied to vanilla Monte Carlo settings.

\bibliographystyle{unsrt}  


\begin{thebibliography}{10}

\bibitem{hasofer1974exact}
Abraham~M Hasofer.
\newblock An exact and invarient first order reliability format.
\newblock {\em J. Eng. Mech. Div., Proc. ASCE}, 100(1):111--121, 1974.

\bibitem{bichon2008efficient}
Barron~J Bichon, Michael~S Eldred, Laura~Painton Swiler, Sandaran Mahadevan,
  and John~M McFarland.
\newblock Efficient global reliability analysis for nonlinear implicit
  performance functions.
\newblock {\em AIAA journal}, 46(10):2459--2468, 2008.

\bibitem{echard2011ak}
Benjamin Echard, Nicolas Gayton, and Maurice Lemaire.
\newblock Ak-mcs: an active learning reliability method combining kriging and
  monte carlo simulation.
\newblock {\em Structural safety}, 33(2):145--154, 2011.

\bibitem{sudret2012meta}
Bruno Sudret.
\newblock Meta-models for structural reliability and uncertainty
  quantification.
\newblock {\em arXiv preprint arXiv:1203.2062}, 2012.

\bibitem{marelli2018active}
Stefano Marelli and Bruno Sudret.
\newblock An active-learning algorithm that combines sparse polynomial chaos
  expansions and bootstrap for structural reliability analysis.
\newblock {\em Structural Safety}, 75:67--74, 2018.

\bibitem{hurtado2001neural}
Jorge~E Hurtado and Diego~A Alvarez.
\newblock Neural-network-based reliability analysis: a comparative study.
\newblock {\em Computer methods in applied mechanics and engineering},
  191(1-2):113--132, 2001.

\bibitem{papadopoulos2012accelerated}
Vissarion Papadopoulos, Dimitris~G Giovanis, Nikos~D Lagaros, and Manolis
  Papadrakakis.
\newblock Accelerated subset simulation with neural networks for reliability
  analysis.
\newblock {\em Computer Methods in Applied Mechanics and Engineering},
  223:70--80, 2012.

\bibitem{au2001estimation}
Siu-Kui Au and James~L Beck.
\newblock Estimation of small failure probabilities in high dimensions by
  subset simulation.
\newblock {\em Probabilistic engineering mechanics}, 16(4):263--277, 2001.

\bibitem{au1999new}
Siu-Kui Au and James~L Beck.
\newblock A new adaptive importance sampling scheme for reliability
  calculations.
\newblock {\em Structural safety}, 21(2):135--158, 1999.

\bibitem{bjerager1988probability}
Peter Bjerager.
\newblock Probability integration by directional simulation.
\newblock {\em Journal of Engineering Mechanics}, 114(8):1285--1302, 1988.

\bibitem{olsson2003latin}
Anders Olsson, G{\"o}ran Sandberg, and Ola Dahlblom.
\newblock On latin hypercube sampling for structural reliability analysis.
\newblock {\em Structural safety}, 25(1):47--68, 2003.

\bibitem{papaioannou2015mcmc}
Iason Papaioannou, Wolfgang Betz, Kilian Zwirglmaier, and Daniel Straub.
\newblock Mcmc algorithms for subset simulation.
\newblock {\em Probabilistic Engineering Mechanics}, 41:89--103, 2015.

\bibitem{wang2019hamiltonian}
Ziqi Wang, Marco Broccardo, and Junho Song.
\newblock Hamiltonian monte carlo methods for subset simulation in reliability
  analysis.
\newblock {\em Structural Safety}, 76:51--67, 2019.

\bibitem{shields2021subset}
Michael~D Shields, Dimitris~G Giovanis, and VS~Sundar.
\newblock Subset simulation for problems with strongly non-gaussian, highly
  anisotropic, and degenerate distributions.
\newblock {\em Computers \& Structures}, 245:106431, 2021.

\bibitem{breitung2019geometry}
Karl Breitung.
\newblock The geometry of limit state function graphs and subset simulation:
  Counterexamples.
\newblock {\em Reliability Engineering \& System Safety}, 182:98--106, 2019.

\bibitem{vovrechovsky2022reliability}
Miroslav Vo{\v{r}}echovsk{\`y}.
\newblock Reliability analysis of discrete-state performance functions via
  adaptive sequential sampling with detection of failure surfaces.
\newblock {\em Computer Methods in Applied Mechanics and Engineering},
  401:115606, 2022.

\bibitem{papaioannou2016sequential}
Iason Papaioannou, Costas Papadimitriou, and Daniel Straub.
\newblock Sequential importance sampling for structural reliability analysis.
\newblock {\em Structural safety}, 62:66--75, 2016.

\bibitem{katafygiotis2008geometric}
Lambros~S Katafygiotis and Konstantin~M Zuev.
\newblock Geometric insight into the challenges of solving high-dimensional
  reliability problems.
\newblock {\em Probabilistic Engineering Mechanics}, 23(2-3):208--218, 2008.

\bibitem{au2003important}
Siu-Kui Au and James~L Beck.
\newblock Important sampling in high dimensions.
\newblock {\em Structural safety}, 25(2):139--163, 2003.

\bibitem{wang2016cross}
Ziqi Wang and Junho Song.
\newblock Cross-entropy-based adaptive importance sampling using von
  mises-fisher mixture for high dimensional reliability analysis.
\newblock {\em Structural Safety}, 59:42--52, 2016.

\bibitem{tocher1967art}
Keith~Douglas Tocher.
\newblock {\em The art of simulation}.
\newblock English Universities Press, 1967.

\bibitem{mckay2000comparison}
Michael~D McKay, Richard~J Beckman, and William~J Conover.
\newblock A comparison of three methods for selecting values of input variables
  in the analysis of output from a computer code.
\newblock {\em Technometrics}, 42(1):55--61, 2000.

\bibitem{helton2003latin}
Jon~C Helton and Freddie~Joe Davis.
\newblock Latin hypercube sampling and the propagation of uncertainty in
  analyses of complex systems.
\newblock {\em Reliability Engineering \& System Safety}, 81(1):23--69, 2003.

\bibitem{shields2016generalization}
Michael~D Shields and Jiaxin Zhang.
\newblock The generalization of latin hypercube sampling.
\newblock {\em Reliability Engineering \& System Safety}, 148:96--108, 2016.

\bibitem{shields2015targeted}
Michael~D Shields and VS~Sundar.
\newblock Targeted random sampling: a new approach for efficient reliability
  estimation for complex systems.
\newblock {\em International Journal of Reliability and safety},
  9(2-3):174--190, 2015.

\bibitem{arunachalamspence}
Srinivasan Arunachalam and Seymour M.~J. Spence.
\newblock Generalized stratified sampling for efficient reliability assessment
  of structures against natural hazards.
\newblock {\em Journal of Engineering Mechanics}, 149(7):04023042, 2023.

\bibitem{WANGSong201642}
Ziqi Wang and Junho Song.
\newblock Cross-entropy-based adaptive importance sampling using von
  mises-fisher mixture for high dimensional reliability analysis.
\newblock {\em Structural Safety}, 59:42--52, 2016.

\bibitem{song2023adaptive}
Chenxiao Song and Reiichiro Kawai.
\newblock Adaptive radial importance sampling under directional stratification.
\newblock {\em Probabilistic Engineering Mechanics}, 72:103443, 2023.

\bibitem{grooteman2011adaptive}
Frank Grooteman.
\newblock An adaptive directional importance sampling method for structural
  reliability.
\newblock {\em Probabilistic Engineering Mechanics}, 26(2):134--141, 2011.

\bibitem{lebrun2009innovating}
Regis Lebrun and Anne Dutfoy.
\newblock An innovating analysis of the nataf transformation from the copula
  viewpoint.
\newblock {\em Probabilistic Engineering Mechanics}, 24(3):312--320, 2009.

\bibitem{HONG2025102546}
Fangqi Hong, Jingwen Song, Pengfei Wei, Ziteng Huang, and Michael Beer.
\newblock A stratified beta-sphere sampling method combined with important
  sampling and active learning for rare event analysis.
\newblock {\em Structural Safety}, 112:102546, 2025.

\bibitem{rackwitz1978structural}
R{\"u}diger Rackwitz and Bernd Flessler.
\newblock Structural reliability under combined random load sequences.
\newblock {\em Computers \& structures}, 9(5):489--494, 1978.

\bibitem{liu1991optimization}
Pei-Ling Liu and Armen Der~Kiureghian.
\newblock Optimization algorithms for structural reliability.
\newblock {\em Structural safety}, 9(3):161--177, 1991.

\bibitem{haukaas2006strategies}
Terje Haukaas and Armen Der~Kiureghian.
\newblock Strategies for finding the design point in non-linear finite element
  reliability analysis.
\newblock {\em Probabilistic Engineering Mechanics}, 21(2):133--147, 2006.

\bibitem{brooks2011handbook}
Steve Brooks, Andrew Gelman, Galin Jones, and Xiao-Li Meng.
\newblock {\em Handbook of markov chain monte carlo}.
\newblock CRC press, 2011.

\bibitem{vrugt2009accelerating}
Jasper~A Vrugt, Cajo~JF ter Braak, Cees~GH Diks, Bruce~A Robinson, James~M
  Hyman, and Dave Higdon.
\newblock Accelerating markov chain monte carlo simulation by differential
  evolution with self-adaptive randomized subspace sampling.
\newblock {\em International journal of nonlinear sciences and numerical
  simulation}, 10(3):273--290, 2009.

\bibitem{goodman2010ensemble}
Jonathan Goodman and Jonathan Weare.
\newblock Ensemble samplers with affine invariance.
\newblock {\em Communications in applied mathematics and computational
  science}, 5(1):65--80, 2010.

\bibitem{chakroborty2024intrepid}
Promit Chakroborty and Michael~D Shields.
\newblock Intrepid mcmc: Metropolis-hastings with exploration.
\newblock {\em arXiv preprint arXiv:2411.17639}, 2024.

\bibitem{sharma2023modified}
Adwait Sharma and CS~Manohar.
\newblock Modified replica exchange-based mcmc algorithm for estimation of
  structural reliability based on particle splitting method.
\newblock {\em Probabilistic Engineering Mechanics}, 72:103448, 2023.

\bibitem{ThompsonHuntElasticStability}
J.~M.~T. Thompson and G.~W. Hunt.
\newblock {\em A General Theory of Elastic Stability}.
\newblock John Wiley \& Sons, 1973.

\bibitem{montoya2015physics}
Arturo Montoya, George Deodatis, Raimondo Betti, and Haim Waisman.
\newblock Physics-based stochastic model to determine the failure load of
  suspension bridge main cables.
\newblock {\em Journal of Computing in Civil Engineering}, 29(4):B4014002,
  2015.

\bibitem{moarefzadeh1999directional}
Mohammad~Reza Moarefzadeh and Robert~E Melchers.
\newblock Directional importance sampling for ill-proportioned spaces.
\newblock {\em Structural safety}, 21(1):1--22, 1999.

\bibitem{CHENG2023102291}
Kai Cheng, Iason Papaioannou, Zhenzhou Lu, Xiaobo Zhang, and Yanping Wang.
\newblock Rare event estimation with sequential directional importance
  sampling.
\newblock {\em Structural Safety}, 100:102291, 2023.

\bibitem{SONG2023102292}
Chenxiao Song and Reiichiro Kawai.
\newblock Adaptive stratified sampling for structural reliability analysis.
\newblock {\em Structural Safety}, 101:102292, 2023.

\end{thebibliography}

\appendix

\section{A Formal Definition of the Tail as a Set}
\label{section:TSS_S_uv_theory}

Prior to this point, we have defined the tail only informally. In this section, we formalize this definition through the sets presented in the previous section. We begin by defining the sets $S_{uv}$ as follows. 
\begin{definition}[Definition of $ S_{uv} $]
\label{defn:definition_of_S_uv}
    There are two cases:
    
    \begin{enumerate}
        \item If $ \displaystyle \lim_{\delta u \to 0} \mathcal{V} \left( \left\{ \mathbf{x} \in \Omega : f_{\mathbf{X}} \left( \mathbf{x} \right) \geq u + \delta u \right\} \right) $ exists\footnote{i.e., $ \mathcal{V} \left( \left\{ \mathbf{x} \in \Omega : f_{\mathbf{X}} \left( \mathbf{x} \right) \geq u \right\} \right) = \mathcal{V} \left( \left\{ \mathbf{x} \in \Omega : f_{\mathbf{X}} \left( \mathbf{x} \right) > u \right\} \right) $. Since $ f_{\mathbf{X}} (\mathbf{x}) $ is a density function, stipulating that $ \Omega $ is compact implies $ \displaystyle \lim_{\delta u \to 0^-} \mathcal{V} \left( \left\{ \mathbf{x} \in \Omega : f_{\mathbf{X}} \left( \mathbf{x} \right) \geq u + \delta u \right\} \right) = \mathcal{V} \left( \left\{ \mathbf{x} \in \Omega : f_{\mathbf{X}} \left( \mathbf{x} \right) \geq u \right\} \right) $ and $ \displaystyle \lim_{\delta u \to 0^+} \mathcal{V} \left( \left\{ \mathbf{x} \in \Omega : f_{\mathbf{X}} \left( \mathbf{x} \right) \geq u + \delta u \right\} \right) = \mathcal{V} \left( \left\{ \mathbf{x} \in \Omega : f_{\mathbf{X}} \left( \mathbf{x} \right) > u \right\} \right)$.}, where 
        $ \mathcal{V} \left( W \right) $ is the volume measure of a set $ W \subset \Omega $, then
        \begin{equation}
        \label{eqn:S_uv_when_no_plateau}
            S_{uv} = S_{u0} = \left\{ \mathbf{x} \in \Omega : f_{\mathbf{X}} \left( \mathbf{x} \right) \geq u \right\}
        \end{equation}
        In this case, the auxiliary variable $ v \equiv 0 $, as mentioned previously.
    
        \item Otherwise, when $ \displaystyle \lim_{\delta u \to 0} \mathcal{V} \left( \left\{ \mathbf{x} \in \Omega : f_{\mathbf{X}} \left( \mathbf{x} \right) \geq u + \delta u \right\} \right) $ does not exist, define
        \begin{align}
            S_{u0} &= \lim_{\delta u \to 0^+} \left\{ \mathbf{x} \in \Omega : f_{\mathbf{X}} \left( \mathbf{x} \right) \geq u + \delta u \right\} = \left\{ \mathbf{x} \in \Omega : f_{\mathbf{X}} \left( \mathbf{x} \right) > u \right\} \label{eqn:S_uv_limit_sets_when_plateau_smaller} \\
            S_{u1} &= \lim_{\delta u \to 0^-} \left\{ \mathbf{x} \in \Omega : f_{\mathbf{X}} \left( \mathbf{x} \right) \geq u + \delta u \right\} = \left\{ \mathbf{x} \in \Omega : f_{\mathbf{X}} \left( \mathbf{x} \right) \geq u \right\} \label{eqn:S_uv_limit_sets_when_plateau_larger}
        \end{align}
        and for all intermediate values of $ v \in \left( 0, 1\right) $, $ S_{uv} $ must satisfy the following conditions:
        \begin{enumerate}
            \item $ S_{u0} \subset S_{uv} \subset S_{u1} $
            \item If $ v_1 < v_2 $, then $ S_{uv_1} \subset S_{uv_2} $
            \item $ \displaystyle \lim_{\delta v \to 0} \mathcal{V} \left( S_{u \left( v + \delta v \right) } \right) = \mathcal{V} \left( S_{uv} \right) $
            \item $ \forall W \subset \Omega $ where $ S_{u0} \subset W \subset S_{u1} $ and $ \mathcal{P} \left( W \right) = \mathcal{P} \left( S_{uv} \right) $, $ \mathcal{M} \left( S_{uv} \right) = \displaystyle \min_W \left( \mathcal{M} \left( W \right) \right) $, where $ \mathcal{M} \left( W \right) $ is a measure of the spread of the set $ W $ around a central point within the set.
        \end{enumerate}
    \end{enumerate}
\end{definition}

In regions where the density function $ f_{\mathbf{X}} (\mathbf{x}) $ changes continuously, i.e., $ \displaystyle \lim_{\delta u \to 0} \mathcal{V} \left( \left\{ \mathbf{x} \in \Omega : f_{\mathbf{X}} \left( \mathbf{x} \right) \geq u + \delta u \right\} \right) $ exists, it can easily be shown that the above definition for $ S_{uv} $ leads to sets with the smallest volume for a given probability content. (An informal proof of this result begins by considering a set $ Z \subset S_{u0} $. Any set $ Z' \subset \Omega \setminus S_{u0} $ such that $ \mathcal{P} \left( Z' \right) = \mathcal{P} \left( Z \right) $ must have $ \mathcal{V} \left( Z' \right) > \mathcal{V} \left( Z \right) $ because by definition, $ \forall \mathbf{x} \in Z' $, $ f_{\mathbf{X}} (\mathbf{x}) < u $, while $ \forall \mathbf{x} \in Z $, $ f_{\mathbf{X}} (\mathbf{x}) \geq u $. Thus, $ \mathcal{V} \left( \left( S_{u0} \setminus Z \right) \cup Z' \right) > \mathcal{V} \left( S_{u0} \right) $, for any arbitrary choice of $ Z $ and $ Z' $.) In other words
\begin{equation}
    \label{eqn:volume_optimality_of_S_uv}
    \begin{gathered}
        \forall u \text{ s.t. } \lim_{\delta u \to 0} \mathcal{V} \left( \left\{ \mathbf{x} \in \Omega : f_{\mathbf{X}} \left( \mathbf{x} \right) \geq u + \delta u \right\} \right) \text{ exists, and } W \subset \Omega \text{ s.t. } \mathcal{P} \left( W \right) = \mathcal{P} \left( S_{u0} \right) \\
        \mathcal{V} \left( S_{u0} \right) = \min_{\forall W} \left( \mathcal{V} \left( W \right) \right)
    \end{gathered}
\end{equation}
We then define the tail of the distribution as $ T_{uv}= S_{uv}^C = \Omega \setminus S_{uv} $. The above analysis implies that defining the tails of the distribution in this way is optimal in the sense that $ S_{uv}$ is the set that includes the highest amount of probability in the smallest volume. However, in regions of uniform density all sets of equal volume also have equal probability. As discussed in Section~\ref{section:defining_the_tails_introduction}, we therefore use a measure of distance from a central point in the distribution to construct the measure of spread $ \mathcal{M} (W) $ of the set $ W $, which might potentially (but not necessarily) be of the form
\begin{equation}
\label{eqn:measure_of_spread_template_general}
    \mathcal{M} \left( W \right) = \int_{W} \mathsf{M} (\mathbf{x}) f_{\mathbf{X}} \left( \mathbf{x} \right) d \mathbf{x}
\end{equation}
where $ \mathsf{M} (\mathbf{x}) $ is some general function that prescribes the contribution of a point $ \mathbf{x} \in W $ to the spread of $ W $.

For the sake of generality, we do not prescribe measures of volume, distance, spread, or centrality of the distribution. In Euclidean space, for example, volume is the standard Lebesgue measure, and distance is given by the Minkowski distance, i.e., the $L_k$ norm. Centrality of the distribution is most commonly measured by its mode, although the mean value or other measures of central tendency may be useful in certain contexts.   
Likewise, a measure of spread of the distribution may be given by the generalized $\beta$-order moments of the set. For example,
\begin{equation}
\label{eqn:measure_of_spread_template}
    \mathcal{M} \left( W \right) = \sum_j \lambda_j \left( \mathcal{P} \left( X_W^{(j)} \right) \right) \int_{X_W^{(j)}} \left[ l \left( \mathbf{x}, \mathbf{x}_{c} \right) \right]^{\beta} f_{\mathbf{X}} \left( \mathbf{x} \right) d \mathbf{x}
\end{equation}
Here, $ \left\{ X_W^{(j)} \right\}_{\forall j} $ is a partition of set $ W $ where each $ X_W^{(j)} $ is a connected component of $ W $, $ l \left( \mathbf{x}, \mathbf{x}_{c} \right) $ is a measure of distance from the point $ \mathbf{x} \in X_W^{(j)} $ to a central point $ \mathbf{x}_{c} \in X_W^{(j)} $, 
and $ \lambda_j \left( \mathcal{P} \left( X_W^{(j)} \right) \right) $ is a weight factor
to balance the relative probability contents of the disjoint sets $ X_W^{(j)} $.
By considering each disjoint component individually, the above measure can handle multimodality innately, and incorporating a probability-based penalty term allows for considering different probability concentrations in different modes. Moreover, in the common case of a unimodal symmetric distribution, by selecting $\beta=2$ the measure in Eq.~\eqref{eqn:measure_of_spread_template} simplifies to the second central moment. (Note that a measure of spread of the form in Eq.~\eqref{eqn:measure_of_spread_template} is simply a special case of the more general structure of Eq.~\eqref{eqn:measure_of_spread_template_general}.)

\section{The Unbiased TSS Estimator}
\label{appendix:unbiased_TSS}

In the main body of this manuscript, we discuss a formulation of the TSS estimator that is biased due to being constructed by truncating after a finite number of strata. However, if an unbiased estimator is preferred, it is simple to augment $ {P}_\mathcal{F} $ with additional samples from $ A_t = \left( A_* \cap T_{u_{m-1} v_{m-1}} \right) = \left( A_* \setminus S_{u_{m-1} v_{m-1}} \right) $, treating it as the final strata, i.e.,
\begin{gather}
    P_{\mathcal{F}, \text{ub}}^{\text{TSS}} = \mathcal{P} \left( \Omega_{\mathcal{F}} | A_t \right) \mathcal{P} \left( A_t \right) + \sum_{i=1}^{m-1} \mathcal{P} \left( \Omega_{\mathcal{F}} | A_i \right) \mathcal{P} \left( A_i \right) \label{eqn:unbiased_TSS_estimator} \\
    \Rightarrow \operatorname{Bias} \left( P_{\mathcal{F}, \text{ub}}^{\text{TSS}} , P_{\mathcal{F}} \right) = 0
\end{gather}
This can be simplified somewhat by substituting in Eqs.~\eqref{eqn:general_TSS_stratification_scheme_first_set_definition} and~\eqref{eqn:general_TSS_stratification_scheme_definition}
\begin{equation}
    P_{\mathcal{F}, \text{ub}}^{\text{TSS}} = \mathcal{P} \left( A_* \right) \left[ p_0^{m-1} \mathcal{P} \left( \mathcal{F}_{t} \right) + \left( 1 - p_0 \right) \sum_{i=1}^{m-1} p_0^{i-1} \mathcal{P} \left( \mathcal{F}_i \right) \right] \label{eqn:unbiased_TSS_simplified}
\end{equation}
where $ \mathcal{P} \left( \mathcal{F}_{i} \right) = \mathcal{P} \left( \Omega_{\mathcal{F}} | A_i \right)$ and $ \mathcal{P} \left( \mathcal{F}_{t} \right) = \mathcal{P} \left( \Omega_{\mathcal{F}} | A_t \right) $. By drawing $N_i$ samples independently from each stratum $A_i$ (and $ N_t $ samples from $ A_t $) and applying $\hat{\mathcal{P}}(\mathcal{F}_i) = \dfrac{1}{N_i} \sum_{i=1}^{N_i} \text{I}_{\left\{ g(\mathbf{x}) \leq 0 \right\}} \left( \mathbf{x}_i | A_i\right)$ (similarly estimating $ \hat{\mathcal{P}}(\mathcal{F}_t) $), this leads to the following unbiased TSS estimator
\begin{equation}
    \hat{P}_{\mathcal{F}, \text{ub}}^{\text{TSS}} = \mathcal{P} \left( A_* \right) \left[ p_0^{m-1} \dfrac{1}{N_t} \sum_{k=1}^{N_t}  \text{I}_{\left\{ g(\mathbf{x}) \leq 0 \right\}}\left( \mathbf{x}_k | A_t \right) + \left( 1 - p_0 \right) \sum_{i=1}^{m-1} p_0^{i-1} \dfrac{1}{N_i} \sum_{k=1}^{N_i}  \text{I}_{\left\{ g(\mathbf{x}) \leq 0 \right\}}\left( \mathbf{x}_k | A_i \right) \right] \label{eqn:unbiased_TSS_estimator_simplified}
\end{equation}

The resultant variance estimator, again assuming that i.i.d. samples are drawn from each stratum, is
\begin{equation}
    \operatorname{\mathbb{V}ar} \left[ \hat{P}_{\mathcal{F}, \text{ub}}^{\text{TSS}} \right] = \left[ \mathcal{P} \left( A_* \right) \right]^2 \left[ p_0^{2(m-1)} \frac{\hat{\mathcal{P}} \left( \mathcal{F}_t \right) \left( 1 - \hat{\mathcal{P}} \left( \mathcal{F}_t \right) \right)}{N_t} + \left( 1 - p_0 \right)^2 \sum_{i=1}^{m-1} p_0^{2 (i-1)} \frac{ \hat{\mathcal{P}} \left( \mathcal{F}_i \right) \left( 1 - \hat{\mathcal{P}} \left( \mathcal{F}_i \right) \right) }{N_i} \right] \label{eqn:unbiased_TSS_estimator_variance}
\end{equation}


\section{Importance Sampling-based Sample Allocation for TSS}
\label{app:is_allocation}

We develop here the preliminary structure of an Importance Sampling-based scheme for adaptively defining the stratum boundaries (similar to the procedure from Section~\ref{section:strata_boundaries_in_practice}) while simultaneously achieving a sample allocation with the potential to outperform the proportional sample allocation discussed in Section~\ref{section:sample_Allocation}. The reasons to attempt such a procedure are discussed in Section~\ref{section:discussion}, where we argue that the efficiency of TSS can be further improved by more efficient estimation of $ \mathcal{P} \left( \mathcal{F}_i \right) $, especially in high-dimensional cases.

The procedure begins by imposing an importance sampling distribution on $ \Omega $, with some density function $ q_{\mathbf{X}} (\mathbf{x}) $. Once again, we would like to avoid sampling from $ A_0 $, and since $ A_* $ is known in closed form, we can construct
\begin{align}
    q_{\mathbf{X} | A_*} \left( \mathbf{x} | A_* \right) &= \begin{cases}
        \frac{q_{\mathbf{X}} (\mathbf{x})}{\mathcal{P}_{q} \left( A_* \right)} & \mathbf{x} \in A_* \\
        0 & \mathbf{x} \notin A_*
    \end{cases} \label{eqn:A*_conditional_alternate_distribution} \\
    \text{where } \mathcal{P}_{q} \left( A_* \right) &= \int_{A_*} q_{\mathbf{X}} (\mathbf{x}) d \mathbf{x} \label{eqn:A*_probability_alternate_distribution}
\end{align}
This alternate distribution can now be sampled from, and used (through Importance Sampling (Eqs.~\eqref{eqn:IS_general_defn},~\eqref{eqn:IS_monte_carlo_estimator})) to estimate $ \mathcal{P} (W) $ and $ \mathcal{M} (W) $\footnote{These are the probability measure and measure of spread of the set $ W $, respectively, under the original distribution, with density function $ f_{\mathbf{X}} \left( \mathbf{x} \right) $. See Eq~\eqref{eqn:measure_of_spread_template_general} for the general form of $ \mathcal{M} (W) $.} for any set $ W \subseteq A_* $ as
\begin{align}
    \hat{\mathcal{P}}^{\text{IS}} (W) &= \frac{\mathcal{P}_{q} \left( A_* \right)}{N} \sum_{k=1}^N \text{I}_W (\mathbf{x}_k) \frac{f_{\mathbf{X}} \left( \mathbf{x} \right)}{q_{\mathbf{X}} \left( \mathbf{x} \right)} &\quad &\text{where } \mathbf{x}_k \sim q_{\mathbf{X} | A_*} \left( \mathbf{x} | A_* \right) \label{eqn:probability_estimate_alternate_distribution} \\
    \hat{\mathcal{M}}^{\text{IS}} (W) &= \frac{\mathcal{P}_{q} \left( A_* \right)}{N} \sum_{k=1}^N \mathsf{M} (\mathbf{x}_k) \frac{f_{\mathbf{X}} \left( \mathbf{x} \right)}{q_{\mathbf{X}} \left( \mathbf{x} \right)} &\quad &\text{where } \mathbf{x}_k \sim q_{\mathbf{X} | A_*} \left( \mathbf{x} | A_* \right) \label{eqn:spread_estimate_alternate_distribution}
\end{align}

Thus, $ N $ samples can be drawn from $ q_{\mathbf{X} | A_*} \left( \mathbf{x} | A_* \right) $ and be used to adaptively define the stratification scheme, as described in Algorithm~\ref{Algo:TSS_strata_construction_alternate_sampling_distribution}. This procedure results in epirical defined strata $ \tilde{A}_i $, and a sample allocation of the form
\begin{equation}
    N_i \approx N \frac{\mathcal{P}_q (A_i)}{\mathcal{P}_q (A_*)}
\end{equation}
which guarantees a smaller variance of the resultant TSS estimator than proportional allocation if Eq.~\eqref{eqn:condition_general_allocation_better} is satisfied. 
Further, the spread of samples within each strata also follows $ q_{\mathbf{X}|A_i} \left( \mathbf{x} | A_i \right) $ instead of $ f_{\mathbf{X}|A_i} \left( \mathbf{x} | A_i \right) $, which implies that $ \mathcal{P} \left( \mathcal{F}_i \right) $ is estimated using IS as well. This may lead to further efficiency for well-constructed ISDs, when $ \operatorname{\mathbb{V}ar} \left[ \hat{\mathcal{P}}^{\text{IS}} \left( \mathcal{F}_i \right) \right] < \operatorname{\mathbb{V}ar} \left[ \hat{\mathcal{P}} \left( \mathcal{F}_i \right) \right] $.

If explicit strata bounds are desired, they can be defined by placing thresholds on the functions $ f_{\mathbf{X}} \left( \mathbf{x} \right) $ and $ \mathsf{M} \left( \mathbf{x} \right) $. (This is true for the standard procedure in Section~\ref{section:strata_boundaries_in_practice} as well.) Let $ \mathbf{x}_{f,i} $ and $ \mathbf{x}_{l,i} $ be the first and last samples in $ \tilde{A}_i $, respectively, when the samples are placed in monotonic increasing order according to the conditions discussed in Section~\ref{section:strata_boundaries_in_practice}. Then, the bounds for $ A_i $ are as follows, with the inner bound for $ A_1 $ being controlled by the boundary of $ A_* $ and the outer bound for $ A_m $ being $ f_{\mathbf{X}} \left( \mathbf{x} \right) > f_{\mathbf{X}} \left( \mathbf{x}_{l, m} \right) - \epsilon $ for some small $ \epsilon > 0 $.
\begin{enumerate}
    \item Inner bound:
    \begin{equation}
        \begin{cases}
            f_{\mathbf{X}} \left( \mathbf{x} \right) < \frac{f_{\mathbf{X}} \left( \mathbf{x}_{l, i-1} \right) + f_{\mathbf{X}} \left( \mathbf{x}_{f, i} \right)}{2} & \text{if } f_{\mathbf{X}} \left( \mathbf{x}_{l, i-1} \right) \neq f_{\mathbf{X}} \left( \mathbf{x}_{f, i} \right) \\
            f_{\mathbf{X}} \left( \mathbf{x} \right) < f_{\mathbf{X}} \left( \mathbf{x}_{f, i} \right)  \; , \; \mathsf{M} \left( \mathbf{x} \right) > \frac{\mathsf{M}_{\mathbf{X}} \left( \mathbf{x}_{l, i-1} \right) + \mathsf{M}_{\mathbf{X}} \left( \mathbf{x}_{f, i} \right)}{2} & \text{otherwise}
        \end{cases}
    \end{equation}
    \item Outer bound:
    \begin{equation}
        \begin{cases}
            f_{\mathbf{X}} \left( \mathbf{x} \right) \geq \frac{f_{\mathbf{X}} \left( \mathbf{x}_{l, i} \right) + f_{\mathbf{X}} \left( \mathbf{x}_{f, i+1} \right)}{2} & \text{if } f_{\mathbf{X}} \left( \mathbf{x}_{l, i} \right) \neq f_{\mathbf{X}} \left( \mathbf{x}_{f, i+1} \right) \\
            f_{\mathbf{X}} \left( \mathbf{x} \right) \geq f_{\mathbf{X}} \left( \mathbf{x}_{l, i} \right)  \; , \; \mathsf{M} \left( \mathbf{x} \right) < \frac{\mathsf{M}_{\mathbf{X}} \left( \mathbf{x}_{l, i} \right) + \mathsf{M}_{\mathbf{X}} \left( \mathbf{x}_{f, i+1} \right)}{2} & \text{otherwise}
        \end{cases}
    \end{equation}
\end{enumerate}

\begin{algorithm}[!ht]
\caption{Adaptive Stratification Construction for TSS with Alternative Sample Allocation}
\label{Algo:TSS_strata_construction_alternate_sampling_distribution}
\begin{codefont}
\begin{algorithmic}[1]
\Require $ f_{\mathbf{X}} \left(\mathbf{x} \right) $, $ A_* $, $ \mathcal{P} (A_*) $, $ f_{\mathbf{X}|A_*} \left(\mathbf{x} | A_* \right) $, $ p_0 $, $ m $, $ N $, $ q_{\mathbf{X}} \left(\mathbf{x} \right) $, $ q_{\mathbf{X}|A_*} \left(\mathbf{x} | A_* \right) $
\For{k = 1:$N$}
\State Draw $ \mathbf{x}_k \sim q_{\mathbf{X}|A_*} \left(\mathbf{x} | A_* \right) $
\State Evaluate $ f_{\mathbf{X}} \left( \mathbf{x}_k \right) $,  $ \mathsf{M} \left( \mathbf{x}_k \right) $, and $ q_{\mathbf{X}} \left( \mathbf{x}_k \right) $
\EndFor
\State Re-index the samples $ \mathbf{x}_k $ to be in monotone increasing order such that for all $ 1 \leq i < j \leq N $, $ f_{\mathbf{X}} \left(\mathbf{x}_i \right) > f_{\mathbf{X}} \left(\mathbf{x}_j \right) $ or $ \displaystyle \mathsf{M} (\mathbf{x}_i) \frac{f_{\mathbf{X}} \left( \mathbf{x}_k \right)}{q_{\mathbf{X}} \left( \mathbf{x}_k \right)} < \mathsf{M} (\mathbf{x}_j) \frac{f_{\mathbf{X}} \left( \mathbf{x}_k \right)}{q_{\mathbf{X}} \left( \mathbf{x}_k \right)} $ if $ f_{\mathbf{X}} \left(\mathbf{x}_i \right) = f_{\mathbf{X}} \left(\mathbf{x}_j \right) $
\State Set $ k = 0 $
\For{i=1:$m$}
\State Initialize $ \tilde{A}_i \gets \emptyset $
\While{$ \displaystyle \sum_{\mathbf{x} \in \tilde{A}_i} \frac{f_{\mathbf{X}} \left( \mathbf{x}_k \right)}{q_{\mathbf{X}} \left( \mathbf{x}_k \right)} < N \left( 1 - p_0 \right) p_0^{i-1} $}
\State $ \tilde{A}_i \gets \tilde{A}_i \cup \left\{ \mathbf{x}_k \right\} $
\State $ k \gets k+1 $
\EndWhile
\EndFor
\end{algorithmic}
\end{codefont}
\end{algorithm}

\end{document}